\documentclass[prd,aps,twocolumn,a4paper,floatfix,superscriptaddress]{revtex4}

\usepackage{graphicx,psfrag}
\usepackage{mathrsfs}
\usepackage{amsmath,amsfonts,amssymb}
\usepackage{ifpdf}
\usepackage{ulem}
\usepackage{comment}
\usepackage{enumitem}

\def\p{\partial}
\def\Lie{{\cal L}}

\ifpdf
\fi

\usepackage{color}

\definecolor{gray}{rgb}{0.8,0.8,0.8}
\definecolor{cyan}{rgb}{0,0.9,0.9}
\definecolor{orange}{rgb}{0.9,0.5,0}
\definecolor{magenta}{rgb}{1,0,1}
\definecolor{purple}{rgb}{0.8,0.4,0.8}

\begin{document}

\title{Compact binary evolutions with the Z4c formulation}

\author{David Hilditch}
\affiliation{Theoretical Physics Institute, University of
Jena, 07743 Jena, Germany}

\author{Sebastiano Bernuzzi}
\affiliation{Theoretical Physics Institute, University of
Jena, 07743 Jena, Germany}

\author{Marcus Thierfelder}
\affiliation{Theoretical Physics Institute, University of
Jena, 07743 Jena, Germany}

\author{Zhoujian Cao}
\affiliation{Institute of Applied Mathematics, Academy of
Mathematics and Systems Science, Chinese Academy of Sciences,
Beijing 100190, China}

\author{Wolfgang Tichy}
\affiliation{Department of Physics, Florida Atlantic University,
Boca Raton, FL 33431 USA}

\author{Bernd Br\"ugmann}
\affiliation{Theoretical Physics Institute, University of
Jena, 07743 Jena, Germany}

\date{\today}

\begin{abstract}
Numerical relativity simulations of compact binaries with the Z4c and
BSSNOK formulations are compared. The Z4c formulation is advantageous
in every case considered. In simulations of non-vacuum spacetimes the 
constraint violations due to truncation errors are between one and three 
orders of magnitude lower in the Z4c evolutions. Improvements are also 
found in the accuracy of the computed gravitational radiation. For 
equal-mass irrotational binary neutron star evolutions we find that the 
absolute errors in phase and amplitude of the waveforms can be up to a 
factor of four smaller. The quality of the Z4c numerical data is also 
demonstrated by a remarkably accurate computation of the ADM mass from 
surface integrals.
For equal-mass non-spinning binary puncture black hole evolutions we
find that the absolute errors in phase and amplitude of the waveforms
can be up to a factor of two smaller. In the same evolutions we find
that away from the punctures the Hamiltonian constraint violation is
reduced by between one and two orders of magnitude.
Furthermore, the utility of gravitational radiation controlling, 
constraint preserving boundary conditions for the Z4c formulation is 
demonstrated. The evolution of spacetimes containing a
single compact object confirm earlier results in spherical symmetry. The
boundary conditions avoid spurious and non-convergent effects present in
high resolution runs with either formulation with a more naive boundary
treatment.
We conclude that Z4c is preferable to BSSNOK for the numerical solution 
of the 3+1 Einstein equations with the puncture gauge.
\end{abstract}

\maketitle

\bigskip

\section{Introduction}
\label{sec:intro}

This is the conclusion of a series of
papers~\cite{BerHil09,RuiHilBer10,WeyBerHil11,CaoHil11,HilRui12}
about the development of a formulation of general relativity (GR),
called Z4c, that attempts to combine the strengths of two popular
evolution systems for applications in free-evolution numerical
relativity. Here we summarize the logical development of the
formulation.

In the BSSNOK formulation~\cite{BauSha98,ShiNak95,NakOohKoj87} there
is a zero-speed characteristic variable in the constraint subsystem,
which can result in large Hamiltonian constraint violations in numerical
applications; the removal of this mode is one of the key advantages
of the generalized harmonic formulation~\cite{Fri86,Gar01,Pre04,LinSchKid05} 
over BSSNOK. The generalized harmonic formulation also possesses a 
constraint damping scheme~\cite{GunGarCal05}, which exponentially 
damps away small, high-frequency constraint violations at the continuum 
level. Furthermore the trivial wave-like nature of the generalized harmonic 
subsystem allows for the convenient construction of constraint preserving 
boundary conditions~\cite{KreWin06,Rin06,RuiRinSar07}. On the other 
hand, the key advantages in the BSSNOK formulation are the choice of conformal 
variables, and the fact that the formulation does not come tied to 
a particular gauge, which allows for the selection of the moving 
puncture gauge~\cite{BonMasSei94a,Alc02,BakCenCho05,CamLouMar05,
MetBakKop06,GunGar06}. The combination of these two strengths allows 
for the evolution of black holes represented by coordinate singularities 
on the grid without severe numerical difficulties. For BSSNOK radiation 
controlling constraint preserving boundary conditions have been 
proposed~\cite{NunSar09}, but to our knowledge have not been successfully 
used in numerical applications. With these considerations in mind a 
conformal decomposition of the Z4 formulation~\cite{BonLedPal03,
BonPal04,BonLedPal03a,BonLedLuq04a,AliBonBon08,BonBonPal10,BonBon10} was
proposed in~\cite{BerHil09}. Because of the close relationship between
the Z4 and generalized harmonic formulations, this conformal
decomposition inherits all of the strengths outlined in the previous
discussion. In~\cite{BerHil09} a set of spherically symmetric (1D)
tests involving single black hole and neutron star spacetimes demonstrated 
that, indeed, the Z4c formulation guarantees the robustness of, and 
better constraint preservation than BSSNOK,  especially for non-vacuum 
spacetimes. It was also found that the main advantage of Z4c in the bulk 
of the computational domain, namely the propagating constraint violations, 
presents problems at the outer boundary if the constraints are not 
absorbed but rather reflected.

A first attempt to tackle this problem was presented
in~\cite{RuiHilBer10}, in which high derivative order constraint
preserving boundary conditions based on those of~\cite{RuiRinSar07}
were proposed for Z4c. Well-posedness of the Z4c constraint
subsystem initial boundary value problem was demonstrated, and numerical
results in explicit spherical symmetry demonstrated the efficacy of the
constraint preserving boundary conditions. In this work we use 
constraint preserving boundary conditions, motivated by 
a forthcoming study~\cite{HilRui12}, in which well-posedness of high 
order boundary conditions that are constraint preserving and control 
incoming gravitational radiation~\cite{BucSar06,BucSar07}, is analyzed.

In~\cite{WeyBerHil11}, the performance of the Z4 constraint damping
scheme of~\cite{GunGarCal05} applied to Z4c in black hole and neutron
star spacetimes was studied in detail in spherical symmetry. The
constraint damping scheme is effective, as expected, in the non-linear
system provided that the constraint violation is sufficiently small and
resolved on the numerical grid; in the case of grid noise the combination
of artificial dissipation and damping helps to suppress constraint
violations. But it was found that care should be taken in the choice
of damping parameters. Success in spherical tests does not necessarily 
guarantee that of 3D simulations. Preliminary simulations with Z4c and
Sommerfeld BCs in 3D with pure box in box mesh refinement showed poor 
behavior at the outer boundaries in long evolutions. Therefore to try and 
bridge the gap between the evolutions in spherical symmetry and full 
applications in astrophysical spacetimes, numerical stability of Z4c in 
3D evolutions was studied in~\cite{CaoHil11}, where numerical stability of 
the linearized Z4c system coupled to the puncture gauge with a novel 
discretization was shown, and numerical evidence from the 
Apples-with-Apples~\cite{AlcAllBon03,BabHusAli08,BoyLinPfe06}
tests was presented with both the standard and the novel discretization.
The importance of algebraic constraint projection was highlighted, and 
limitations of the puncture gauge for applications in cosmology were 
observed. 

The first application of Z4c was the study of the end-state of
a collapsing neutron star with the puncture gauge~\cite{ThiBerHil10} in
1D and 3D simulations of spherical configurations. The puncture gauge
handles collapsing matter without the need for (matter or metric) excision
because, during the process, the shift condition pushes the spatial
coordinates off of the matter region. This study was followed up with a
similar discussion of dust collapse in~\cite{StaBauBro11}.

A variation of the conformal decomposition, CCZ4, was proposed
in~\cite{AliBonBon11}. The difference between Z4c and CCZ4 is that
CCZ4 includes parametrized constraint addition that, for some choice,
correspond to the original four-covariant Z4 formulation. In the generalized 
harmonic formulation it is known that the inclusion of these non-principal 
terms can be problematic in some test cases~\cite{BabSziWin06}, but 
constraint growth can be mitigated by the use of constraint damping, which 
is the approach of~\cite{AliBonBon11}. For a particular choice of the 
constraint addition parameters {\it not} corresponding to the 
four-covariant Z4 system, evolutions of binary black hole spacetimes 
were presented and shown to reduce Hamiltonian constraint violation, by 
a factor of around four or five (see Fig.~4 of~\cite{AliBonBon11}), relative 
to such simulations for the BSSNOK formulation. On the other hand, in the 
Z4c system non-principal constraint addition that makes the equations of 
motion as close as possible to those of the BSSNOK formulation whilst 
still obtaining the desired PDE properties in the constraint subsystem 
is chosen. It would be interesting to know in more generality how the 
addition of non-principal constraints affects their evolution. 

In this paper we present the first long-term 3D evolutions of black hole
and neutron star binaries with the Z4c formulation. In section~\ref{sec:z4c}
the Z4c equations of motion and boundary conditions are presented. We then
discuss, in section~\ref{sec:numerics}, changes to the BAM numerical code
since~\cite{BruGonHan06}. In particular we describe the implementation of
spherical patches for the wave zone~\cite{PolReiSch11} and the radiation
controlling constraint preserving conditions of~\cite{HilRui12}. In
sections~\ref{sec:single} and~\ref{sec:bin} we present our simulations of
single and binary compact objects, respectively. Appendices~\ref{app:shells} 
and~\ref{app:teukolsky} contain, respectively, descriptions of the spherical 
patch implementation and evolution of Teukolsky waves, the latter of which 
we use for code validation. We conclude in section~\ref{sec:conc}.

We use units~$G=c=1$ throughout, unless otherwise stated.

\section{The Z4c equations of motion}
\label{sec:z4c}

In this section we summarize for completeness the Z4c equations of
motion, constraints, and boundary conditions. The evolved quantities
of the formulation are the conformal spatial 
metric~$\tilde{\gamma}_{ij}$, the lapse~$\alpha$, the shift vector~$\beta^i$, 
the conformal tracefree part of the extrinsic curvature~$\tilde{A}_{ij}$, 
the constraint~$\Theta$ and, up to constraint addition, the trace of 
the extrinsic curvature~$\hat{K}=K-2\Theta$. Finally we evolve the 
the conformal contracted Christoffel symbols~$\tilde{\Gamma}^i$, which 
are initially set according to~$\tilde{\Gamma}^i=-\p_j\tilde{\gamma}^{ij}$.

\paragraph*{Evolution equations.} The equations of motion for the
Z4c formulation are
\begin{align}
\p_t \chi &= \frac{2}{3}\,\chi\,
\left[\alpha\,(\hat{K}+2\Theta) - D_i\beta^i\right]\,,\\
\p_t \tilde{\gamma}_{ij} &= -2\,\alpha\,\tilde{A}_{ij}+\beta^k\p_k
\tilde{\gamma}_{ij}+2\,\tilde{\gamma}_{k(i}\p_{j)}\beta^k\nonumber\\
&\quad-\frac{2}{3}\,\tilde{\gamma}_{ij}\p_k\beta^k\,,
\end{align}
for the metric components,
\begin{align}
\p_t \hat{K}&= -D^iD_i\alpha + \alpha\,\left[\tilde{A}_{ij}\tilde{A}^{ij}
+\frac{1}{3}(\hat{K}+2\Theta)^2\right]\nonumber\\
&+4\,\pi\,\alpha\,\left[S+\rho\,\right]
+\alpha\,\kappa_1\,(1-\kappa_2)\,\Theta+\beta^i\p_i\hat{K}\,,\\
\p_t \tilde{A}_{ij} &= \chi[-D_iD_j\alpha
+\alpha\,(R_{ij}-8\,\pi\,S_{ij})]^{\textrm{tf}}\nonumber\\
& +\alpha\,\left[(\hat{K}+2\,\Theta)\tilde{A}_{ij}-
2\,\tilde{A}^k{}_i\tilde{A}_{kj}\right]\nonumber\\
& +\beta^k\,\p_k\tilde{A}_{ij}+2\,\tilde{A}_{k(i}\,\p_{j)}\beta^k
-\frac{2}{3}\,\tilde{A}_{ij}\,\p_k\beta^{k}\,,
\end{align}
for the extrinsic curvature components and
\begin{align}
\p_t\tilde{\Gamma}^{i} &= -2\,\tilde{A}^{ij}\,\p_j\alpha+2\,\alpha
\left[\tilde{\Gamma}^i{}_{jk}\,\tilde{A}^{jk}
-\frac{3}{2}\,\tilde{A}^{ij}\,\p_j\ln(\chi)\right.
\nonumber\\
&\left.-\frac{1}{3}\,\tilde{\gamma}^{ij}\,\p_j(2\,\hat{K}+\Theta)
-8\,\pi\,\tilde{\gamma}^{ij}\,S_j\right]
+\tilde{\gamma}^{jk}\,\p_j\p_k\beta^i\nonumber\\&
+\frac{1}{3}\,\tilde{\gamma}
^{ij}\p_j\p_k\beta^k+\beta^j\,\p_j\tilde{\Gamma}^i
-(\tilde{\Gamma}_{\textrm{d}})^j\,\p_j\beta^i\nonumber\\
&+\frac{2}{3}\,(\tilde{\Gamma}_{\textrm{d}})^i\,\p_j\beta^j
-2\,\alpha\,\kappa_1\,\big[\tilde{\Gamma}^i
-(\tilde{\Gamma}_{\textrm{d}})^i\big]\,,\\
\p_t\Theta &=\frac{1}{2}\,\alpha\,\big[R - 
\tilde{A}_{ij}\,\tilde{A}^{ij}
+\frac{2}{3}\,(\hat{K}+2\,\Theta)^2\big]\nonumber\\
&-\alpha\,\big[8\,\pi\,\rho+\kappa_1\,(2+\kappa_2)\,
\Theta\big]+\beta^i\p_i\Theta\,,
\end{align}
for the remaining variables. Here the intrinsic curvature 
associated with the ADM metric~$\gamma_{ij}=\chi^{-1}\tilde{\gamma}_{ij}$ is 
written as
\begin{align}
R_{ij} &= R^{\chi}{}_{ij} + \tilde{R}_{ij},\\
\tilde{R}^{\chi}{}_{ij}&=
\frac{1}{2\chi}\tilde{D}_i\tilde{D}_j\chi+\frac{1}{2\chi}
\tilde{\gamma}_{ij}\tilde{D}^l\tilde{D}_l\chi\nonumber\\
&-\frac{1}{4\chi^2}\tilde{D}_i\chi\tilde{D}_j\chi-\frac{3}{4\chi^2}
\tilde{\gamma}_{ij}\tilde{D}^l\chi\tilde{D}_l\chi,\\
\label{eqn:conf_curv}
\tilde{R}_{ij} &=
 - \frac{1}{2}\tilde{\gamma}^{lm}\p_l\p_m\tilde{\gamma}_{ij}
+\tilde{\gamma}_{k(i}\p_{j)}\tilde{\Gamma}^k+(\tilde{\Gamma}_{\textrm{d}})^k
\tilde{\Gamma}_{(ij)k}\nonumber\\
&+\tilde{\gamma}^{lm}\left(2\tilde{\Gamma}^k{}_
{l(i}\tilde{\Gamma}_{j)km}+\tilde{\Gamma}^k{}_{im}
\tilde{\Gamma}_{klj}\right),
\end{align}
and we employ the shorthand
\begin{align}
(\tilde{\Gamma}_{\textrm{d}})^i&=\tilde{\gamma}^{jk}\tilde{\Gamma}^i{}_{jk}.
\end{align}
The derivative operator~$D_i$ is that compatible with the ADM metric.
Numerical evolutions are performed with a particular flavor of the
puncture gauge~\cite{BonMasSei94a,Alc02,MetBakKop06,GunGar06}
\begin{align}
\label{eqn:BMlapse}
\p_t\alpha&=-\alpha^2\mu_L\hat{K}+\beta^i\p_i\alpha,\\
\label{eqn:Gamdriver}
\p_t\beta^i&=\alpha^2\mu_S\tilde{\Gamma}^i-\eta\beta^i
+\beta^j\p_j\beta^i\,.
\end{align}

\paragraph*{Constraints.} The system is subject to constraints
\begin{align}
\Theta&=0,\quad\quad\quad
2\tilde{Z}^i=\tilde{\Gamma}^i-(\tilde{\Gamma}_{\textrm{d}})^i=0,\\
H&=R+\tilde{A}_{ij}\tilde{A}^{ij}-\tfrac{2}{3}(\hat{K}+2\Theta)^2-16\pi\rho=0,\\
\tilde{M}^i&=\p_j\tilde{A}^{ij}+\tilde{\Gamma}^i{}_{jk}\tilde{A}^{jk}
-\tfrac{2}{3}\tilde{\gamma}^{ij}\p_j(\hat{K}+2\Theta)\nonumber\\
&-\tfrac{2}{3}\tilde{A}^{ij}\p_j(\log\chi)
-8\pi\tilde{\gamma}^{ij}S_j=0,\\
\ln(\det\tilde{\gamma})&=0,\quad\quad\quad
\tilde{\gamma}^{ij}\tilde{A}_{ij}=0,\label{eq:algebraic}
\end{align}
of which the latter two, the algebraic constraints, are explicitly imposed
in numerical integration.

\paragraph*{Boundary conditions.} When the spacetime manifold has
a smooth boundary with spacelike unit (with respect to the ADM
metric) normal~$s^i$ we choose for the trace of the extrinsic 
curvature the boundary condition
\begin{align}
\label{eq:lapse_BC}
\p_t\hat{K}\,&\hat{=}-\alpha\,\sqrt{\mu_L}\,
\big(\p_s\hat{K}+\tfrac{1}{r}\,\hat{K}\big)-\p^A\p_A\alpha
+\beta^i\p_i\hat{K}\,,
\end{align}
where~$\hat{=}$ denotes equality only in the boundary, we use the
shorthand~$\p_s\equiv s^i\p_i$ and we use upper case Latin
indices~$A,B,C$ to denote those that have been projected with the
operator~$q^i{}_j=\delta^i{}_j-s^is_j$. We can alternatively write 
this as a second order derivative boundary condition on the lapse.
Note that in the expressions for the boundary conditions we {\it never} 
commute the unit normal, projection operator~$q^i{}_j$, or physical 
projection operator~$q^{(P)\,kl}{}_{ij}=q^k{}_iq^l{}_j-\tfrac{1}{2}q_{ij}q^{kl}$
through any derivative operator. For example, 
$\p_t\tilde{\Gamma}^s=s_i\p_t\tilde{\Gamma}^i$. So we have
\begin{align}
\p_tv^i&=s^i\p_tv^s+q^i{}_A\p_tv^A,\\
\p_tS_{ij}&=s_is_j\p_tS_{ss}+\tfrac{1}{2}q_{ij}\p_tS_{qq}
+2s_{(i}q_{j)}{}^A\p_tS_{sA}\nonumber\\
&+q^{(P)\,AB}{}_{ij}\p_tS_{AB}^{TF}\,,
\end{align}
for vectors~$v^i$ and symmetric tensors~$S_{ij}$, respectively. We refer
to the various components of the time derivative under this~$2+1$
decomposition as the scalar, vector and tensor sectors in the obvious
way. If we are given conformally flat initial data with~$\beta^is_i=0$
and~$\alpha^2\mu_S$ is constant, as is always the case for the data
evolved in this work, we choose the boundary conditions
\begin{align}
\p_t\tilde{\Gamma}^s&\hat{=}\,\eta\,\beta^i\p_i\tilde{\Gamma}^s
+\frac{\eta}{\alpha^2\mu_S}\beta^i\beta^j\p_i\p_j\beta^s\nonumber\\
&+\frac{\eta}{\alpha^2\mu_S}\beta^i[\p_i\beta^j]\p_j\beta^s
-\frac{\eta^2}{\alpha^2\mu_S}\beta^i\p_i\beta^s\,,
\label{eq:long_shift_BC}
\end{align}
for the longitudinal part of the shift. We take
\begin{align}
\p_t\Theta&\,\hat{=}\,-\alpha\,\p_s\Theta+\beta^i\,\p_i\Theta\,,\\
\p_t\tilde{A}_{ss}&\,\hat{=}\,
-\alpha\,\chi\,\left\{2\,\tilde{D}^i\tilde{A}_{is}
-\frac{2}{3}\,\tilde{D}_s(2\,\hat{K}+\Theta)
-\frac{2}{3}\,R_{ss}\right.\nonumber\\
&+\frac{2}{3}\chi\,\p_s\left[\tilde{\Gamma}^s
-(\tilde{\Gamma}_{\textrm{d}})^s\right]
-\frac{1}{3}\,\chi\,\p_A\left[\tilde{\Gamma}^A
-(\tilde{\Gamma}_{\textrm{d}})^A\right]
\nonumber\\
&\left.+\frac{1}{3}\,R_{qq}-3\,\tilde{D}^i(\ln\chi)\tilde{A}_{is}
-\kappa_1\,\left[\tilde{\Gamma}_s-(\tilde{\Gamma}_{\textrm{d}})_s
\right]\right\}
\nonumber\\
&+\alpha\,\left[\tilde{A}_{ss}\,(\hat{K}+2\,\Theta)
-2\,\tilde{A}^i{}_{s}\,\tilde{A}_{is}\right]
-\frac{2}{3}\,\chi\,D_sD_s\alpha\nonumber\\
&+\frac{1}{3}\,\chi\,D^AD_A\alpha+\Lie_\beta\tilde{A}_{ss}\,,
\end{align}
for constraint preservation in the scalar sector, 
where for example~$\tilde{D}^i=\tilde{\gamma}^{ij}\p_j$ when acting 
on scalars, so that the tilde denotes that the conformal metric was 
used in the contraction. In the vector sector we have
\begin{align}
\p_t\tilde{\Gamma}^A\,&\hat{=}
-\alpha\,\sqrt{\tilde{\mu}_{S_T}}\,
\left[\p_s\tilde{\Gamma}^A-\tilde{\p}^A\tilde{\Gamma}^s\right]
+\tilde{\p}^B\p_B\beta^A\nonumber\\
&+\frac{4}{3}\,\tilde{\p}^A\p_s\beta^s+\frac{1}{3}\,
\tilde{\p}^A\p_B\beta^B-\frac{2}{3}\,\alpha\,
\tilde{\p}^A(2\hat{K}+\Theta)
\nonumber\\
&+\beta^j\p_j\tilde{\Gamma}^A\,,
\end{align}
for the gauge, and
\begin{align}
\p_t\tilde{A}_{sA}&\,\hat{=}\,-\alpha\,\chi
\left\{\tilde{D}^i\tilde{A}_{iA}
-\frac{2}{3}\,\tilde{D}_A\hat{K}
-\frac{1}{3}\,\tilde{D}_A\tilde{\Theta}\right.\nonumber\\
&-\frac{3}{2}\,\tilde{D}^i(\ln\chi)\,\tilde{A}_{iA}
-\frac{1}{2}\,\kappa_1\,\left[\tilde{\Gamma}_A
-(\tilde{\Gamma}_{\textrm{d}})_A\right]
\nonumber\\
&\left.-R_{sA}+\frac{1}{2}\chi\,q_{Ai}\,
\p_s\left[\tilde{\Gamma}^i
-(\tilde{\Gamma}_{\textrm{d}})^i\right]\right\}-\chi\,
D_AD_s\alpha\nonumber\\
&+\alpha\,\left[\tilde{A}_{sA}\,(\hat{K}+2\,\Theta)
-2\,\tilde{A}^i{}_A\tilde{A}_{is}\right]
+\Lie_\beta\tilde{A}_{sA}\,,
\label{eq:Psi0_BC}
\end{align}
for the constraints, where we denote projected indices by upper
case characters starting from the beginning of the alphabet.
Here we denote~$R_{qq}=q^{ij}R_{ij}$. Finally we take
\begin{align}
\p_t\tilde{A}^{\textrm{TF}}_{AB}&\,\hat{=}\,-\alpha\Big[
\tilde{D}_s\tilde{A}_{AB}-\tilde{D}_{(A}\tilde{A}_{B)s}
+\frac{1}{2}\tilde{A}_{s(A}\tilde{D}_{B)}(\ln\chi)
\nonumber\\
&-\frac{1}{2}\,\tilde{A}_{AB}\tilde{D}_s(\ln\chi)
+\tilde{A}^i{}_A\,\tilde{A}_{iB}
-\frac{2}{3}\,\tilde{A}_{AB}\,(\hat{K}+2\Theta)\Big]^{\textrm{TF}}
\nonumber\\
&-\chi\,D_AD_B^{\textrm{TF}}\alpha
+\Lie_\beta\tilde{A}_{AB}^{\textrm{TF}}\,,\label{eqn:BC_last}
\end{align}
for the tensor sector. The boundary conditions in the scalar and vector 
sectors are designed to absorb outgoing constraint violations. The last of the
conditions~\eqref{eqn:BC_last} are equivalent to the requirement
that~$\Psi_0\,\hat{=}\,0$ (see~\cite{SarTig04,BucSar06,BucSar07} 
for more details). This condition could also be used with the BSSNOK 
formulation, but for constraint preservation more work may be needed 
to adapt the other conditions. In the development of this work we have 
tried alternative conditions on~$\hat{K}$ and~$\tilde{\Gamma}^i$ with only 
small differences in the outcome. We do not claim that these gauge boundary 
conditions are optimal. In our numerical experiments we sometimes also 
employ the more naively constructed Sommerfeld conditions
\begin{align}
\p_t\hat{K}&\hat{=}-\sqrt{\mu_L}\,\alpha
\big(\p_s\hat{K}+\tfrac{1}{r}\,\hat{K}\big)+\beta^i\p_i\hat{K}
\,,\label{eqn:Sommerfeld_first}\\
\p_t\tilde{\Gamma}^s&\hat{=}-\tfrac{2}{\sqrt{3}}\sqrt{\mu_S}\,\alpha
\big(\p_s\tilde{\Gamma}^s+\tfrac{1}{r}\,\tilde{\Gamma}^s\big)
+\beta^i\p_i\tilde{\Gamma}^s\,,\\
\p_t\tilde{\Gamma}^A&\hat{=}-\sqrt{\mu_S}\,\alpha\,
\big(\p_s\tilde{\Gamma}^A+\tfrac{1}{r}\,\tilde{\Gamma}^A\big)
+\beta^i\p_i\tilde{\Gamma}^A\,,\\
\p_t\Theta&\hat{=}-\alpha\big(\p_s\Theta+\tfrac{1}{r}\,\Theta\big)
+\beta^i\p_i\Theta\,,\\
\p_t\tilde{A}_{ij}&\hat{=}-\alpha\big(\p_s\tilde{A}_{ij}
+\tfrac{1}{r}\,\tilde{A}_{ij}\big)
+\beta^k\p_k\tilde{A}_{ij}\,.\label{eqn:Sommerfeld_last}
\end{align}
Note that since the trace constraint on~$\tilde{A}_{ij}$ is constantly
imposed, the last of these conditions~\eqref{eqn:Sommerfeld_last}
constitute only five boundary conditions. Since the Z4c formulation
coupled to the puncture gauge has ten incoming characteristics in
the weak field region, these conditions are not overdetermined,
in contrast to the standard conditions used with BSSNOK with box-in-box
mesh refinement. Regardless of the formulation, or whether
Sommerfeld conditions are taken for every evolved field or just the
subset~$\hat{K},\tilde{\Gamma}^i,\Theta,\tilde{A}_{ij}$, they are
not constraint preserving and do not control incoming gravitational
radiation.

\section{Numerical method and parameters}
\label{sec:numerics}

In this section we describe the numerical technique employed in this
work. We use the BAM
code~\cite{ThiBerBru11,BruGonHan06,BruTicJan03,Bru97}, a Cartesian-based
adaptive mesh refinement (AMR) infrastructure optimized for the 
evolution of BBH and BNS spacetimes in 3+1 GR. Vacuum spacetime 
evolutions have also been performed with the AMSS-NCKU 
code~\cite{CaoYoYu08}, which employs the same methods but with an 
independent implementation.

\paragraph*{AMSS-NCKU and BAM basics.} Before discussing the upgrades
to the codes used in this work, we summarize the main points of the
numerical methods used by AMSS-NCKU and BAM. The evolution algorithm
is based on the method-of-lines with explicit Runge-Kutta (RK) time
integrators (in this work we employed fourth order RK for vacuum
spacetimes and third order RK for non-vacuum spacetimes) and finite
differences approximation of the spatial derivatives. The numerical
domain is made of a hierarchy of cell-centered nested Cartesian grids
(nested boxes centered on the punctures~\cite{Bru96,Bru97}).
The hierarchy consists of $L$ levels of refinement labeled
by~$l = 0,...,L-1$. A refinement level consists of one or more
Cartesian grids with constant grid spacing~$h_l$ on level~$l$. A
refinement factor of two is used such that~$h_l = h_0/2^l$. The grids
are properly nested in that the coordinate extent of any grid at
level~$l$, $l > 0$, is completely covered by the grids at level~$l-1$.
Some of the mesh refinement levels can be dynamically moved and adapted
during the time evolution according to the technique of~``moving boxes''.
The Berger-Oliger algorithm is employed for the time stepping~\cite{BerOli84}, 
though only on the inner levels~\cite{BruTicJan03}. Interpolation in 
Berger-Oliger time stepping is performed at second order. A 
Courant-Friedrich-Lewy factor of~$0.25$ is employed in all the runs. 
We refer the reader to~\cite{BruGonHan06,CaoYoYu08} for more details.

\paragraph*{Numerical treatment of the field equations.} The BSSNOK
and Z4c equations of motion are implemented numerically in the same way.
Fields derivatives are approximated by centered finite difference
expressions (fourth order in this work), except for the shift advection
terms which are instead computed with lop-sided
expressions~\cite{ZloBakCam05,BruGonHan06,HusGonHan07,ChiHus10}. Algebraic 
constraints are enforced after every time step in BAM and after every 
Runge-Kutta substep in AMSS-NCKU. The gauge parameters in our numerical 
simulations are fixed to~$\mu_L=2/\alpha$,~$\mu_S=1/\alpha^2$ 
and~$\eta=2/M_{\rm ADM}$, unless otherwise stated. Note that this is 
{\it not} a choice of parameters for which the calculations 
of~\cite{HilRui12} are expected to guarantee well-posedness of the initial 
boundary value problem because we have not carefully taken~$\mu_S$ in such 
a way as to avoid either some generically distinct speeds in the system 
clashing, or sets of measure zero on which the evolution equations may 
be only weakly hyperbolic. In earlier studies~\cite{AliBonBon11} the 
choice~$c>0$ with~$\mu_S\simeq c/\alpha^2$ was not found to greatly 
affect the behavior of the scheme in applications. As highlighted 
in Appendix A of~\cite{CaoHil11} it is challenging to identify problems 
caused by weak hyperbolicity in applications, even if the system is weakly 
hyperbolic everywhere in space. If the degeneracy happens on sets 
of measure zero we therefore expect that in practical applications 
it will be nearly impossible to identify as the cause of any 
concrete numerical problem, although in principle such degeneracy 
should of course be avoided. In the Z4c simulations presented in this 
work the constraint damping scheme with the values~$\kappa_1=0.02$ 
and $\kappa_2=0$ is used. These values have been suggested in the detailed 
1D numerical analysis of~\cite{WeyBerHil11}. Preliminary exploratory 
runs in 3D indicated the combined use of artificial dissipation and 
constraint damping terms is important (in some cases essential) to 
avoid instabilities at the interfaces between boxes and spherical patches, 
thus confirming some of the expectations from the 1D runs (see also 
the discussion in Appendix~\ref{app:shells}). OpenMP support 
in terms of a hybrid OpenMP/MPI implementation has been added
inside BAM to key functions with high computational cost, such as 
evolution equations, interpolation and wave extraction, which 
improves the efficiency of memory management of the code.

\paragraph*{Hydrodynamics treatment (BAM only).} The algorithm
implemented for the general relativistic hydrodynamics
(GRHD, referred to hereafter as ``matter'' for brevity) is a robust
high-resolution-shock-capturing (HRSC)
method~\cite{ThiBerBru11} based on primitive reconstruction
and the Local-Lax-Friedrichs (LLF) central scheme for the numerical
fluxes, see e.g.~\cite{DelBuc02}. Metric variables are interpolated
in space by means of sixth order Lagrangian polynomials, while matter
ones by a fourth order weighted-essentially-non-oscillatory scheme.
Primitive reconstruction is performed here with the 5th order WENO
scheme of~\cite{BorCarCos08}, which has been found
important for long term accuracy~\cite{BerNagThi12,BerThiBru11}.

\paragraph*{Spherical patches for the wave zone.} Both the AMSS-NCKU
and BAM codes have been upgraded for this work. The Cartesian
box-in-box mesh refinement has been extended with spherical patches
(``cubed spheres'')~\cite{Tho00short,Tho04,PolReiSch11} for the wave zone.
They provide us with adapted coordinates for waves, and, as demonstrated 
in appendix~\ref{app:shells}, improved accuracy in GW extraction. The
presence of a spherical outer boundary furthermore allows a 
straightforward implementation of BCs (either Sommerfeld or CP) due 
to the absence of corners. When the spherical patches are being used 
the Cartesian moving boxes, as previously implemented, are employed 
only in the strong-field region for the simulation of the binary 
orbital motion, or a collapsing star, while the propagation of 
gravitational and possibly also electromagnetic waves distant from 
the source is simulated on cubed spheres~\cite{Tho04,PolReiSch11} (see
also~\cite{LehReuTig05,DieDorSch05,SchDieDor06}). The technique is based
on the covering of the sphere by six patches, each patch having local
coordinates that are then mapped to Cartesian ones in such a way to avoid
pathologies associated with standard spherical polar coordinates. As opposed 
to the box-in-box setup, spherical patches allow constant radial resolution 
with linear scaling in the number of grid points, while the boxes result in 
effectively constant angular resolution as well. In practice, the~$l=0$ 
Cartesian box is substituted with six patches overlapping with the Cartesian 
box at level~$l=1$ and among themselves. The resolution of~$h_1$ is also the 
radial resolution employed in the patches. A grid configuration is specified 
by the number of:
i)~levels~$L$,  
ii)~grid points in each non-moving box per direction~$n$, 
iii)~grid points in each moving box per direction~$n^{\rm mv}$, 
iv)~the coarsest box per direction~$h_1=h$, 
v)~grid points in each patch,~$n_r$ and~$n_{\theta,\,\phi}$, which are
typically chosen as~$n_{\theta,\,\phi}=n/2$. More details on the implementation 
are given in Appendix~\ref{app:shells}). Finally, GWs are extracted using 
the Newman-Penrose formalism, in particular by computing the~$\psi_4$ scalar 
on coordinate spheres in the wave zone (see Sec.~III of~\cite{BruGonHan06}). 
Mode decomposition of~$\psi_4$ is performed by projections onto spin weighted 
spherical harmonics and integration on the spheres with a Simpson 
algorithm.

\paragraph*{Boundary conditions.} The radiation controlling, constraint
preserving boundary conditions~(\ref{eq:lapse_BC}-\ref{eqn:BC_last})
are implemented according to the following simple recipe. Inside every
Runge-Kutta substep Lagrange extrapolation, of sixth order in our
experiments, is used to populate enough ghostzones in a neighborhood
of the boundary, so that the same finite difference and dissipation
operators used in the bulk may be evaluated at the boundary.
The metric components~$\alpha,\beta^i,\chi,\tilde{\gamma}_{ij}$ are
updated at the boundary with their standard equations of motion from
the bulk, whereas the boundary
conditions~(\ref{eq:lapse_BC}-\ref{eqn:BC_last}) are used in place of
the standard equations of motion 
for~$\hat{K},\tilde{\Gamma}^i,\Theta,\tilde{A}_{ij}$. Since the evolution
system is not symmetric hyperbolic, at least inside a large class of
symmetrizers~\cite{CaoHil11}, with the standard puncture gauge, we
can not rely on a discrete energy method to guarantee numerical
stability, even in the linear constant coefficient approximation. We
will see however that this implementation of the boundary conditions
is numerically well-behaved in the experiments we perform.
Sommerfeld boundary conditions are implemented in a similar way;
instead of replacing the equations of motion for~$\hat{K},
\tilde{\Gamma}^i,\Theta,\tilde{A}_{ij}$ 
by~(\ref{eq:lapse_BC}-\ref{eqn:BC_last}) we
choose~(\ref{eqn:Sommerfeld_first}-\ref{eqn:Sommerfeld_last}) and
likewise for BSSNOK, but without the~$\Theta$ boundary
condition.

\section{Long-term evolution of single compact objects}
\label{sec:single}

In this section we present 3D evolutions of single isolated compact
objects and compare systematically BSSNOK and Z4c runs. The main
focus is on the behavior of the Hamiltonian constraint violation,
due to truncation and artificial BCs errors. We assess long term
stability of the Z4c formulation in 3D evolutions of puncture black
holes (both non-spinning and rapidly spinning) and compact stars,
and demonstrate an overall improvement in constraint preservation
and in some instances accuracy of physical quantities (absolute 
numerical errors at finite resolution) when the Z4c formulation is
employed. The spurious effect of Sommerfeld BCs and the improvement
obtained with the new BCs are discussed in detail. Our results for
spherically symmetric spacetimes are interpreted in view of previous
results obtained in~\cite{BerHil09,RuiHilBer10} by means of 1D
simulations. The use of Z4c does not significantly improve the
computation of the gravitational radiation emitted by a rapidly 
spinning Bowen-York puncture~\cite{BowYor80}. In particular we find 
that in both cases we do not obtain a clear pointwise convergence 
of the waves at the resolutions used in these tests. This result 
is possibly related to a lack of resolution, or to the use of 
punctures in the black hole description, rather than to deficiencies 
in either formulation.

\subsection{Non-spinning puncture}

Evolutions of a non-spinning puncture of mass~$M=1$ with BSSNOK and
Z4c and Sommerfeld and constraint preserving BCs are
compared. We confirm qualitatively the results obtained with 1D
simulations~\cite{BerHil09,RuiHilBer10}. In particular the new BCs
reduce significantly the spurious constraint violation incoming from
the boundary in the case of Z4c. The largest single constraint 
violation however occurs at the puncture, where both formulations 
give similar results.

\begin{figure}[t]
  \begin{center}
    \includegraphics[width=0.49\textwidth]{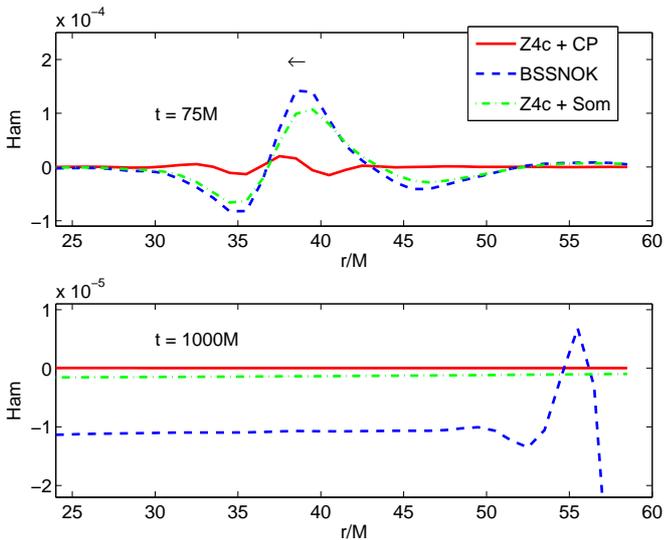}
    \caption{ \label{fig:po:ham_d}
    Constraint violations in space at~$t=75M$ (upper panel)
    and~$t=1000M$ (lower panel) in an evolution of a Schwarzschild
    puncture. At~$t=75M$ we see an incoming pulse of Hamiltonian
    constraint violation in both the BSSNOK and Z4c evolutions. The fact
    that the violation propagates in the BSSNOK evolution test is not
    in contradiction with the PDE properties of the BSSNOK constraint
    subsystem, because the Hamiltonian constraint itself is not a
    zero-speed characteristic variable of the constraint subsystem. At
    this resolution, the incoming constraint violation with the Z4c
    constraint preserving boundary conditions is roughly three times
    smaller than that of Z4c with the Sommerfeld boundary condition,
    but the violations with the constraint preserving boundary conditions
    converge away with resolution whilst those of the Sommerfeld conditions
    do not. In the lower panel we can see the effect that the zero-speed
    mode of the BSSNOK constraint subsystem has on the Hamiltonian
    constraint violation at the outer boundary at late times. }
  \end{center}
\end{figure}

\paragraph*{Setup.}
We employ five levels of box-in-box mesh refinement, and attach
the shells at~$r\sim 21.5\,M$. Each box has~$n=40$, and the
resolution of level~$l=5$ is~$h_5= 0.0625\,M$ per direction. We
choose~$n_{\theta,\, \phi}=40$ angular and~$n_r=40$ radial points in
each spherical patch so that the outer boundary is located
at~$r=58.5\,M$. No symmetries are imposed. The puncture is evolved
with a precollapsed lapse as discussed in~\cite{BruGonHan06}.
In real applications the outer boundary is typically placed further
out, perhaps at~$500\,M$ or~$1000\,M$. Although this does not solve
issues caused by the boundary in principle, and is not an efficient
treatment, in practice it reduces some of the features we will 
encounter here. By design however, in these tests,  we aim to see 
the behavior of the constraint preserving conditions.

\paragraph*{Constraint violation in the strong-field region.}
In each test we find that the Hamiltonian constraint violation in a
neighborhood of the puncture is large, taking a maximum value, at 
this resolution of $\sim 10^3$ at around $t=5\,M$, before being 
rapidly suppressed to $\sim 3$. These values should not be taken 
particularly seriously because of the finite regularity of the solution 
at the puncture. Since much of the physics is concentrated around 
the puncture large numerical error and therefore large constraint 
violation are to be expected in this region. These violations converge 
away more slowly than others in the simulations. The Z4c evolutions 
do not reduce this violation. As observed in spherical
symmetry (see Fig.~10 of~\cite{WeyBerHil11}) we see small amounts of
the puncture constraint violation propagating out of the horizon with 
either formulation. Unsurprisingly, the rectangular mesh-refinement 
boundaries in the 3D evolutions obscure the feature. The Z4c evolutions 
do not qualitatively affect this behavior, which seems to improve 
with resolution.

\paragraph*{Constraint violation at large radii.}
In Fig.~\ref{fig:po:ham_d} the Hamiltonian constraint violation on
the spherical shells in space at times $t=75\,M$ and $t=1000\,M$ is plotted.
We find that the incoming constraint violation of the Z4c Sommerfeld
evolution is comparable to that of the BSSNOK evolution, although at late
times most of the constraint violation caused by the outer boundary has
been absorbed by the boundary; the problem is that the constraint
violation induced by the Sommerfeld boundary condition, with
either BSSNOK or Z4c, does not converge away with resolution. Note that 
comparing the BSSNOK Sommerfeld evolution with the Z4c CPBC data
is not entirely fair, because there is every possibility that constraint
preserving boundary conditions for BSSNOK, see for example~\cite{NunSar09},
could also be implemented. In any case, it is evident that the Z4c
constraint preserving boundary conditions are helpful in reducing
this violation. It is possible to reduce further the constraint
violation by using a large constraint damping term, as for example
in~\cite{AliBonBon11}, but in spherical symmetry we found that values
of~$\kappa_1>0.1\,M$ can adversely affect the dynamics of the evolutions
with Z4c at late times. We are therefore wary of using large damping
parameters, but make no claim about the effect of such damping terms
on other conformal decompositions of Z4.

\subsection{Non-rotating star}
\label{sec:ss}

Our comparison of the new three dimensional numerics with earlier
findings in spherical symmetry is completed by the evolution of
a single stable neutron star. We find Z4c advantageous in reducing
constraint violation that accumulates in grid regions occupied by the
matter. In particular the $L_2$ norm of the Hamiltonian constraint is
found to be three orders of magnitude smaller for standard resolutions.
The spurious ringing effect due to Sommerfeld BCs pointed out in spherical
symmetry is visible also in 3D simulations, but does not dominate the
error budget at typical resolutions. In our earlier study~\cite{BerHil09}
we also evolved an unstable single star in a so-called migration test.
We suppress such an experiment here; our tests of strong field
dynamics are instead performed with binary spacetimes in
section~\ref{sec:bin}.

\begin{figure*}[t]
  \begin{center}
    \includegraphics[width=0.49\textwidth]{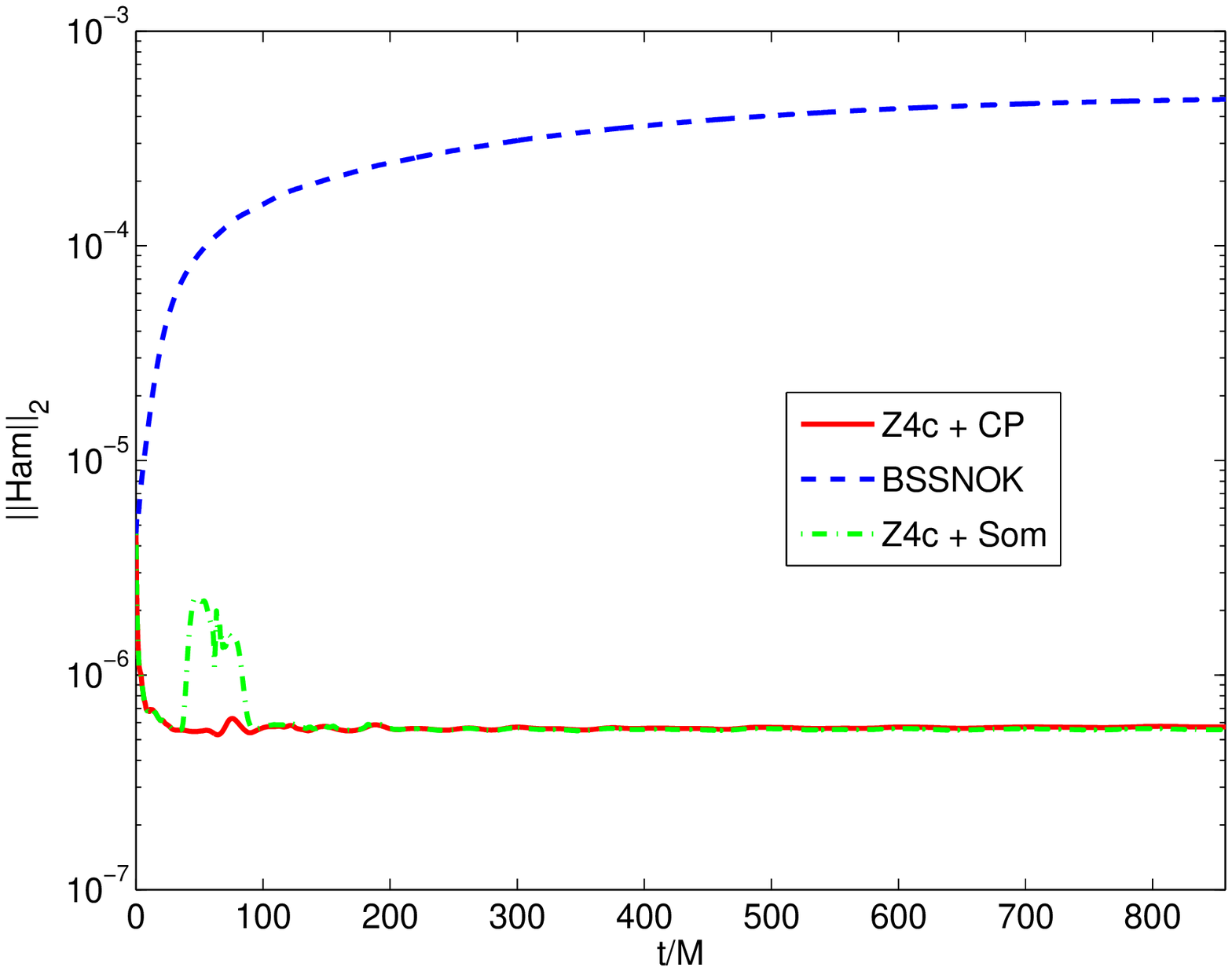}
    \includegraphics[width=0.49\textwidth]{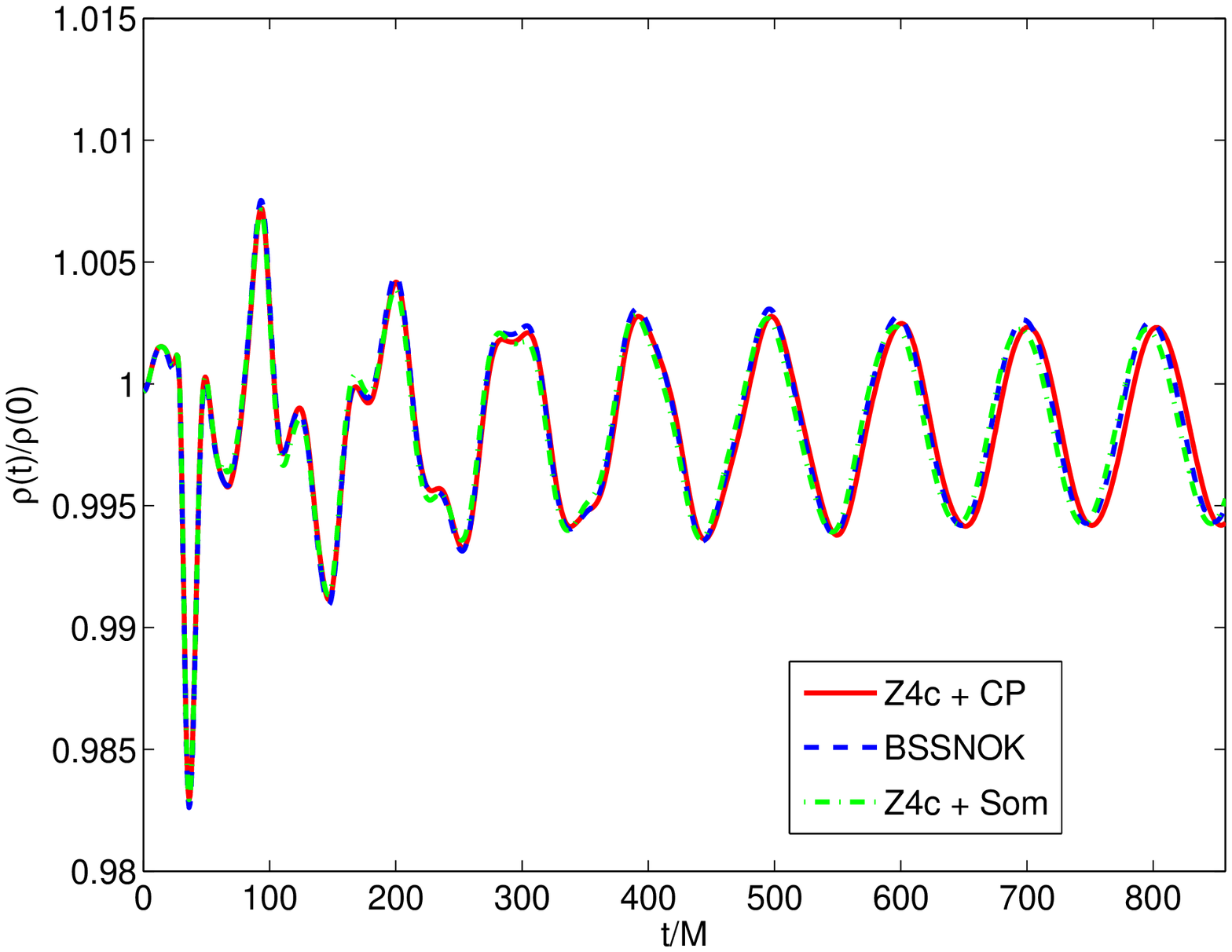}
    \caption{ \label{fig:ss:rho_ham} The~$L_2$ norm of the
      Hamiltonian constraint violation (left) and
      the central density (right) as a function of time
      for evolution of a single stable star with polytropic EOS
      of index~$\Gamma=2$. The Hamiltonian constraint violation is
      approximately two orders of magnitude smaller at the end of
      the experiment when using Z4c. With Z4c, the Sommerfeld
      boundary
      conditions~~(\ref{eqn:Sommerfeld_first}-\ref{eqn:Sommerfeld_last})
      cause large violation when the outer boundary becomes causally
      connected to the central body. The dynamics of all of the
      evolutions, the oscillations in the central density, are very
      similar. The star rings at its radial mode proper frequency.}
  \end{center}
\end{figure*}

\paragraph*{Setup.}
We use three levels of box mesh refinement, and attach the spherical
grids at~$r\sim 18\,M$. The lowest resolution runs use~$n=48$, the
star is completely covered by the innermost grid level $l=2$ with a resolution
of $h_2=0.36\,M$ per direction. We choose $n_{\theta,\,\phi}=48$
angular and~$n_r=48$ radial points in each spherical patch so that
the outer boundary is located at~$r=50\,M$.
Additional runs at twice the resolution~($n=96$) are performed.
No symmetries are imposed. The star is described by a~$\Gamma=2$
polytropic EOS with gravitational mass~$M=1.4\,M_\odot$ and 
radius~$5.7\,M,$ which is exactly the same model evolved in~\cite{BerHil09} 
in bespoke spherical symmetry.  The evolutions are characterized by 
oscillations triggered by truncation errors.

\paragraph*{Constraint violation and truncation error dynamics.}
The key findings of the earlier numerical experiments in spherical
symmetry were:
(i)~in the BSSNOK region containing matter, the Hamiltonian constraint
inside the  star grows throughout the evolution, whereas in the Z4c
evolution  it does not;
(ii)~the BSSNOK simulations have larger oscillations (larger
truncation errors) than Z4c;
(iii)~if Sommerfeld BCs and sufficient resolution are employed,
incoming modes from the boundary perturb the star amplifying
unphysically the oscillations (see Fig.~1~\cite{RuiHilBer10}).
In 3D simulations we find some similar features. The growth of the
Hamiltonian constraint is the dominant one, because of lack of
resolution the oscillations are not significantly affected, and the
effect of Sommerfeld BCs is less pronounced but visible.
The norm of the Hamiltonian constraint is plotted as a function of
time in the left panel of Fig.~\ref{fig:ss:rho_ham}. By the end of
the simulation the constraint violation in the Z4c evolution is about
two orders of magnitude smaller than the BSSNOK violation and 
interestingly smaller even than the initial violation. The behavior
is similar to that shown in Fig.~5 of~\cite{BerHil09}. We will
demonstrate the growth of the Hamiltonian constraint also pointwise
on the grid in section~\ref{sec:bin}, in the case of binary spacetimes.
Considering momentum constraint violations one finds, as in spherical
symmetry, that the differences are far less dramatic. The main
difference in this case is that the momentum constraint violations with BSSNOK
are more dynamical, and slightly larger, although this is probably just
because of reflections from the Sommerfeld outer boundary condition,
which propagate many times over the computational domain. With either
BSSNOK or Z4c, the largest persistent violation in the momentum constraint
occurs at the surface of the star. In the right panel of
Fig.~\ref{fig:ss:rho_ham} we show the oscillations of the central
density during the simulation time. It is not possible to distinguish
significant differences, probably because of the low resolution
employed. The large effect seen in~\cite{BerHil09} was with 
a resolution ten times higher than here.

\begin{figure}[t]
  \begin{center}
    \includegraphics[width=0.49\textwidth]{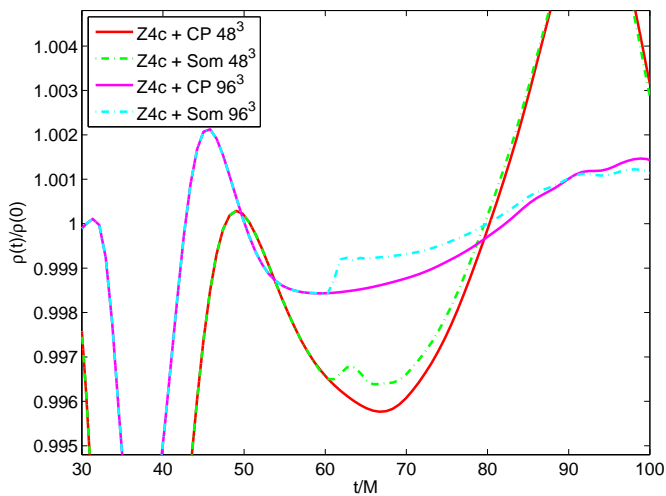}
    \caption{\label{fig:ss:rho_kick}
      A closer inspection of the oscillation of the central
      density at early times, both at the original and twice the
      resolution, for the Z4c evolutions. The constraint violation
      from the Sommerfeld boundary causes a jump in the central
      density, as observed in earlier
      work~\cite{BerHil09,RuiHilBer10}. This effect does not
      converge away with resolution, but at these resolutions
      is not the dominant effect. }
  \end{center}
\end{figure}

\paragraph*{The Sommerfeld boundary kick.} In the left panel of
Fig.~\ref{fig:ss:rho_ham} one sees the effect of the Sommerfeld BCs
with Z4c on the norm of the Hamiltonian constraint. As in the single
puncture evolution, roughly when the outer boundary becomes causally
connected to the central body, there is a large incoming pulse of
constraint violation. This pulse perturbs the central object, but
unlike in the spherical case (see Fig.~1 of~\cite{RuiHilBer10}), the
violation is not the dominating effect on the dynamics. In the right panel
of Fig.~\ref{fig:ss:rho_ham} we do not see a significant jump in the
central density. A blow-up of the central density plot is however shown in
Fig.~\ref{fig:ss:rho_kick}, together with data from a shorter run at
twice the resolution. The figure demonstrates that at approximately~$60M$
the star is slightly perturbed by the Sommerfeld BCs. More importantly,
the figure shows that the size of the perturbation is not converging away
with resolution, whereas the amplitude of the oscillations does, so we
expect this error to be dominant at higher resolutions. By contrast CPBCs
are characterized by smaller reflections. The Hamiltonian constraint
violations propagating out from the star appear to converge at approximately
second order, in line with our expectation for our hydrodynamics scheme.
This rate of convergence is maintained by the constraint preserving
boundary conditions (see also Fig.~5 of~\cite{RuiHilBer10}).

\subsection{Rapidly spinning puncture}

Here we compare the evolutions of a single Bowen-York spinning
puncture~\cite{BowYor80,ChoUnr86,BraSei96} with
spin~$a=S/M^2\simeq0.92$. We choose such a comparatively large value of the spin 
parameter (see also~\cite{LouNakZlo12}) to test performance for large values 
of the conformal factor. For~$a\simeq0.92$ the puncture contributes 
only~$24\%$ of the mass, while the remainder is in the Brill wave 
contribution of the conformal factor. Both formulations give comparable 
results in terms of stability, the norms of the constraint violation, the
majority of which occurs at a few points near the puncture, and
gravitational waves. With either system we observe that pointwise
fourth order convergence is {\it not} achieved in the GWs at the
resolutions at which we performed the tests, although global errors
scale between third and fourth order. We expect that this behavior
is related to inaccuracies caused either by lack of resolution, or
intrinsic to the puncture description of the black hole.

\paragraph*{Setup.} We use five levels of box mesh refinement, and
attach the spherical grids at~$r\sim 50\,M$. In the lowest resolution
runs each box has~$n=48$ points per direction, and the resolution
of level~$l=5$ is~$h=0.065\,M$. We choose~$n_{\theta,\,\phi}=48$ angular
and~$n_r=48$ radial points in each spherical patch so that the outer
boundary is located at~$r=150\,M$. Runs at
resolutions~$n=64,\,96,\,128$ (with the grid spacing scaled in order
to maintain the same grid setup) are performed. Bitant symmetry is
imposed. The initial data for this test is constructed in with an 
ad-hoc method, used elsewhere in the literature, in which the BAM 
spectral binary puncture initial data solver is applied to a single 
puncture with large spin and another, unboosted and unspinning,  
located very close by with a relative uncorrected mass of~$10^{-12}$. 
There is no sign of the second puncture on the grid. Problems 
with convergence do however make this construction a point to 
address.

\begin{figure}[t]
  \begin{center}
    \includegraphics[width=0.49\textwidth]{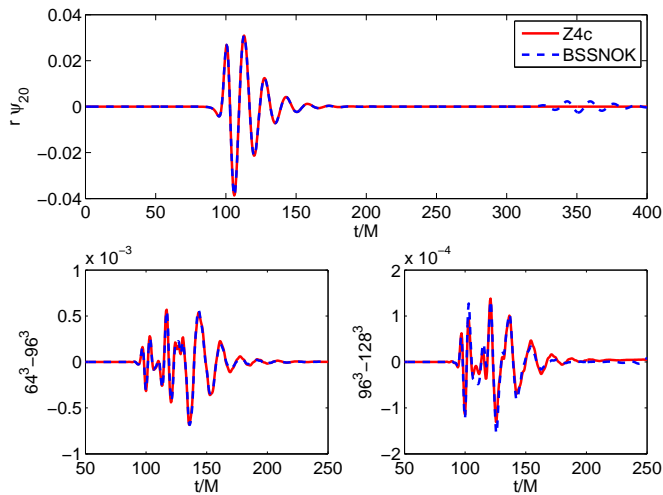}
    \caption{ \label{fig:sp:wav} Comparison of the waves with
      Z4c and BSSNOK from the single spinning puncture. The upper panel
      shows the~$(2,0)$ multipole mode of the real part of~$r\,\psi_4$
      emitted during the evolution. The data is taken from the lowest
      resolution~($48^3$ points per direction) test. The Z4c and BSSNOK waves
      agree extremely well until about~$t\sim300\,M$, roughly the time when
      the incoming constraint violation from the BSSNOK Sommerfeld boundary
      condition is reflected from the central body and reaches
      the extraction sphere at~$r=90\,M$. 
      The lower two panels show the difference between the same quantity
      for the~$64$ ($96$) and~$96$ ($128$) point runs, respectively.
      The two systems are almost perfectly comparable.
    }
  \end{center}
\end{figure}

\paragraph*{Basic features of the dynamics.} The~$(2,0)$ multipolar
mode of~$r\,\psi^4$ emitted during the evolution of this initial data
are shown in Fig.~\ref{fig:sp:wav}, for the lowest resolution runs.
The waves lie on top of each other; as resolution is increased they
converge to each other. At late times,~$t=350\,M$, a
boundary effect is visible in the BSSNOK data. The BCs of Z4c improves
this behavior significantly. Note that~$t=350\,M$ corresponds roughly
to the time needed by a wave initially near the puncture to propagate
to the outer boundary, be reflected from the strong field region and
travel out once again to the extraction sphere at~$90\,M$. No feature
is visible as the wave passes from the outer boundary at
around~$t=210\,M$, perhaps because~$\psi^4$ is by construction
insensitive to incoming gravitational radiation. We computed 
the ADM mass integral (see Eq.~\eqref{eq:ADM_Ei} later) and find 
that in the BSSNOK runs a drift begins exactly when the outer boundary 
becomes causally connected with each observer, although this drift is then 
swamped by the effect of the outgoing gravitational waves. The effect of 
the Sommerfeld boundary condition on physical quantities is discussed 
in more detail for binary neutron star simulations later. 

\begin{figure}[t]
  \begin{center}
    \includegraphics[width=0.49\textwidth]{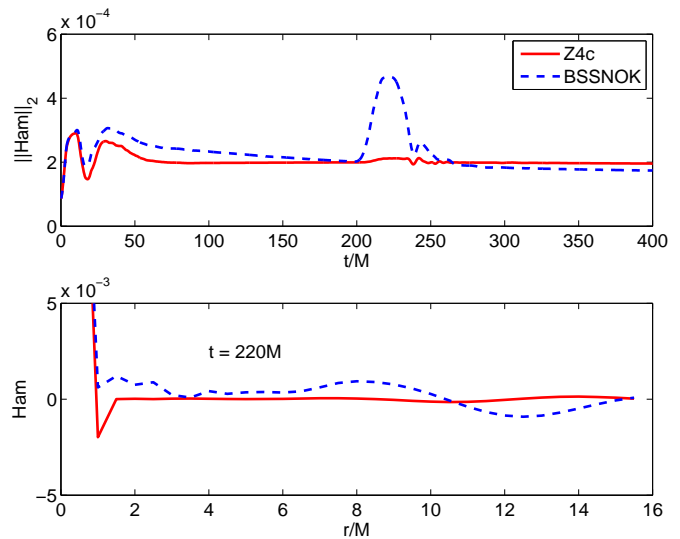}
    \caption{ \label{fig:sp:ham}
    Hamiltonian constraint violation for the spinning puncture evolution.
    As in the upper panel of~Fig.~\ref{fig:sp:wav} data is taken from the
    lowest resolution~($n=48$ points per direction) test. The upper panel
    shows the~$L_2$ norm of the constraint as a function of time. The jump
    in violation at around~$t=200\,M$ in the BSSNOK data appears to be
    caused by the Sommerfeld boundary conditions, and does not converge
    away with resolution. The bottom panel shows the Hamiltonian violation
    in space at time~$t=220\,M$, approximately at the peak of the jump in
    the upper plot, in the near-field~($l=2$ box) region.}
  \end{center}
\end{figure}

\paragraph*{Constraint violation.} As in the non-spinning case the
largest Hamiltonian constraint violation is at the puncture. The
evolution of the~$L_2$ norm of the Hamiltonian constraint is shown
in the top panel of Fig.~\ref{fig:sp:ham}, for the lowest resolution
runs. The results from Z4c and BSSNOK are comparable. At
time~$t\sim220\,M$ BSSNOK data are affected by the Sommerfeld boundary
conditions. The bottom panel of Fig.~\ref{fig:sp:ham} shows the
Hamiltonian violation in space at~$t\sim220\,M$. The largest
violation is at the puncture and of similar magnitude, but far
from the grid origin Z4c violation behaves better.

\paragraph*{Convergence.} We looked at self convergence 
of the waves presented in the upper panel of Fig.~\ref{fig:sp:wav}.
At early times, up to~$t\sim100$, only second order pointwise
convergence is observed with either BSSNOK or Z4c. Later, the errors
scale at third-to-fourth order rate in magnitude (norm) but pointwise
convergence is lost. The use of either more resolution, properly 
constructed initial data, or simply more refinement levels, i.e.~high 
resolution at the puncture, might improve this behavior. 
In~\cite{BruGonHan06} similar convergence tests were performed but with 
a lower spin~$a=0.194$, with two more, a 
total of seven, levels of mesh refinement, but with a smaller range in
resolutions, and a lower maximum resolution in a neighborhood of the
puncture. Although the plots presented in~\cite{BruGonHan06} (see
Fig.~6) are scaled for fourth order convergence, the difference in
resolutions makes it hard to distinguish between different convergence
factors with confidence. The earlier study also found pointwise
convergence is not maintained in the gravitational waves, thus
our findings are consistent. A detailed discussion of spin and higher 
order finite differencing can be found in~\cite{LouNakZlo12}.

\section{Evolution of compact binary initial data}
\label{sec:bin}

In this section we assess the performance of the Z4c formulation in
the simulation of the merger of two compact objects. Two standard
initial configurations are studied, with a set of approximately~$60$
simulations, in detail: an equal-mass, non spinning, black hole binary
(BBH), and an equal-mass, irrotational neutron star binary (BNS). We
discuss evolutions of about three orbits and compare systematically
the results from several convergence series with the corresponding
ones obtained with BSSNOK.

\paragraph*{Presentation rubric.} In the presentation of the results,
we describe the main features of the dynamics, then the Hamiltonian
constraint violations are compared and finally the accuracy
of the GWs and other physical quantities are discussed in relation
with the Hamiltonian constraint violations. For brevity we discuss
only the main emission channel, the~$(2,2)$ multipole of the
radiation. As is standard, the (complex)~$\psi^4$ 
projection~$r\,\psi^4_{22}$, extracted on a coordinate sphere $S_r$ of 
radius~$r$, is decomposed into amplitude and phase according to
\begin{equation}
r\,\psi^4_{22} = A_{22} \ e^{- i \phi_{22}} \ .
\end{equation}
On the same coordinate sphere we compute the integral
\begin{equation}
\label{eq:ADM_Ei}
E_{\rm ADM}(r)=\int_{S_r} d s_l
\sqrt{\gamma}\,\gamma_{ij} \gamma^{kl}\left( \gamma_{ik,j} - \gamma_{ij,k}\right) \ ,
\end{equation}
the spatial metric~$\gamma_{ij}=\chi^{-1}\tilde{\gamma}_{ij}$ which approximately 
represents the ADM energy. The ADM energy (or mass) of the system is given 
by
\begin{equation}
M_{\rm ADM}=\lim_{r\to\infty} E_{\rm ADM}(r) \ ,
\end{equation}
and it is a conserved quantity. On a given sphere in the wave zone, however,
$E_{\rm ADM}(r)$ is expected to deviate from~$M_{\rm ADM}$ due to the
gravitational energy radiated away from the sphere, 
\begin{equation}
\label{eq:Erad_psi4}
E_{\rm rad}(r)=\dfrac{r^2}{16\pi} \int^t_0 dt' \int\, d\Omega\, \left|
  \int^{t'}_{0} dt'' \psi^4\right|^2 \ . 
\end{equation}
Note that the outer integral in Eq.~\eqref{eq:Erad_psi4} is performed by a 
simple Riemann sum. The angular momentum is computed with similar
ADM-like integrals, see e.g.~\cite{RuiAlcNun07} and references
therein, but note that  
this quantity remains ambiguously defined (gauge dependent) also in the
asymptotic limit $r\to\infty$ for a generic asymptotically flat spacetime.  
We stress that the observed differences, up to outer boundary effects, seem 
to converge to the same continuum solution. They are however significant 
at fairly high resolutions in most of the cases we studied. Particularly 
relevant are the differences in Hamiltonian constraint violation in the 
evolution of non-vacuum spacetimes. We show evidence that these violations 
are correlated with the quality of the numerical waveforms.

\paragraph*{Convergence tests.} Standard three level self-convergence
tests are performed using simulations at different grid resolutions.
These tests can be biased by the choice of the triplet. As
discussed in~\cite{BerThiBru11} we consider only triplets in which
i)~the ratios between the low and medium and medium and high
resolutions are~$h_L/h_M\simeq h_M/h_H>1$ (ideally~$>2$); and 
ii)~the scaling factor is at least of order two,
$s=(h_L^p-h_M^p)/(h_M^p-h_H^p)\gtrsim2$, where~$p$ is the convergence
rate. We find that if these criteria are not met then the measured 
convergence order from different triplets is not consistent. 
Additionally, in our experience we found it important to verify
that the use of different triplets gives consistent results. 
Even when the criteria above are satisfied, different triplets can give 
differing convergence factors. The result is acceptable if the rates 
consistently improve as higher resolution triplets are taken.

\begin{table}[t]
  \centering
  \begin{ruledtabular}
  \caption{\label{tab:gridsetup}
    Summary of the grid configurations and of the runs.
    Columns: 
    name of the configuration, 
    maximum refinement level, 
    moving levels are those with $l>L^{\rm mv}$, 
    number of points per direction in the moving levels, 
    resolution in the level $l=L-1$, 
    number of points per direction in the non-moving levels, 
    resolution in the levels $l=1$ (radial resolution in
    the shells), 
    number of radial points in the shells, 
    number of angular points in the shells, 
    outer boundary. Note that the resolution is given in units of $M$
    for BBH runs, but in units of $M_{\odot}$ for BNS runs. Runs
    marked with ``*'' were reproduced with the AMSS-NCKU code.}
  \begin{tabular}{ccccccccccc}
    \hline
    Name & $L$ & $L^{\rm mv}$ & $n^{\rm mv}$ &
    $h_{L-1}$ & $n$ & $h_1$ &
    $n_{\rm r}$ & $n_{\theta,\,\phi}$ & $r_{\rm out}$\\
    \hline
    BBH0* & 9 & 2 & 48 & 0.0182 &  72 & 2.33 &  865 & 24 & 2092 \\
    BBH1 & 9 & 2 & 56 & 0.0156 &  84 & 2.0  & 1008 & 28 & 2091 \\
    BBH2* & 9 & 2 & 64 & 0.0137 &  96 & 1.75  & 1150 & 32 & 2089 \\
    BBH3 & 9 & 2 & 72 & 0.0122 & 108 & 1.556 & 1293 & 36 & 2088 \\
    BBH4* & 9 & 2 & 80 & 0.0109 & 120 & 1.4   & 1436 & 40 & 2088 \\
    BBH5 & 9 & 2 & 88 & 0.0099 & 132 & 1.273 & 1581 & 44 & 2090 \\
    BBH6 & 9 & 2 & 96 & 0.0091 & 144 & 1.167 & 1718 & 48 & 2087 \\
    BBH7 & 9 & 2 &112 & 0.0078 & 168 & 1.0   & 2008 & 56 & 2088 \\
    BBH8 & 9 & 2 &128 & 0.0068 & 192 & 0.875 & 2293 & 64 & 2086 \\
    BBH9 & 9 & 2 &144 & 0.0061 & 216 & 0.777 & 2579 & 72 & 2086 \\
    \hline
    BNS0  & 4 & 1 & 48 & 0.5   &  72 & 2.0  & 212 & 24 &  482 \\
    BNS0r & 4 & 1 & 48 & 0.5   &  72 & 2.0  & 412 & 24 &  885 \\
    BNS1  & 4 & 1 & 56 & 0.429 &  84 & 1.71 & 245 & 28 &  482 \\
    BNS1r & 4 & 1 & 56 & 0.429 &  84 & 1.71 & 478 & 28 &  882 \\
    BNS2  & 4 & 1 & 64 & 0.375 &  96 & 1.5  & 278 & 32 &  481 \\
    BNS2r & 4 & 1 & 64 & 0.375 &  96 & 1.5  & 545 & 32 &  881 \\
    BNS2a & 4 & 1 & 64 & 0.375 &  96 & 1.5  & 678 & 32 & 1081 \\
    BNS3  & 4 & 1 & 72 & 0.333 & 108 & 1.33 & 312 & 36 &  477 \\
    BNS3r & 4 & 1 & 72 & 0.333 & 108 & 1.33 & 612 & 36 &  877 \\
    BNS4  & 4 & 1 & 80 & 0.3   & 120 & 1.2  & 345 & 40 &  476 \\
    BNS4r & 4 & 1 & 80 & 0.3   & 120 & 1.2  & 678 & 40 &  875 \\
    BNS5  & 4 & 1 & 88 & 0.273 & 132 & 1.09 & 378 & 44 &  475 \\
    BNS5r & 4 & 1 & 88 & 0.273 & 132 & 1.09 & 745 & 44 &  875 \\
    BNS6  & 4 & 1 & 96 & 0.25  & 144 & 1.0  & 412 & 48 &  476 \\
    BNS6r & 4 & 1 & 96 & 0.25  & 144 & 1.0  & 812 & 48 &  876 \\
    \hline
  \end{tabular}
  \end{ruledtabular}
\end{table}

\paragraph*{Initial data.} Before discussing the results we summarize the 
initial data and grids employed. Initial data for BBHs are conformally
flat puncture initial data constructed using the Brandt-Br\"ugmann
puncture method~\cite{BraBru97} with the BAM implementation 
of the spectral puncture initial data solver~\cite{AnsBruTic04}. The 
holes have equal mass, an initial separation of~$d=7\,M$ and are 
placed in a quasi-circular configuration, on which the Pad\'e resummed 
eccentricity reduction algorithm of~\cite{WalBruMue09} was applied. 
The initial data are interpolated onto the AMSS-NCKU and BAM grids 
by eighth order Lagrangian barycentric interpolation. Initial data for 
BNS assume a conformally flat metric and irrotational flow. The initial 
separation is~$D\sim 10\,M$, ADM mass and angular momentum
are~$M=3.005\,M_\odot$ and~$J=8.3\,M^2$, respectively. The stars are
described by a~$\Gamma=2$ polytropic EOS, each has baryonic
mass of~$M_b=1.625\,M_\odot$. These initial data have been produced with the
LORENE~\cite{LORENE} library and have been already evolved in
several places~\cite{BaiGiaRez09,ThiBerBru11,BerThiBru11}. The data
are interpolated onto the BAM grid by spectral interpolation.

\paragraph*{Grid setup.} All of the runs performed at different
resolutions are listed in Tab.~\ref{tab:gridsetup}, together with
details of the grid setup. Note that several of the BBH simulations,
those marked with a~``$*$'' in Tab.~\ref{tab:gridsetup}, have been
performed with both the BAM and AMSS-NCKU codes, indicating, at least
for the chosen grid settings, the robustness of our findings. Note 
that our lowest resolution vacuum setup BBH0 has maximum resolution
comparable to that of the highest resolution in the earlier BAM 
calibration paper~\cite{BruGonHan06}, although the boxes 
in~\cite{BruGonHan06} were larger which also affects accuracy.
The resolutions of the setups BBH2 and BBH3 are comparable to the ones of
recent BBH simulations~\cite{ReiTic12,TicMar10} that required only moderate
accuracy, while the maximum BBH9 resolution is similar to what was used to
obtain some of the BAM waveforms for the NINJA-2 catalog~\cite{AjiBoyBro12}.
The highest resolution in the matter simulations BNS6, BNS6r
approaches those in the accurate
runs of~\cite{BerThiBru11,BerNagThi12}. Other details about gauge
conditions, damping parameters, etc. have been already given in
Sec.~\ref{sec:numerics}.

\subsection{Equal-mass, non-spinning BBH}

\begin{figure}[t]
  \begin{center}
    
    \includegraphics[width=0.49\textwidth]{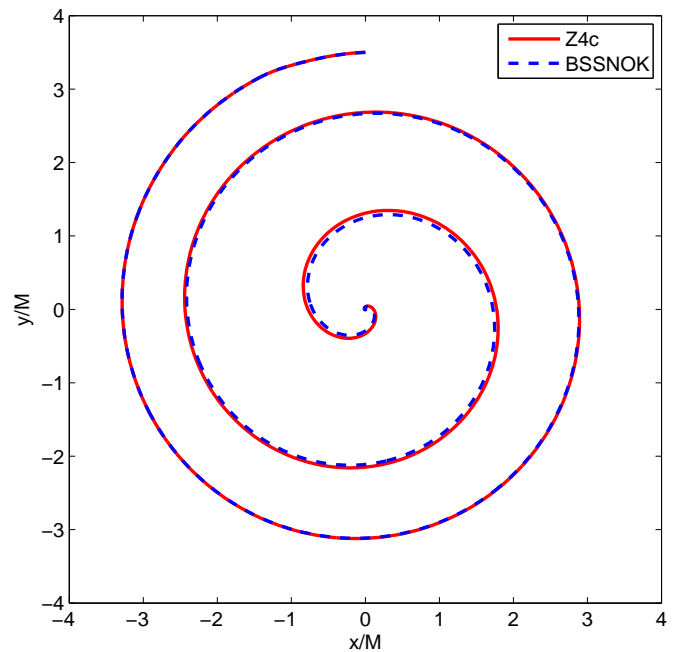}
    \caption{\label{fig:bbh:tracks}
      Tracks of the punctures for binary black hole inspiral for the
      configuration BBH2 in Tab.~\ref{tab:gridsetup}.
      In the continuum limit, upto outer boundary effects, the tracks
      should agree perfectly because we are evolving the same data
      with the same gauge choice regardless of the formulation.
      At finite resolution however different formulations and
      discretizations may give different results. Initially
      the tracks agree, but a difference accumulates; the BSSNOK
      punctures merge slightly sooner than the Z4c ones. This plot does
      not indicate either set of numerical data is better than
      the other, only that there is qualitative agreement between
      the results.}
  \end{center}
\end{figure}

\paragraph*{Basic features of the dynamics.} Let us discuss the
evolution of BBH initial data. The black holes evolve for about~$2.5$
orbits before merging, radiating energy and angular momentum in 
gravitational waves. In Fig.~\ref{fig:bbh:tracks} the puncture 
tracks are plotted for the
grid BBH2 in Tab.~\ref{tab:gridsetup}. The gravitational waves
have exactly the same basic profile, but unlike in the evolution of a
single spinning puncture (see Fig.~\ref{fig:sp:wav}) they {\it are}
visually distinguishable, with the absolute maximum of the~$(2,2)$
multipole mode of the real part of~$r\,\psi_4$ occurring, 
approximately~$2.2\,M$, earlier in the BSSNOK data, in accordance with 
the expectation from Fig.~\ref{fig:bbh:tracks}. 
The differences that accumulate over the
evolution converge away with resolution. In the BBH4 data the delay
is only~$1.1\,M$. Also in contrast to the single spinning puncture
case, presumably because the outer boundary is placed so far
out at~$r\simeq2090\,M$, no boundary feature is visible in the BSSNOK
waves, at least up to a radius of~$r=400\,M$ within the run-time of
the simulation.

\begin{figure*}[t]
  \begin{center}
    \includegraphics[width=0.49\textwidth]{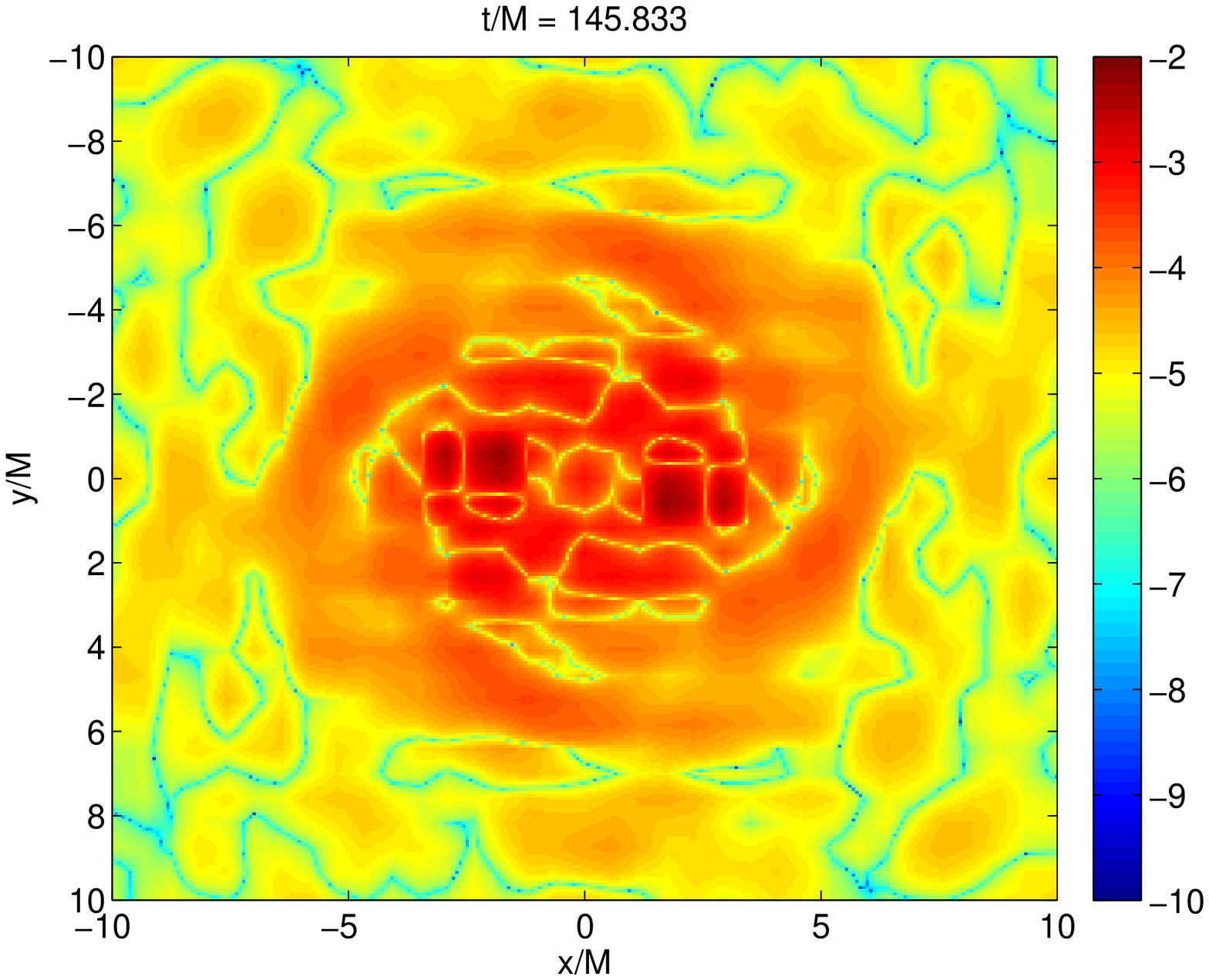}
    \includegraphics[width=0.49\textwidth]{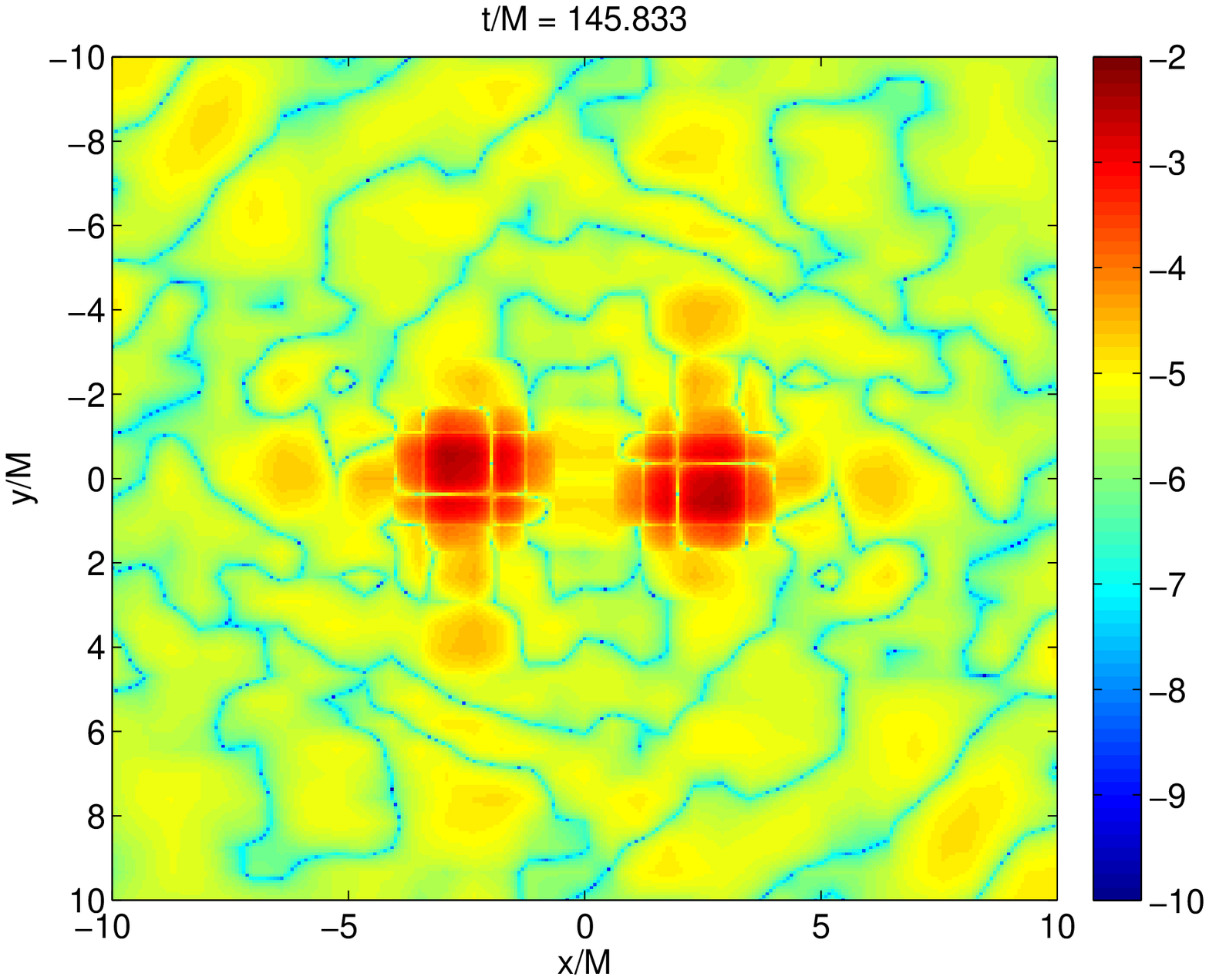}
    \caption{\label{fig:bbh:ham2d} Hamiltonian constraint violation in
      BBH simulations at time~$t\sim146\,M$ (after $\sim 1.5$ orbits)
      on the orbital plane and in the strong field region for the grid
      setup BBH1 in Tab.~\ref{tab:gridsetup}. The plots show~$\log_{10}|H|$
      for levels~$l=5,4,3$; the left panel shows the BSSNOK data, the right
      that of Z4c. Directly at the punctures the violation is similar in
      either case, but the surrounding region has smaller violation in the
      Z4c data. Some aspects of the grid structure are visible in the 
      violation.}
  \end{center}
\end{figure*}

\paragraph*{Constraint violation.} In Fig.~\ref{fig:bbh:ham2d}
the Hamiltonian constraint~$\log_{10}|H|$ is plotted in space for the 
BBH1 runs on the orbital plane at a simulation time $t=146\,M$
(roughly 1.5 orbits) on refinement levels $l=3,4,5$.
As in case of the single puncture evolution the Hamiltonian 
constraint violation is dominated by the punctures and, for this data, 
at this time, has a maximum absolute value of~$\sim10^{-2}$ regardless
of the formulation. However, the Hamiltonian constraint violation differs 
in the strong field region depending on the formulation. In the BSSNOK
case a significant Hamiltonian violation extends to level~$l=3$ with 
an almost spherical pattern. In the Z4c case the violation is mainly
restricted to the highest level around the puncture. Note the effect
of Cartesian mesh refinement in the plot. For the same data the 
momentum constraint away from the puncture is also roughly an 
order of magnitude smaller in the Z4c data. The highest resolution 
runs, BBH9, have the smallest constraint violation even in the strong 
field region, but the violation there is dominated by that of the
puncture. The difference between the Hamiltonian constraint violation 
of the BBH1 and the BBH9 data in this region is at most a factor of 
three for either formulation despite the difference in resolution.

\begin{figure*}[t]
  \begin{center}
    
    \includegraphics[width=0.49\textwidth]{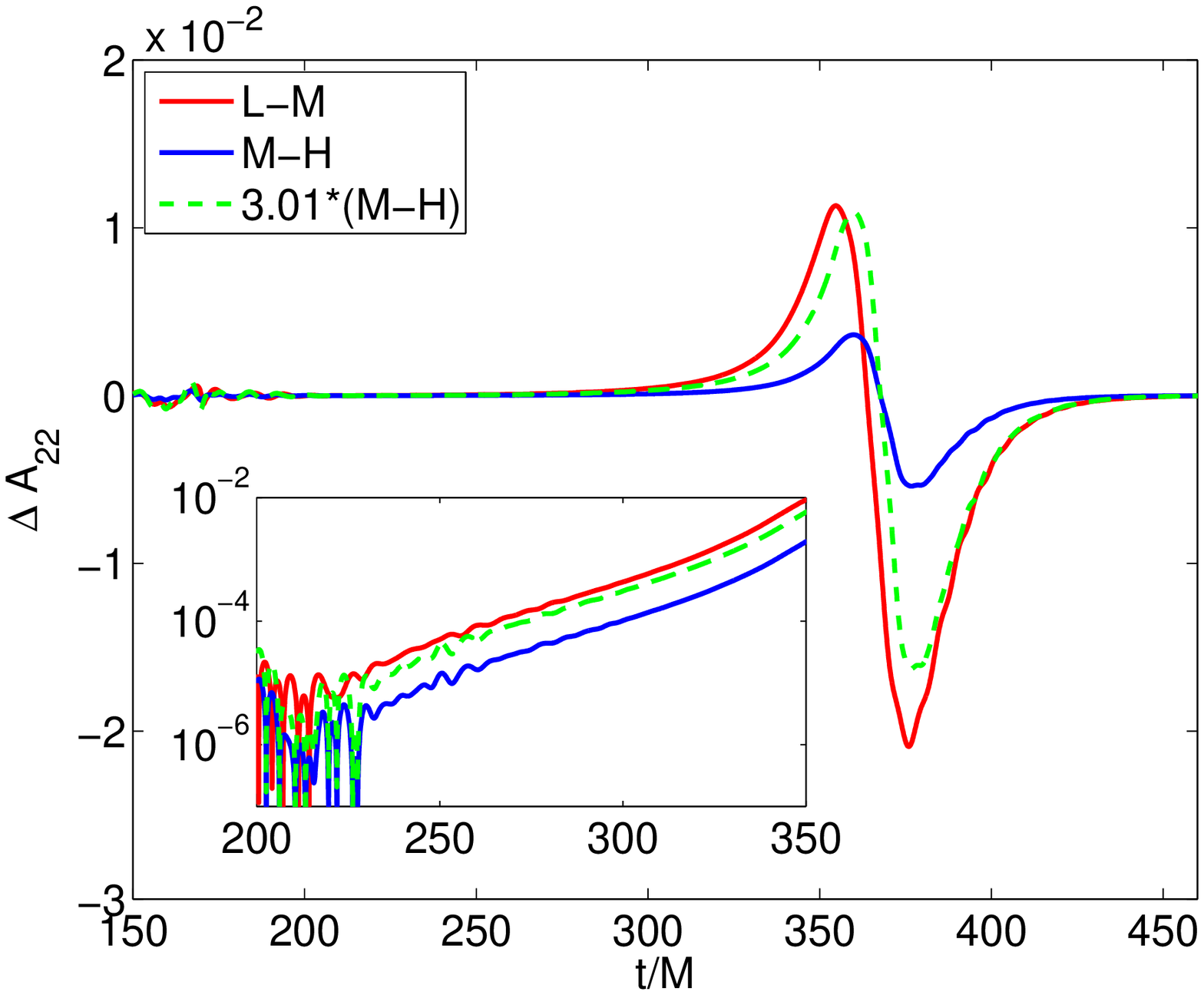}
    \includegraphics[width=0.49\textwidth]{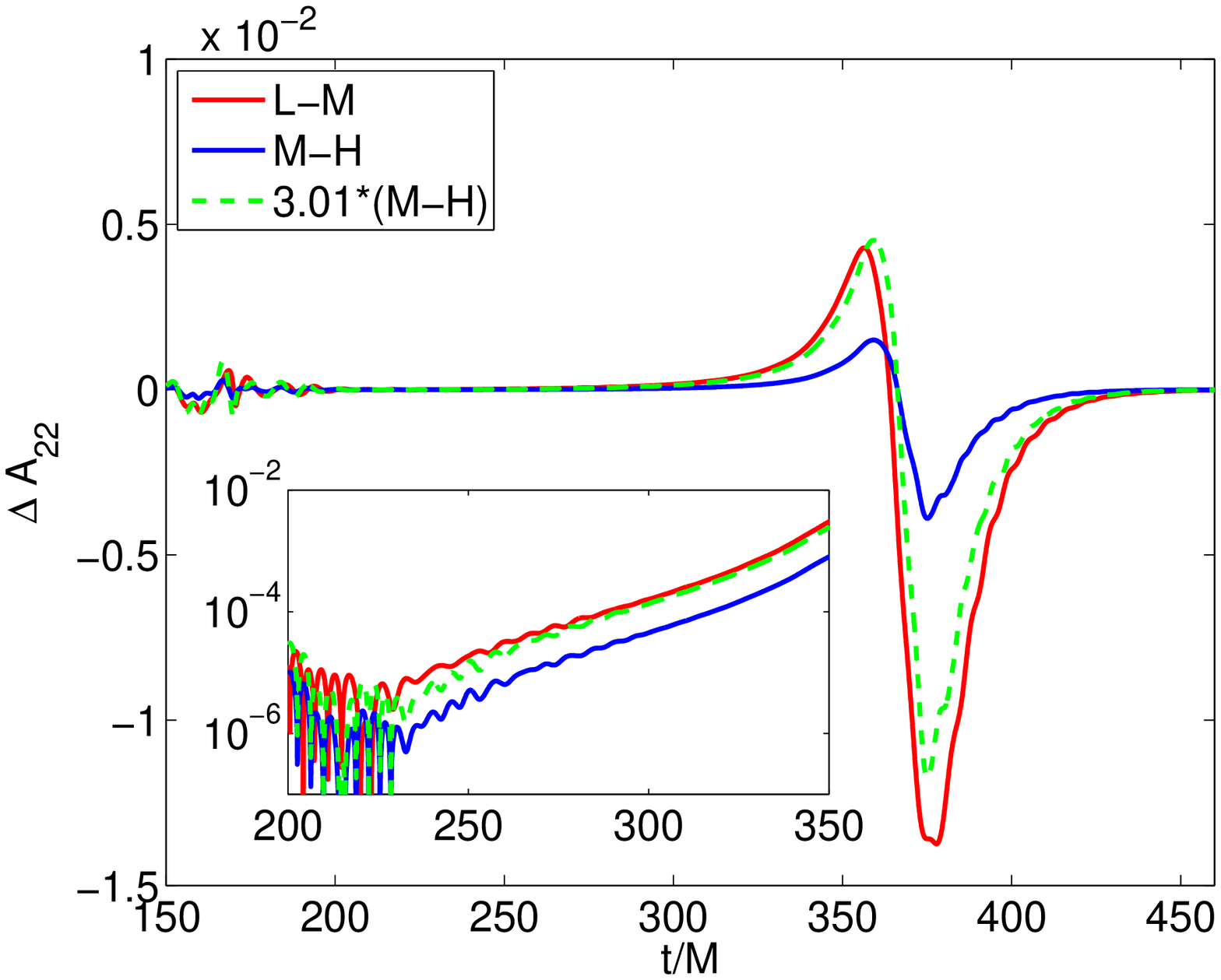}
    \includegraphics[width=0.49\textwidth]{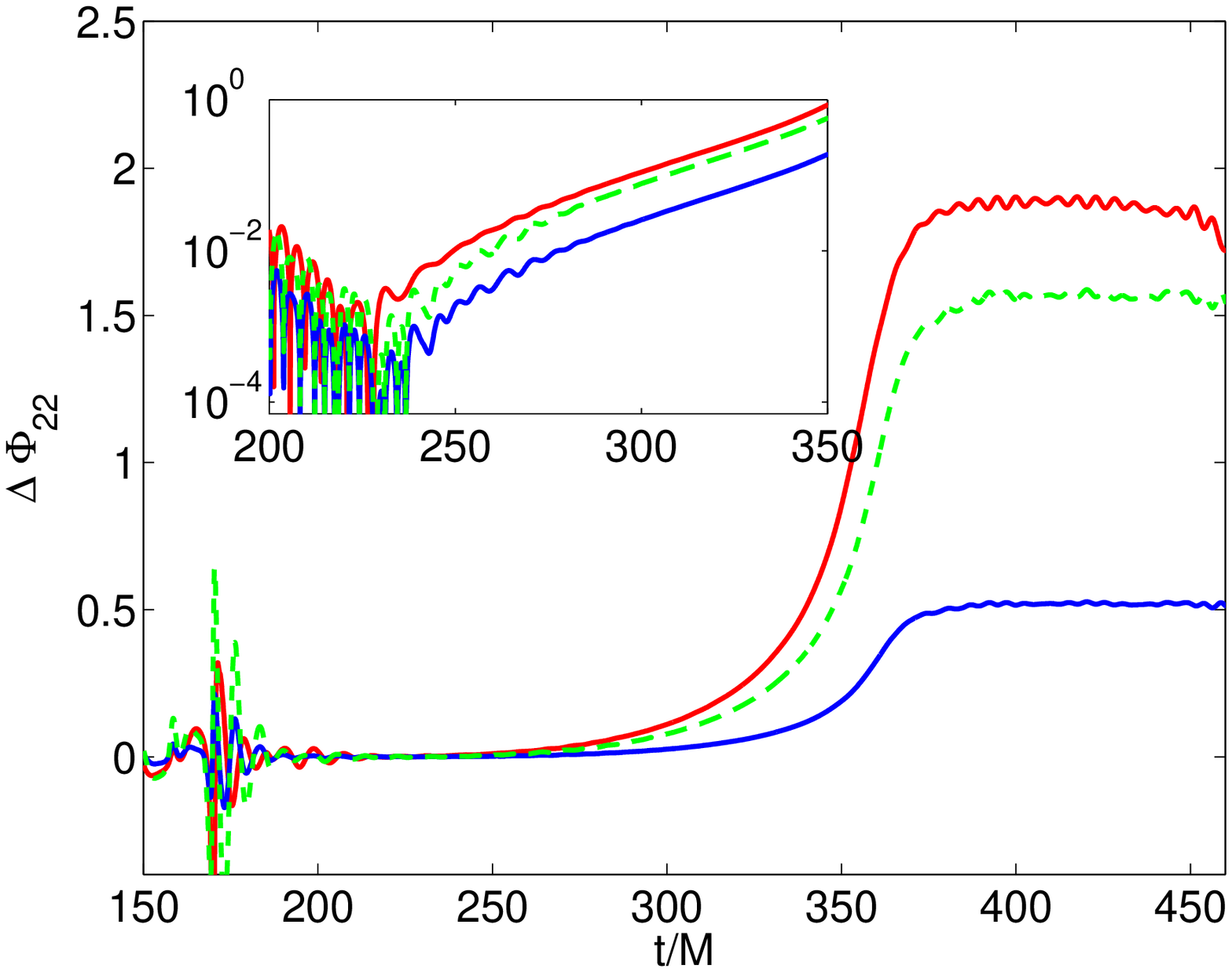}
    \includegraphics[width=0.49\textwidth]{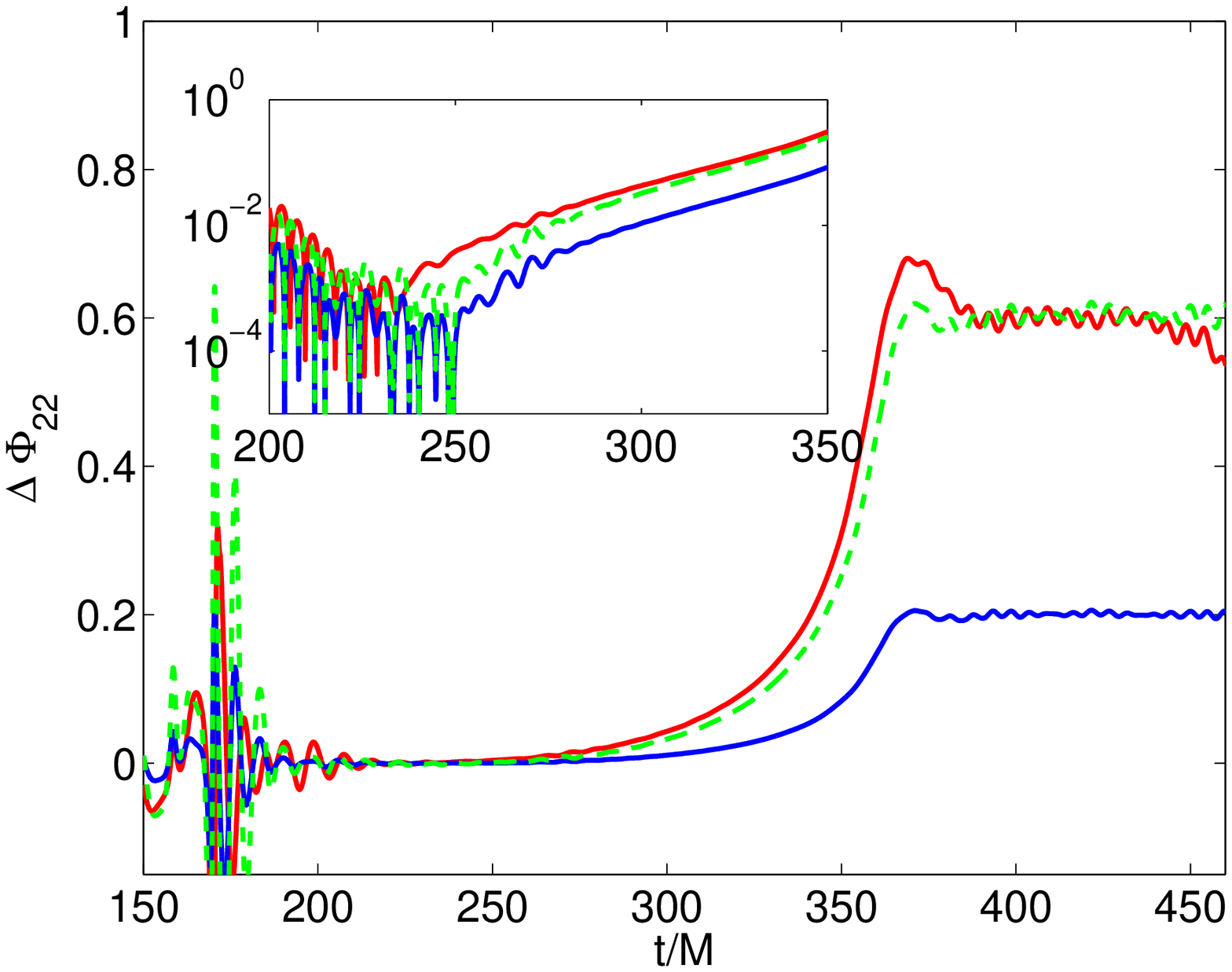}
    \caption{\label{fig:bbh:conv} Convergence plot of binary black hole
      inspiral for the resolutions $h=1/56,1/80,1/112$ (Runs BBH1,
      BBH4, and BBH7). The left
      panel shows results for BSSNOK, the right panel for Z4c. All
      the differences are scaled for 3rd order convergence.
      The extraction radius is at $150M$.}
  \end{center}
\end{figure*}

\paragraph*{Gravitational wave accuracy.} Quantitative differences
are observed in the gravitational radiation computed in BSSNOK and
Z4c simulations. The differences can be appreciated in
self-convergence tests for phase and amplitude. 
Typical results are presented in Fig.~\ref{fig:bbh:conv}.
The phase errors accumulated to merger~($t\sim270\,M$) are shown in
bottom panels. When the Z4c simulation is employed we observe a factor
three of improvement. This behavior is common to all of the simulations
performed. Similarly, the amplitude error improves by a
factor two-to-three, as can be appreciated from the top panel of the
figure. The results obtained with the Z4c are not only characterized
by smaller errors, but also by the fact that convergence is maintained
for longer time, in particular in phase errors. All the errors are
observed to converge with increasing resolution. The Z4c and BSSNOK
data sets also appear to converge to each other, suggesting the same
continuum limit is approached. We conclude that at finite resolution 
the use of Z4c results in the computation of more accurate waveforms.

\paragraph*{Convergence issues.} For the converge tests
presented in Fig.~\ref{fig:bbh:conv} third order convergence is
obtained, despite the use of fourth order operators for the bulk
derivatives (see below). In our experiments with the
described grid settings we find it difficult to achieve fourth order
convergence in GWs from orbiting puncture runs, regardless of
the formulation employed. On the other hand clearer (order approximately 
3.5) convergence is demonstrated in Appendix~\ref{app:teukolsky} for the 
Teukolsky wave, a simpler problem which however retains some of the 
important features, namely, nonlinearity, in the sense that we evolve 
with the full BAM Z4c infrastructure, constraint violation, because 
the wave only satisfies the constraints to linear order in perturbation 
theory, present in the BBH calculations. On the other hand the Teukolsky 
wave test has neither moving boxes or punctures. We were unable to 
identify the precise reason for the behavior in presence of punctures, 
however we point out the following well-known facts:
i)~punctures have finite regularity, which can obviously affect the 
formal order of convergence of high-order finite-differencing 
stencils;
ii)~moving box simulations have several source of errors (second
order time interpolation, regridding, the number of mesh refinement 
buffer zones, 3D interpolation, data between boxes and shells, spheres 
and mode projection), a clearly defined error budget is very 
complicated to construct (for a discussion see
e.g.~\cite{BruGonHan06,PolReiSch11,ZloPonLou12});
iii)~simulations have several freely specifiable parameters (grid 
parameters and/or gauge parameters), and systematically tuning all of them 
is beyond the aim of this work;
iv)~the convergence order observed actually depends on the triplets
chosen in the self-converge test. The experimentally measured 
convergence factors lie between two and four for both the formulations during 
the inspiral. Around the time of the merger the 
BSSNOK convergence rate drifts up to five or six. The drift is 
on average smaller for higher resolution triplets. This does not mean 
that the BSSNOK data are converging at a high order, but rather that 
the errors in the simulation do not allow us to judge the rate in a 
meaningful way with a simple error model. We do not claim that waveforms 
from either formulation are unreliable, but we are reluctant to estimate 
the absolute errors by Richardson extrapolation, which assumes a certain 
rate of convergence.

\paragraph*{AMSS-NCKU-BAM comparison.} The qualitative features of 
these results have also been observed in simulations with the AMSS-NCKU 
code. Starting with the puncture tracks as in Fig.~\ref{fig:bbh:tracks}, 
we find that the BSSNOK also merge slightly earlier at a given 
resolution, roughly~$3.2\,M$ for BBH0 and~$1.0\,M$ for BBH2, in the same 
ball-park as the values obtained with BAM. Note that we do not expect to
obtain identical values from the two codes, because although they share 
many ingredients, some specifics, for example the shells implementation 
and grid placement, differ. As in Fig.~\ref{fig:bbh:ham2d} 
the Hamiltonian constraint appears to be smaller in the Z4c data when using 
AMSS-NCKU. To test the robustness of this finding we have made a number 
of experiments with different precollapsed initial profiles for the 
lapse. We also tried different constraint damping factors for~$\kappa_1$, 
up to~$\kappa_1=0.1$. Neither change seems to make a significant 
difference; the Z4c Hamiltonian constraint is always smaller. Note 
that in the earlier stability work~\cite{CaoHil11}, the Hamiltonian 
constraint was computed directly from the conformal variables. Now, 
as in BAM, we compute it by transforming first to the ADM variables. 
The two versions of the calculation differ by additions of~$\Theta$ 
and~$\tilde{Z}_i$, and obviously the finite difference approximation. 
The AMSS-NCKU code appears to produce slightly larger errors at mesh 
refinement boundaries than those of BAM. The reason for this is currently 
unclear, but may be the cause of the larger errors that we find in the 
AMSS-NCKU wave-forms. Nevertheless we still find, as in 
Fig.~\ref{fig:bbh:conv} that the Z4c wave forms are roughly twice as 
accurate in phase and amplitude.

\subsection{Equal-mass, irrotational BNS}

\begin{figure*}[t]
  \begin{center}
    \includegraphics[width=0.31\textwidth]{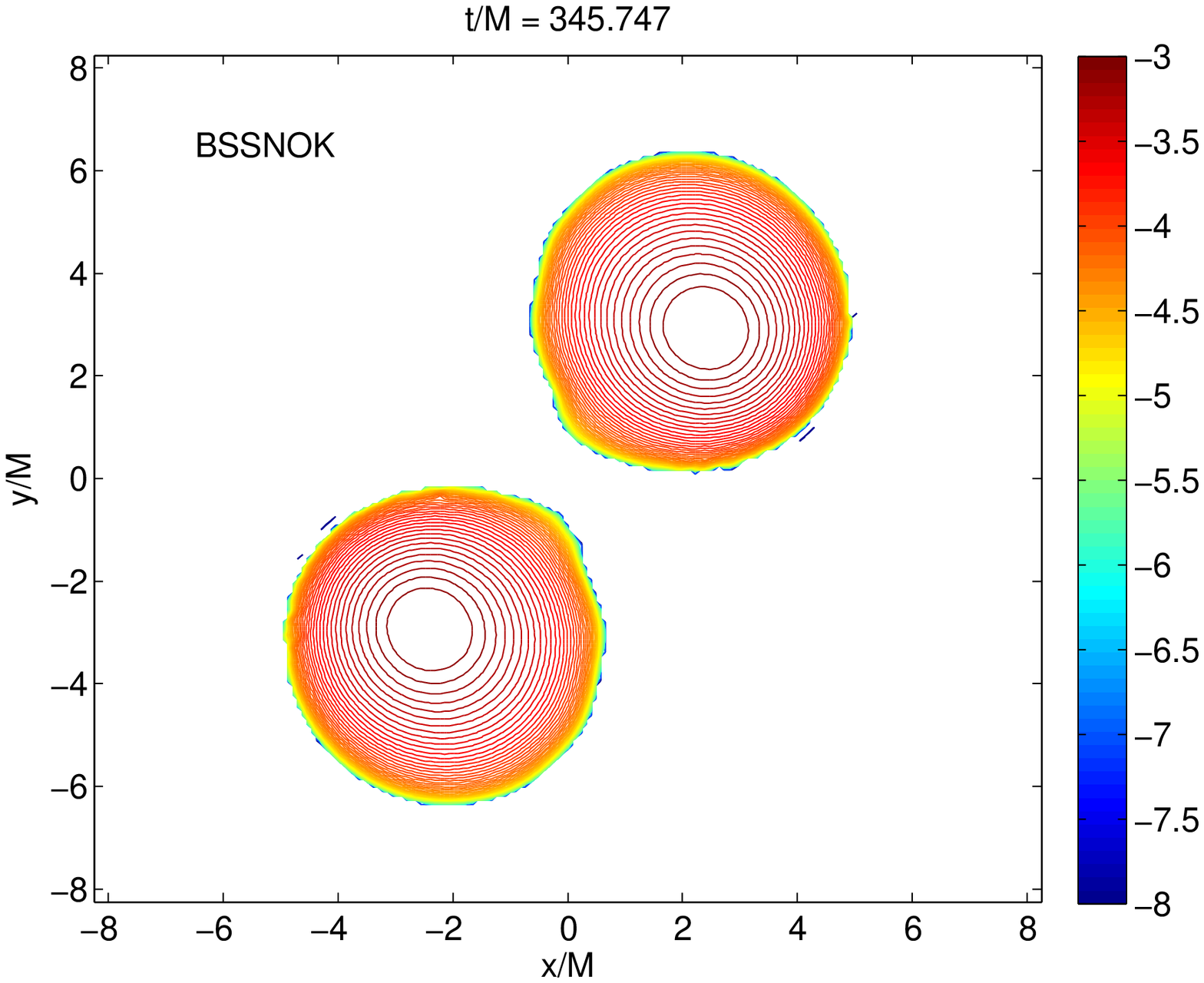}
    \includegraphics[width=0.31\textwidth]{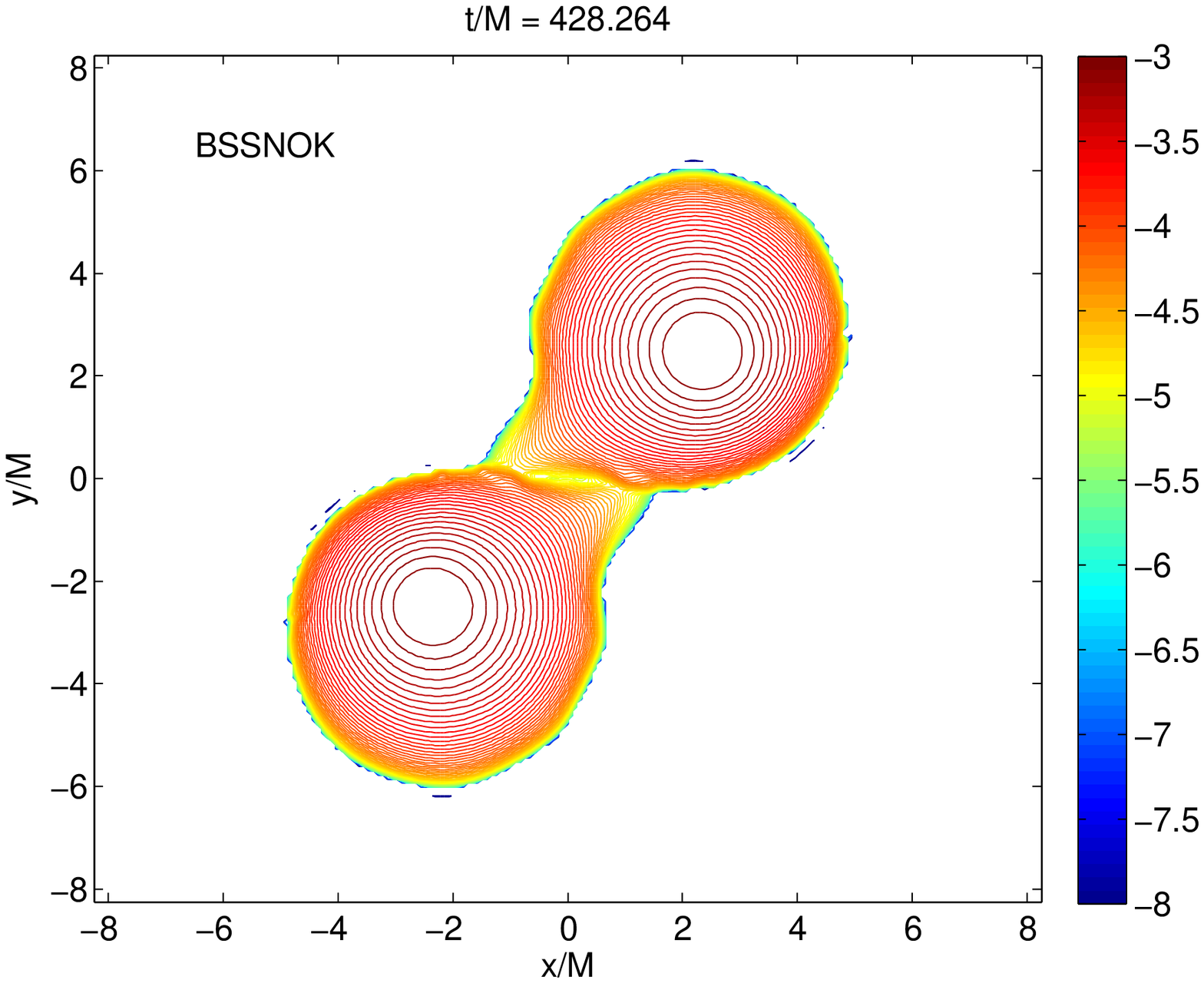}
    \includegraphics[width=0.31\textwidth]{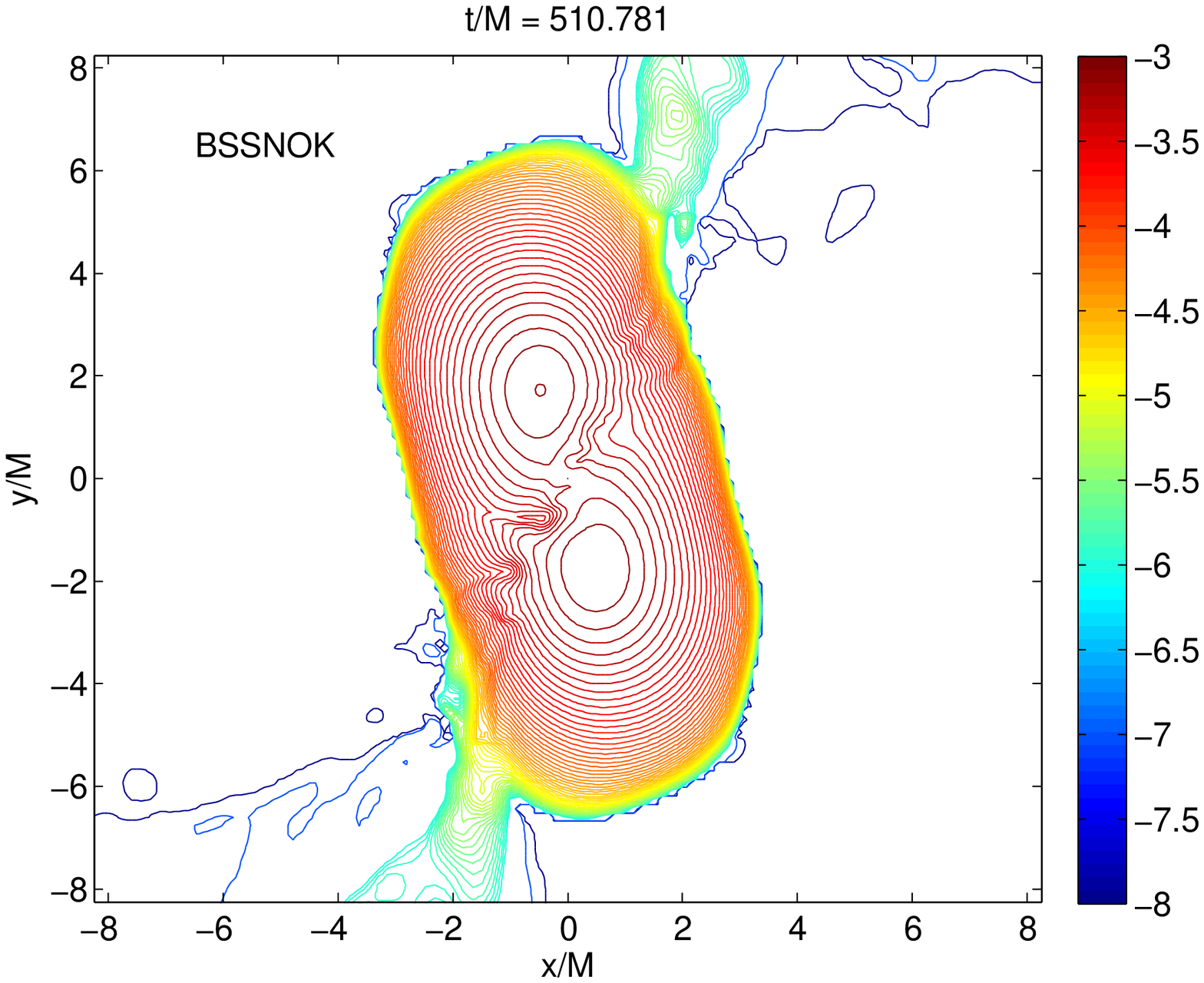}\\
    \includegraphics[width=0.31\textwidth]{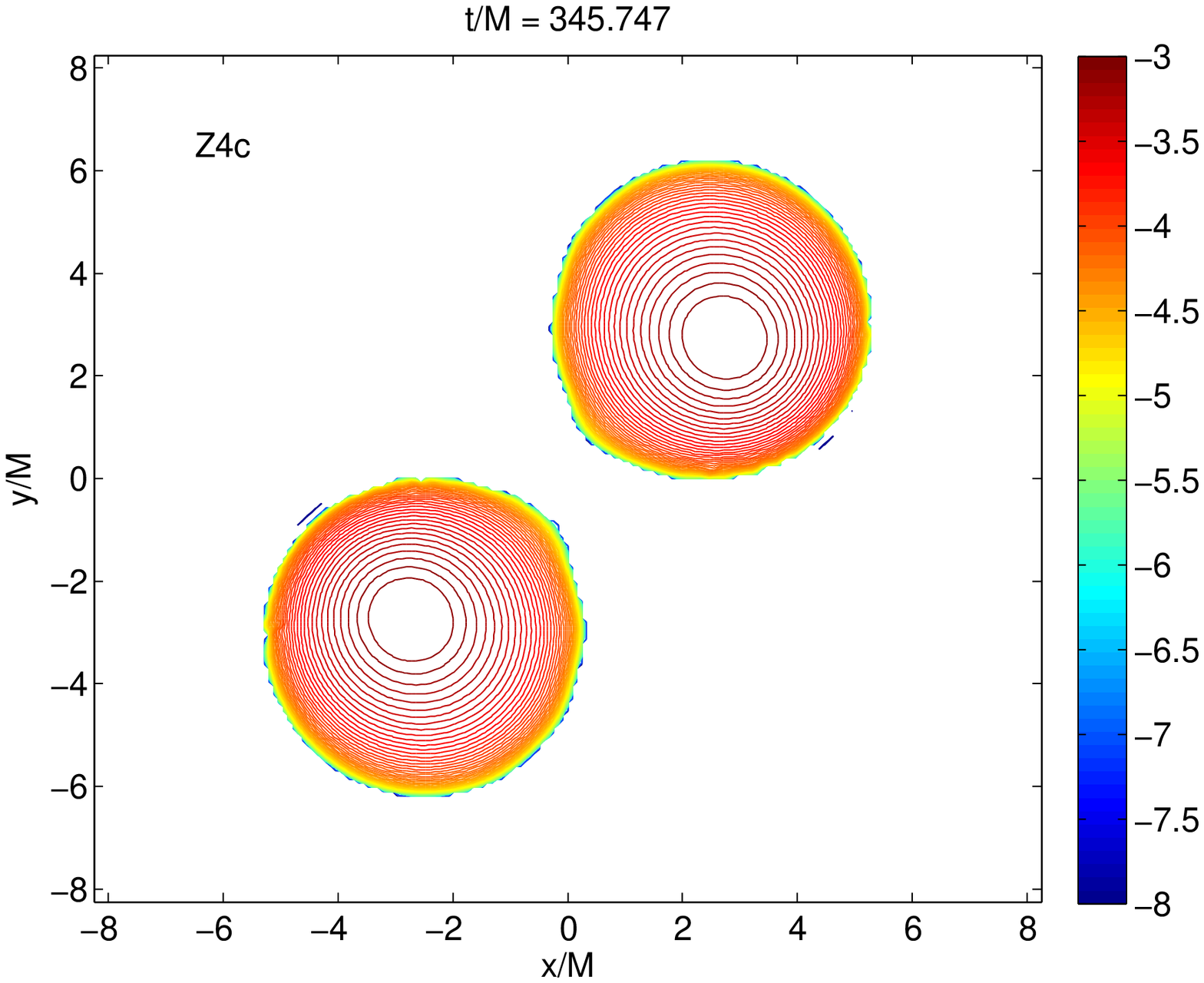}
    \includegraphics[width=0.31\textwidth]{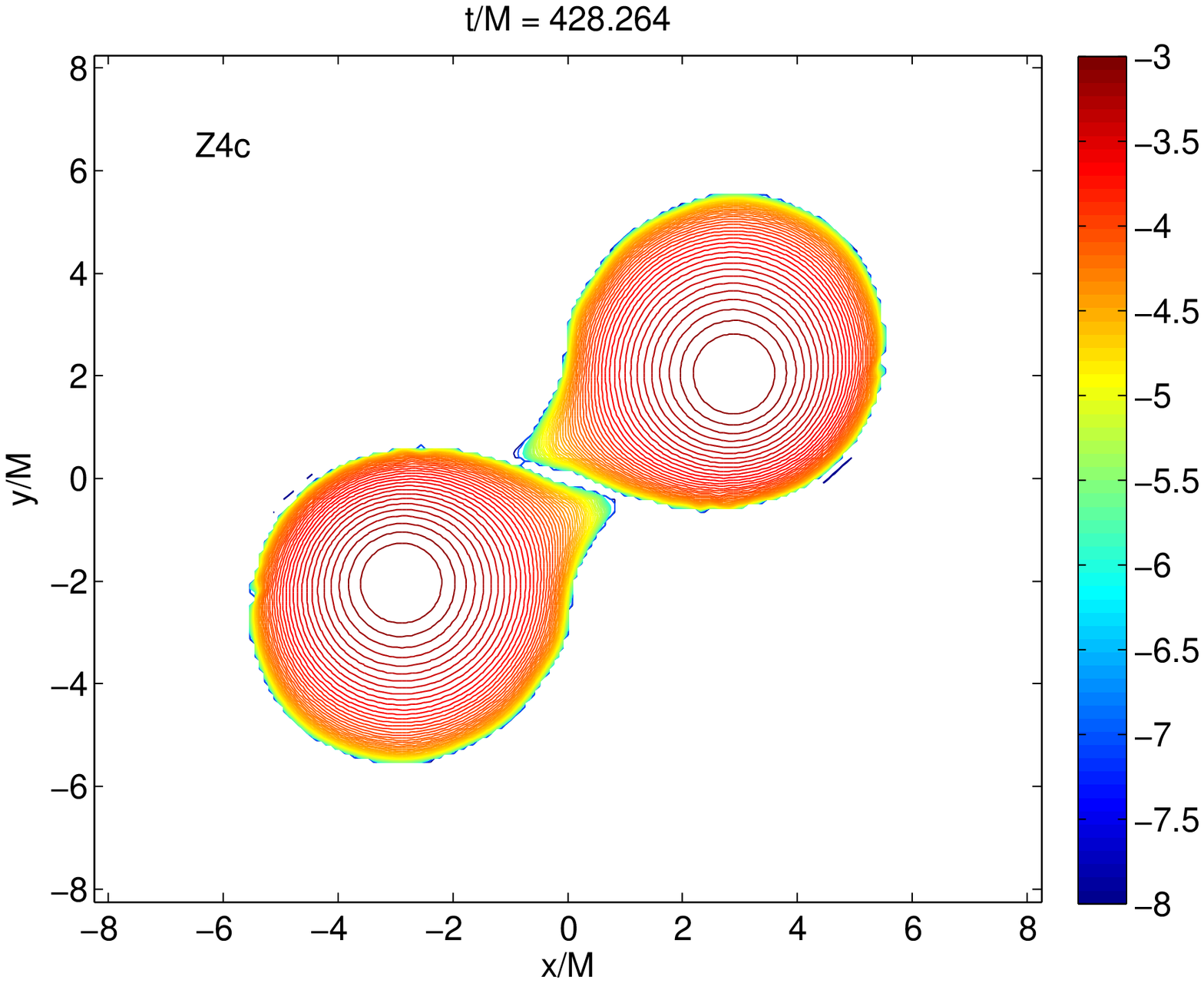}
    \includegraphics[width=0.31\textwidth]{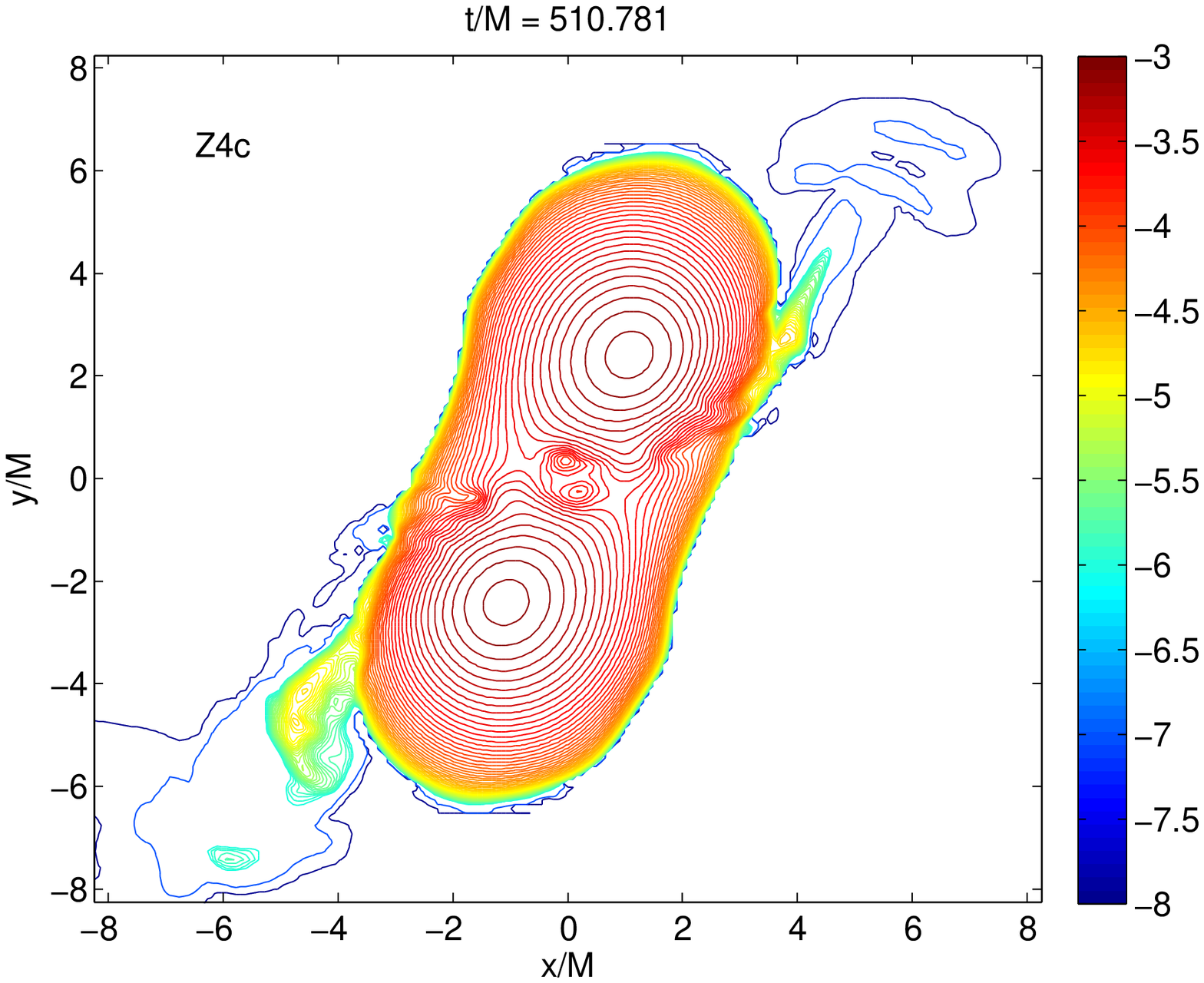}\\
    \caption{\label{fig:bns:rho2d} Evolutions of the rest-mass density
      on the orbital plane as computed with the grid setup BNS6 from
      Tab.~\ref{tab:gridsetup}. The results from the BSSNOK runs are
      plotted in the top panel and those from Z4c underneath. Similar
      comments to those in the caption of Fig.~\ref{fig:bbh:tracks}
      apply; namely, it seems that the compact objects merge earlier
      in the BSSNOK data. In this case however there is already an
      expectation that the Z4c data will be more accurate because in
      earlier work~\cite{ThiBerBru11} it was found that the objects 
      merge later and later as resolution is increased with the BSSNOK 
      formulation.}
  \end{center}
\end{figure*}

\begin{figure}[t]
  \begin{center}
    \includegraphics[width=0.49\textwidth]{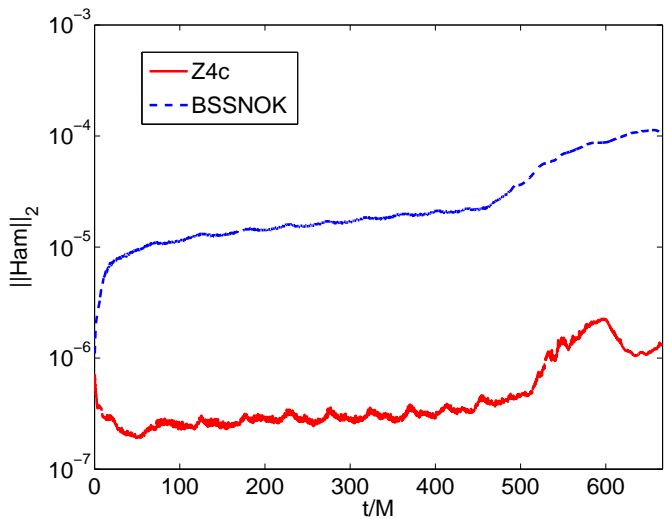}
    \caption{\label{fig:bns:ham_norm} $L_2$ Norm of the Hamiltonian
      constraint computed in the strong field, level~$l=2$, region, with
      the grid setup BNS6 from Tab.~\ref{tab:gridsetup} as in
      Fig.~\ref{fig:bns:rho2d}. At this resolution the Z4c violation
      is between one and two orders of magnitude smaller throughout
      the evolution.
      }
  \end{center}
\end{figure}

\begin{figure*}[t]
  \begin{center}
    \includegraphics[width=0.49\textwidth]{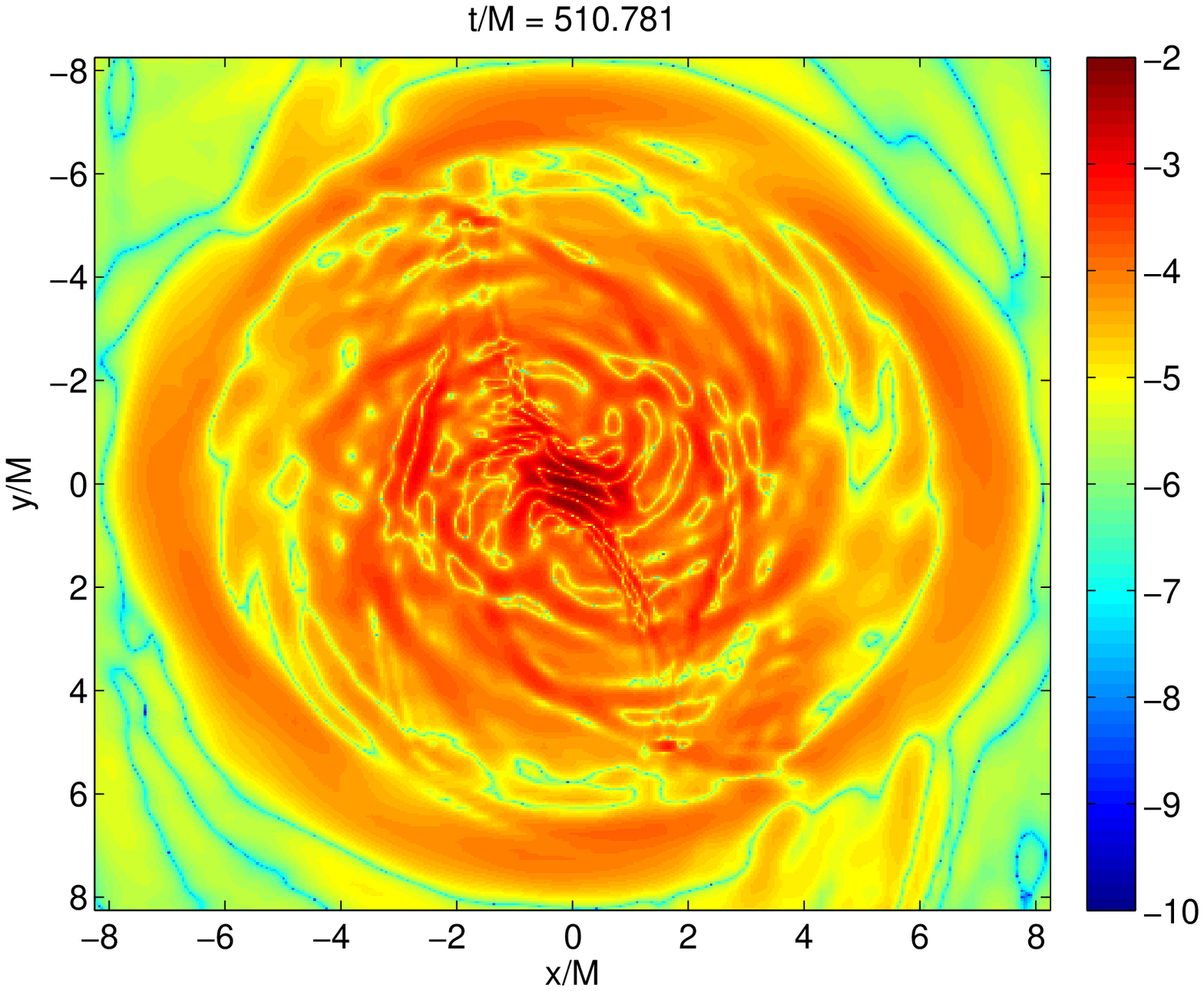}
    \includegraphics[width=0.49\textwidth]{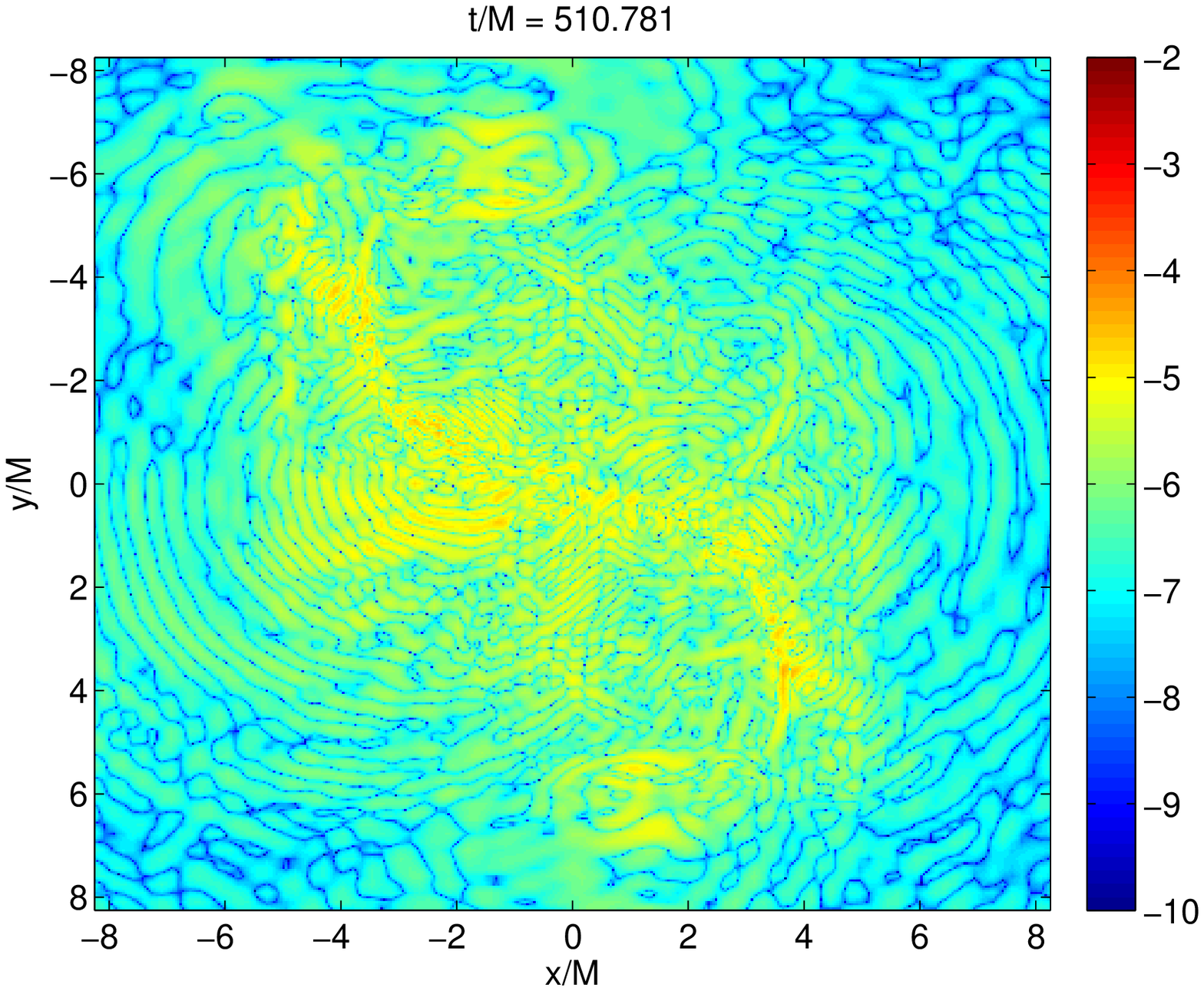}
    \caption{\label{fig:bns:ham2d} Hamiltonian constraint violation in
      BNS simulations at time~$t\sim511\,M$ (around merger) on the orbital
      plane and in the strong field region with the grid setup BNS6 from
      Tab.~\ref{tab:gridsetup} as in Figs.~\ref{fig:bns:rho2d}
      and~\ref{fig:bns:ham_norm}. The plots show~$\log_{10}|H|$; this plot is
      to the BNS data what Fig.~\ref{fig:bbh:ham2d} is to the BBH data.
      In this case there is no sign of the Cartesian grid structure in
      the violation. Most of violation in the BSSNOK simulation appears
      inside the stars. The Z4c violation seems to be on average one or
      two orders of magnitude smaller than the BSSNOK
      violation, as seen more clearly in Fig.~\ref{fig:bns:rho2d}.}
  \end{center}
\end{figure*}

\paragraph*{Basic features of the dynamics.} Let us discuss the evolution
of BNS. The binary evolves approximately two orbits before contact, then
merges forming a hypermassive neutron star (HMNS) which finally collapses
on dynamical timescales. Dynamics and related gravitational wave emission
were described in detail in our previous work~\cite{ThiBerBru11} so we
do not repeat them here. We only mention that the gravitational emission
is characterized by approximately six cycles during which the GW frequency
increases monotonically (even after contact), the merger time is defined
conventionally at peak of the amplitude's~$(2,2)$ mode, nonlinear (quasi-radial
and non-axisymmetric) oscillations of the HMNS generate the post-merger
signal which decays exponentially after collapse. Figure~\ref{fig:bns:rho2d}
shows snapshots of the rest-mass density on the orbital plane around merger
for a typical simulation obtained with the two formulation (and same
setup). There are visible differences in the ``dynamics'', and in the
Z4c run the contact and merger happens at a later simulation time. As
in the case of the BBH puncture tracks, these plots are gauge
dependent but they refer to the same gauge choice up to truncation
errors and so they can be compared. In this respect, note that the
centroids of the stars are offset, this is a coordinate effect due to the
choice~$\eta=2$ in the shift condition and was studied in detail
in~\cite{ThiBerBru11}.

\paragraph*{Constraint violation.} The~$L_2$ norm of the
Hamiltonian constraint violation during the evolution is reported in
Fig.~\ref{fig:bns:ham_norm}. On the refinement level~$l=2$ one can observe
an improvement of a factor~$\sim100$ during the whole evolution if Z4c
is employed. By contrast the norms of each component of the 
momentum constraint agree for both evolutions, an almost constant violation
around~$10^{-7}$ is observed. The Hamiltonian violation on the orbital
plane around merger time is shown in Fig.~\ref{fig:bns:ham2d}, which
includes now refinement levels~$l=2$ and~$l=3$. Even using the highest
resolution run (BNS6) we register a two-to-three order magnitude
difference in the absolute value of the Hamiltonian constraint. It is
noteworthy that the violation has an almost spherical
pattern on and around the strong field region of the binary, which
suggests that the violation is {\it not} dominated by rectangular 
mesh refinement boundaries.

\paragraph*{Gravitational wave accuracy.} The differences at finite
resolution between BSSNOK and Z4c described so far have an impact on
the computation of physical quantities: gravitational waves and ADM mass.
Figure~\ref{fig:bns:conv} shows the result of a standard three-level
self-convergence test for phase and amplitude of the~$(2,2)$ mode of
the gravitational radiation emitted. For the particular triplet shown 
BSSNOK data cease converging before contact. An analogue result was 
already found at similar resolutions in~\cite{ThiBerBru11} (see also
the detailed discussion in~\cite{BerNagThi12,BerThiBru11}). On
the other hand the Z4c data are found to converge at approximately
second order rate {\it beyond} contact and up to merger
($t\sim650\,M$). Similarly to the BBH case, Z4c waveforms are found to
be more accurate by a factor three-to-four in accumulated phase and
amplitude to merger time. We relate this behavior with the
improvement obtained in Hamiltonian constraint preservation. Because
truncation errors and the Hamiltonian violation (especially in the
matter region) in BNS simulations are generically larger, 
the use of Z4c makes a significant difference for the accuracy of
these simulations. 

\begin{figure*}[t]
  \begin{center}
    \includegraphics[width=0.49\textwidth]{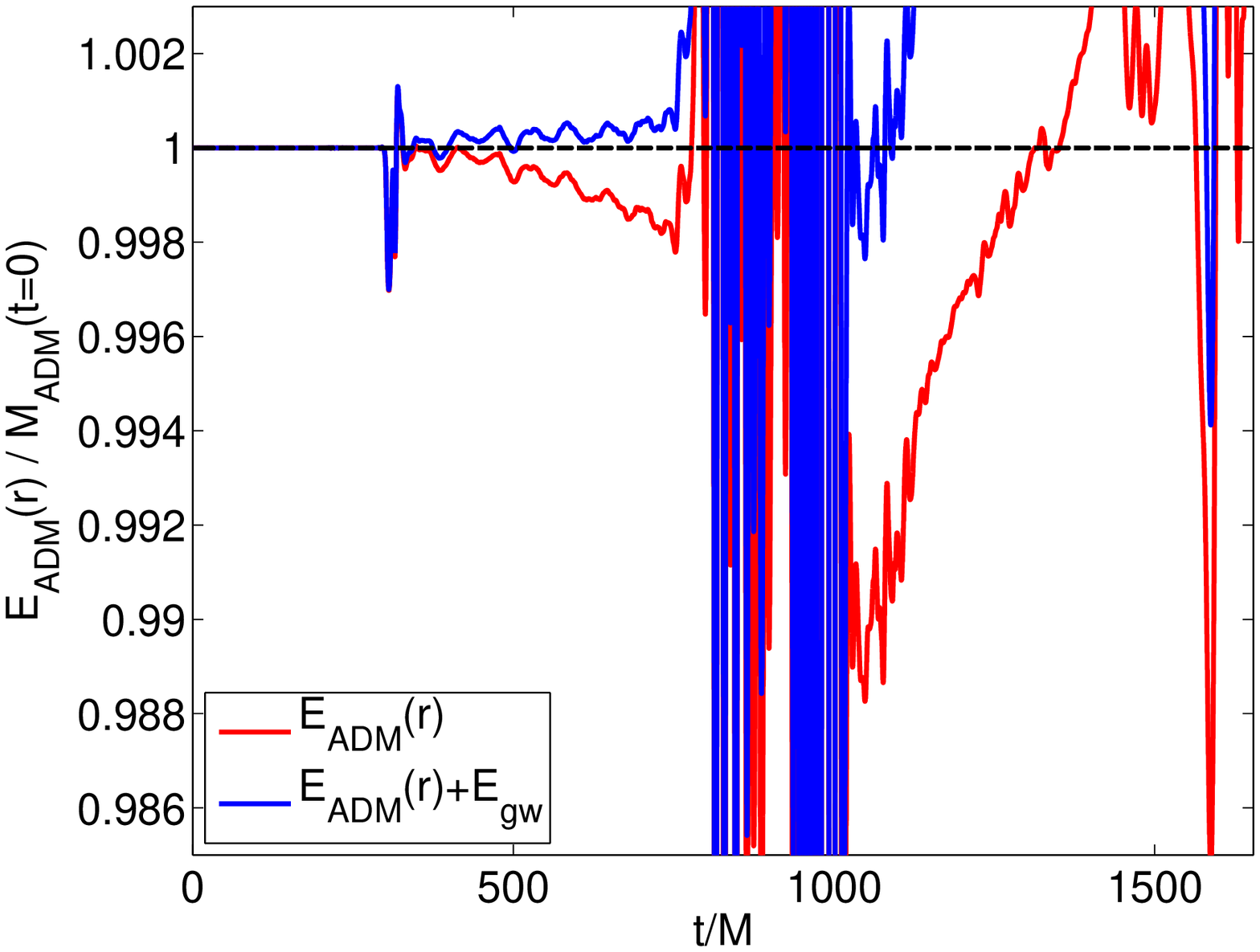}
    \includegraphics[width=0.49\textwidth]{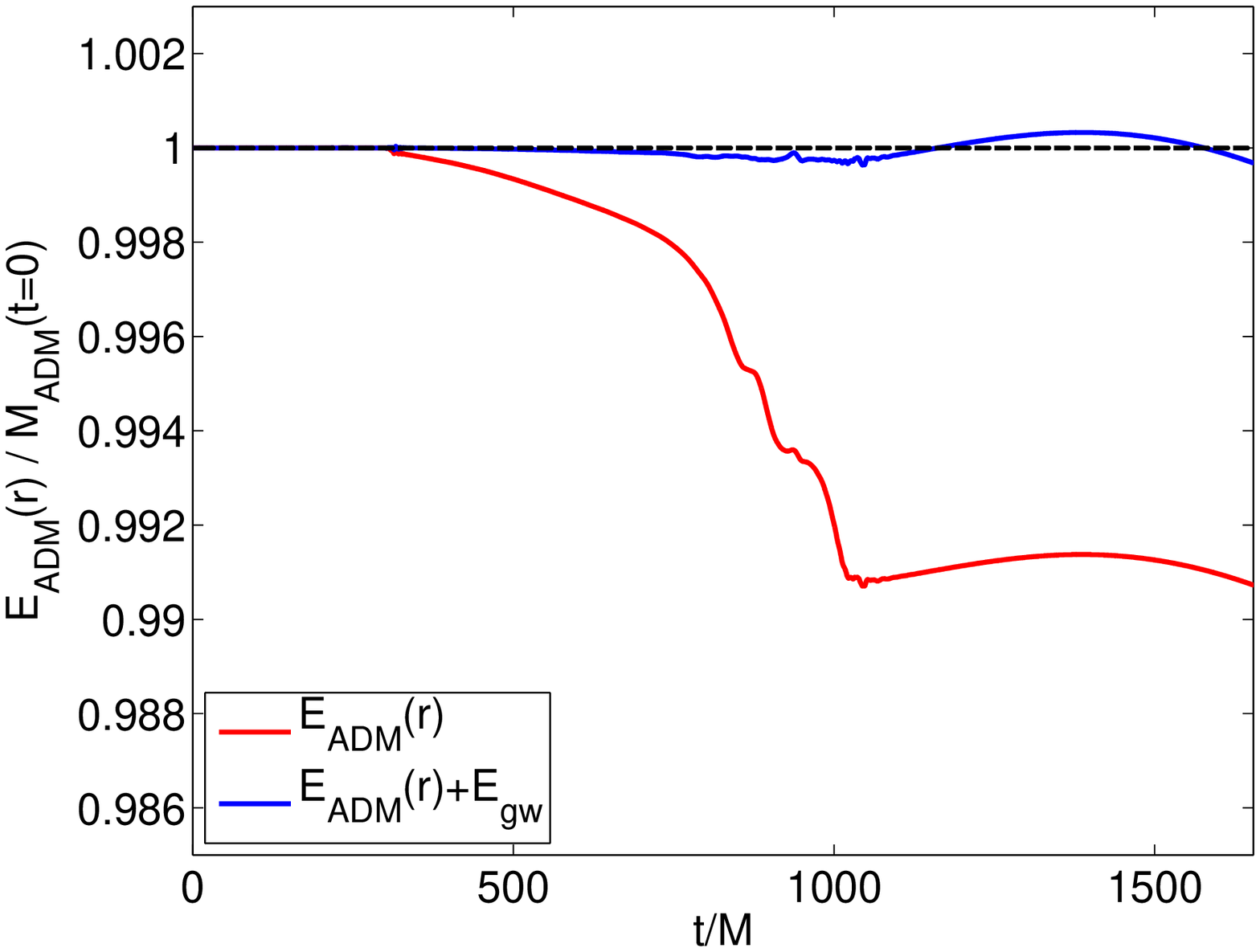}
    \caption{ \label{fig:bns:Madm} ADM mass and radiated energy for BNS
      simulations with the grid setup BNS2a from Tab~\ref{tab:gridsetup}.
      Both are extracted inside the shells at~$R=300M$. The GW energy
      is computed from~$\psi^4$ according to~\cite{BruGonHan06}. The left
      panel shows results for BSSNOK, the right panel for Z4c; in
      comparison the latter demonstrates remarkable conservation.
      }
  \end{center}
\end{figure*}

\begin{figure*}[t]
  \begin{center}
    \includegraphics[width=0.9\textwidth]{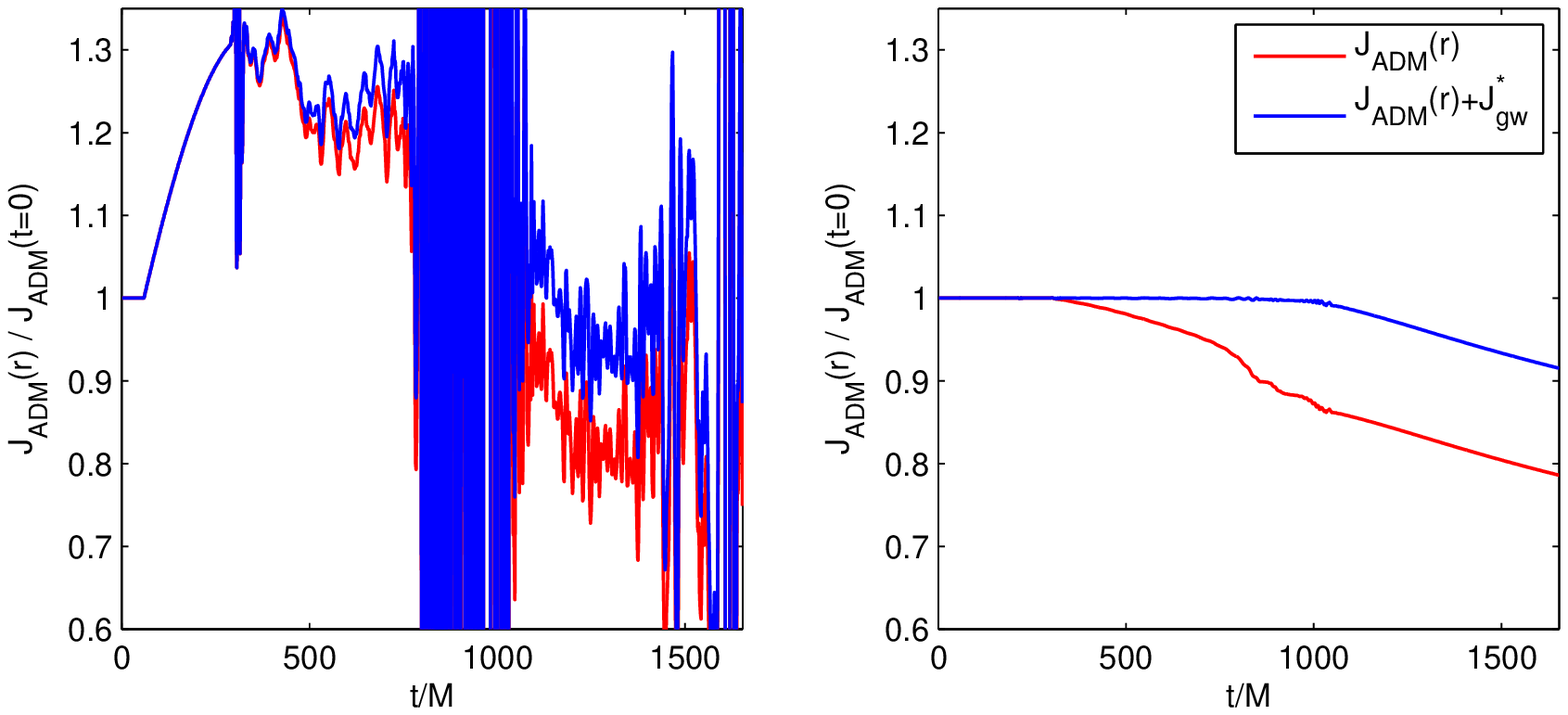}
    \caption{\label{fig:bns:Jadm} ADM angular momentum and radiated energy 
      for BNS simulations with the grid setup BNS2a from 
      Tab.~\ref{tab:gridsetup} computed at an extraction radius 
      of~$R=300M$. The radiated angular momentum $J_{GW}$ is computed 
      according to standard formulas, see for example~\cite{RuiAlcNun07}. The 
      left panel shows results for BSSNOK, the right panel for Z4c. The jumps 
      in the BSSNOK data happen exactly as the extraction spheres become 
      causally connected to the outer boundary 
      at~$360\,M$. Other extraction radii show the same feature.
      }
  \end{center}
\end{figure*}

\paragraph*{Mass and angular momentum conservation.} As in the 
case of a single spinning puncture we consider the ADM mass integral, 
Eq.~\eqref{eq:ADM_Ei}, and the energy radiated in GWs, Eq.~\eqref{eq:Erad_psi4}.
In Fig.~\ref{fig:bns:Madm} we show that Z4c permits a reliable
computation of $E_{\rm ADM}(r)$. When corrected for the GW energy, the
conservation of the ADM mass is of order of $0.1\%$. By contrast
BSSNOK data do not even allow for a reliable estimate of $M_{\rm
  ADM}$. Note that the ADM mass can be also estimated by means of a
volume integral rather than a surface integral, see
e.g.~\cite{YoBauSha02,MarTicBru07}. The more expensive volume integral 
computation can be more accurate and is found to give better results for 
BSSNOK in the case of black hole binaries~\cite{MarTicBru07}.
Figure~\ref{fig:bns:Jadm} shows the ADM angular momentum integral, both 
with and without a correction by radiated angular momentum, at an extraction 
radius of~$300\,M$, for both BSSNOK and Z4c with CPBCs. The outer boundary 
for this run (BNS2a
in Tab.~\ref{tab:gridsetup}) is at approximately~$360\,M$. Evidently at
either radius the BSSNOK data is very poorly behaved, and has large error.
In contrast in the Z4c evolution the corrected angular momentum are well
conserved until the merger signal, which we expect to be much less accurate, 
reaches the extraction sphere. Early in the simulation the BSSNOK data 
are also conserved, but there is a jump at {\it exactly} the time~$t=60\,M$ 
when the outer boundary becomes causally connected to the extraction radius. 
We checked that this feature holds on every extraction sphere. This 
demonstrates clearly that the Sommerfeld boundary condition has an effect 
on the behavior of physical quantities inside BSSNOK simulations. One way 
to avoid the feature would be to place the outer boundary further away, 
which, depending on the physics of interest, may not be prohibitively 
expensive with the spherical shells for the wave zone, but it is not 
desirable to discard every extraction sphere as soon as it becomes causally 
connected to the outer boundary, because a larger domain would ideally 
be used for more reliable wave extraction, rather than as a buffer for poor 
boundary conditions. For comparison it would be very interesting to see data from 
BSSNOK simulations, where there is still a zero-speed mode in the constraints, 
with an implementation of the constraint preserving conditions 
of~\cite{NunSar09}. It is natural to compare the early times in  
Figures~\ref{fig:bns:Madm} and~\ref{fig:bns:Jadm} to Fig.~$11$ 
of~\cite{DueFouKid08}, a similar plot, computed with the spectral Einstein
code, for inspiralling black hole neutron star binary data. 
The Z4c results are competitive, although it remains to be seen if they 
will continue to be so over many orbits. 

\begin{figure*}[t]
  \begin{center}
    \includegraphics[width=0.49\textwidth]{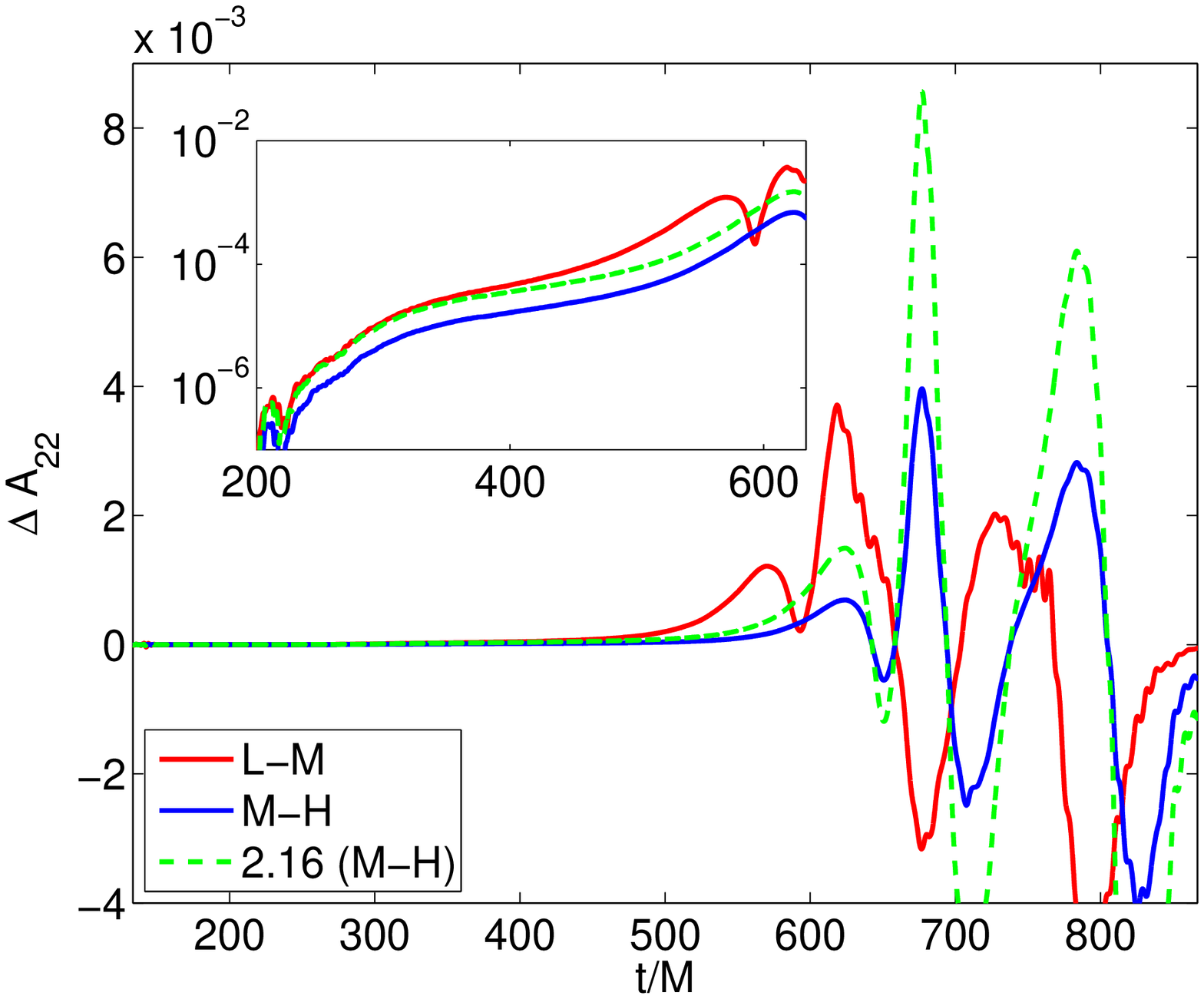}
    \includegraphics[width=0.49\textwidth]{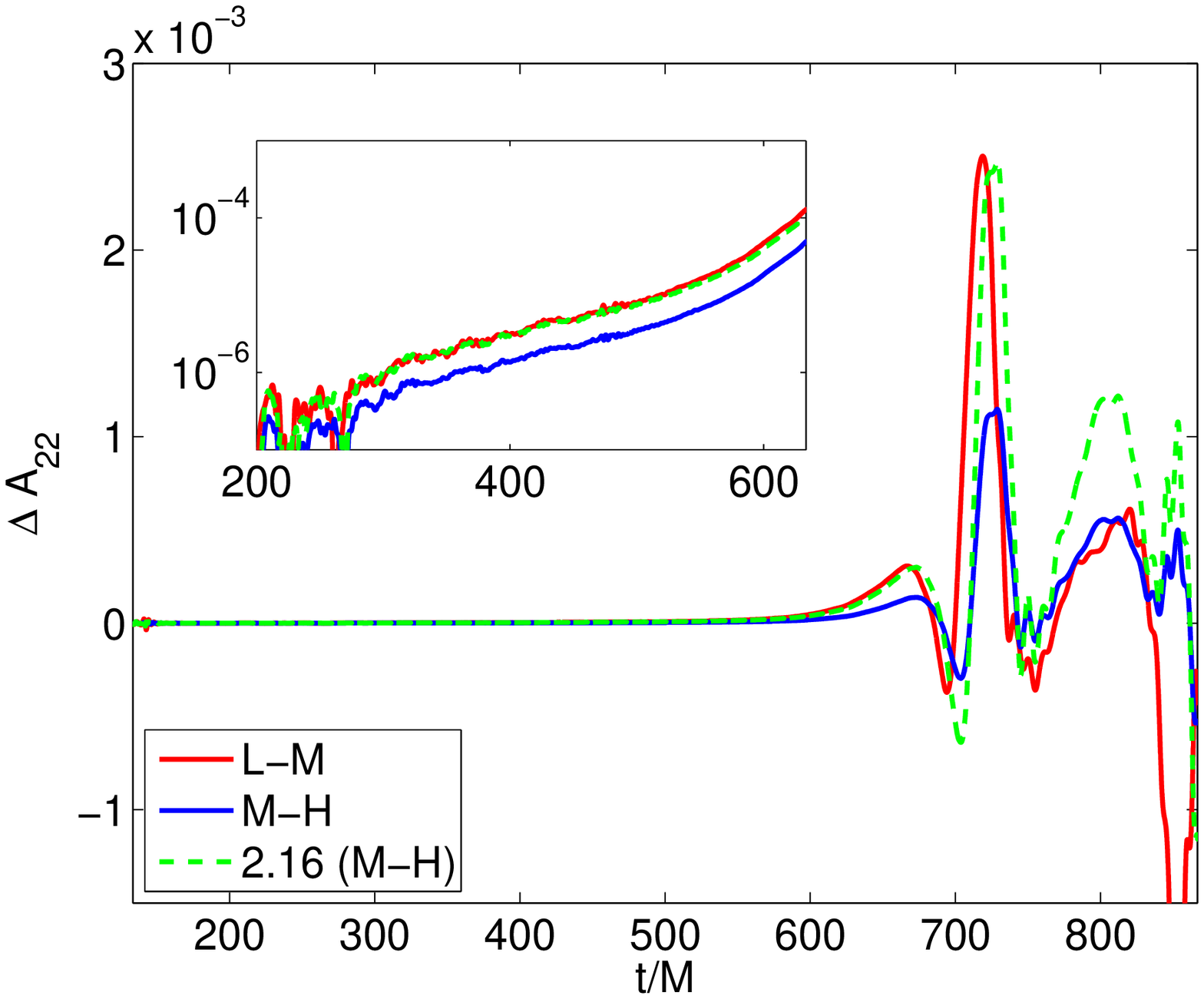}
    \includegraphics[width=0.49\textwidth]{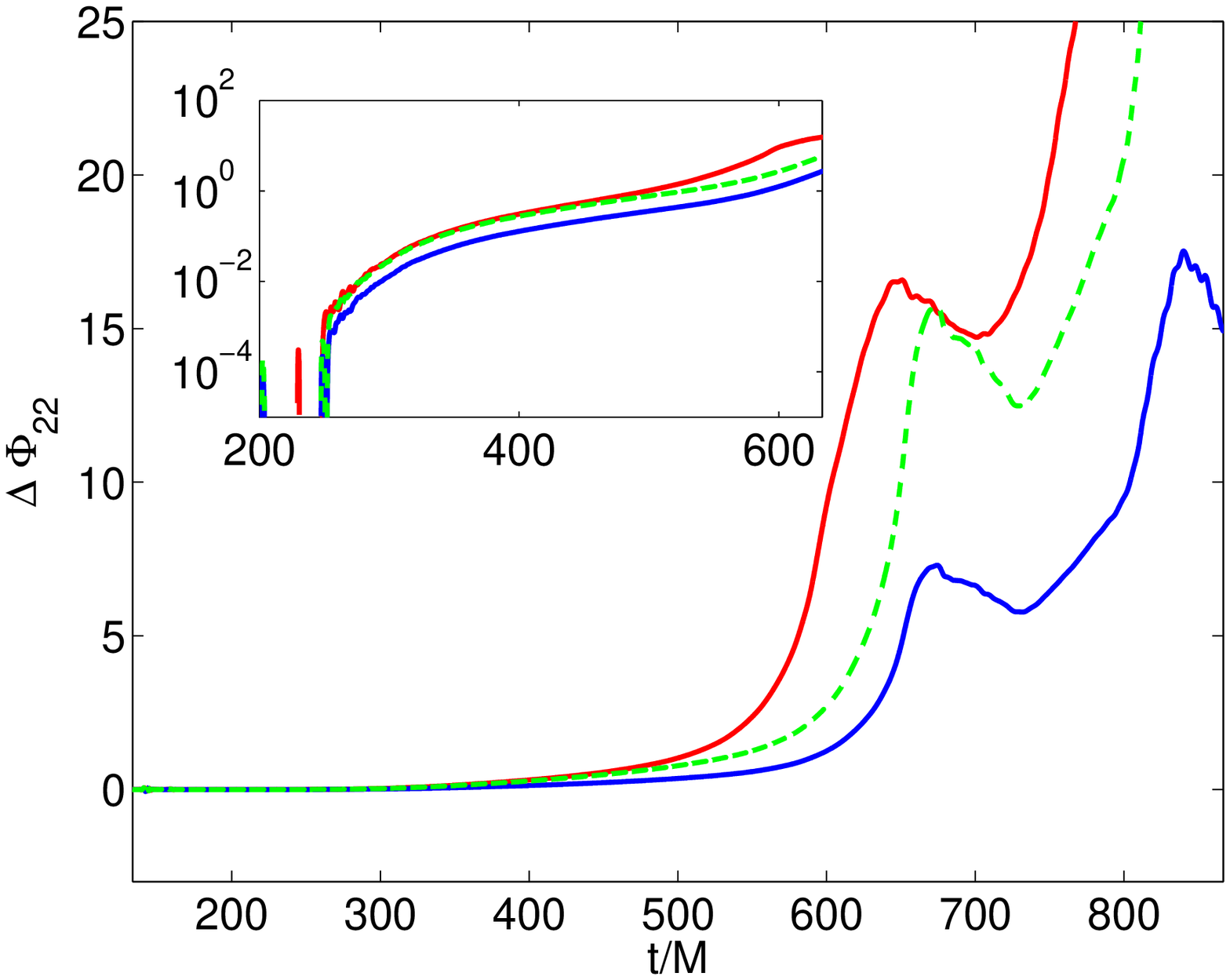}
    \includegraphics[width=0.49\textwidth]{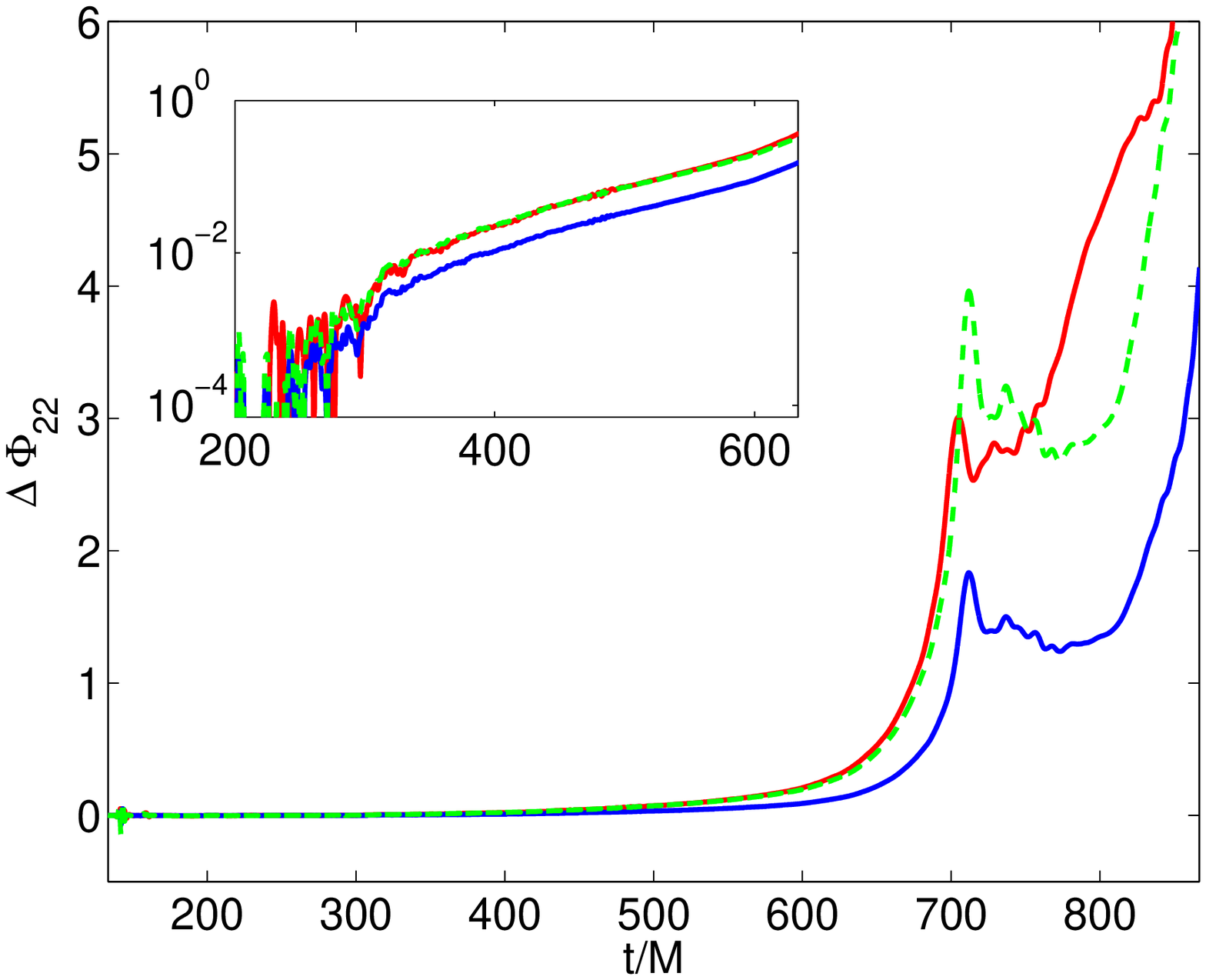}
    \caption{\label{fig:bns:conv} Convergence plot of binary neutron
      star inspiral for the resolutions $h=1/48,1/64,1/80$ (runs BNS0,
      BNS2, and BNS4). The left
      panel shows results for BSSNOK, the right panel for Z4c. All
      the differences are scaled for 2nd order convergence. The
      extraction radius is at $R = 400\,M$ within the shells.
      The qualitative behavior of all curves does not change at
      different extraction radii.}

  \end{center}
\end{figure*}

\section{Conclusions}
\label{sec:conc}

This paper is the conclusion of a 
series~\cite{BerHil09,RuiHilBer10,WeyBerHil11,CaoHil11,HilRui12}, 
the aim of which was to bring the advantages of the generalized 
harmonic formulation to the moving puncture method. We presented 
here for the first time 3D numerical relativity simulations of 
compact binaries performed with Z4c, a conformal decomposition of 
the Z4 formulation.

We started with evolutions of single compact objects and found that
the expectations obtained by earlier work in spherical symmetry are
largely borne out in the 3D numerics. The most striking feature
in these tests is that, in the evolution of a single stable star
by~$t=1000\,M$ at the resolutions used in our tests, is that the
norm of the Hamiltonian constraint is approximately three orders
of magnitude smaller in the Z4c data. For the first time we have
presented results which combine radiation controlling,
constraint preserving outer boundary conditions with the moving
puncture method. These boundary conditions removed a perturbation
to the central rest mass density of the star which is present
when using Sommerfeld boundary conditions. At the resolutions
of our tests however the outer boundary is typically not the
leading order contribution to numerical error. In evolutions of
a single spinning puncture we found that the Z4c with the new outer
boundary conditions remove certain constraint violating features
present in the BSSNOK data, but at least at early times the
qualitative physical picture is unaltered.

We then compared evolutions of compact binary spacetimes. In
these tests we placed the outer boundary much farther away from
the central body so as to simplify the discussion. Throughout the
evolution of binary neutron star initial data, we find that at the
same resolutions the Z4c formulation has between one and two orders
of magnitude less Hamiltonian constraint violation in the norm.
Interestingly the Hamiltonian constraint violation in the Z4c tests in
this case stays at or below the level in the initial data, at least
until the stars merge, and the simulations may therefore even be
competitive with those of a constrained formulation. We find similar,
albeit much less pronounced effects in the constraint violation in binary
black hole simulations, but the change of formulation does nothing to
cure the dominant constraint violation at the punctures themselves. The
higher quality of the Z4c data is also apparent in physically meaningful
quantities. In terms of gravitational wave accuracy we find that with
any triplet, satisfying certain criteria, for either binary neutron star
or binary black hole data, the absolute error in either amplitude or
phase of the extracted gravitational waves is between two and four times
smaller in the Z4c evolutions. The difference, in the evolution of compact
binaries, between conservation of the ADM mass integral with the two
formulations is remarkable. In the BSSNOK simulations one can not reliably
correct the integral with the radiated gravitational wave energy to arrive
at a constant. In the Z4c simulations near perfect conservation is
achieved. Furthermore, despite placing the outer boundary at a large
coordinate radius, we find that the BSSNOK data are corrupted, for example
in the angular momentum, by the Sommerfeld boundary condition, whereas
the Z4c data are free of this problem. If nothing else this motivates
the use of constraint preserving boundary conditions with the BSSNOK
formulation.

In summary we have presented a large suite of numerical experiments
in which the Z4c formulation was shown to give more accurate results
than BSSNOK, both in terms of constraint violation and extracted
physical quantities. We therefore expect that the Z4c formulation
will become a standard tool for numerical relativity.

\begin{acknowledgments}
The authors are grateful to Roman Gold, Mark Hannam, Sascha Husa, 
Nathan Johnson-McDaniel, Doreen M\"uller, Milton Ruiz and Andreas 
Weyhausen for helpful discussions. We are especially grateful to Denis 
Pollney for his patient explanation of the Llama multi-patch communication 
algorithm. We are grateful to Tim Dietrich for providing the initial data 
for the spinning puncture evolutions. This work was supported in part 
by DFG grant SFB/Transregio~7 ``Gravitational Wave Astronomy''
as well as by NSF grants PHY-0855315 and PHY-1204334. Computations 
were performed primarily on JUROPA (JSC, J\"ulich), LRZ (Munich), 
and LOUHI (CSC) cluster. Computer time on LOUHI is granted under 
PRACE-Tier-1 2011.
\end{acknowledgments}

\appendix

\section{Spherical patches}
\label{app:shells}

\begin{figure}[t]
\includegraphics[width=0.49\textwidth]{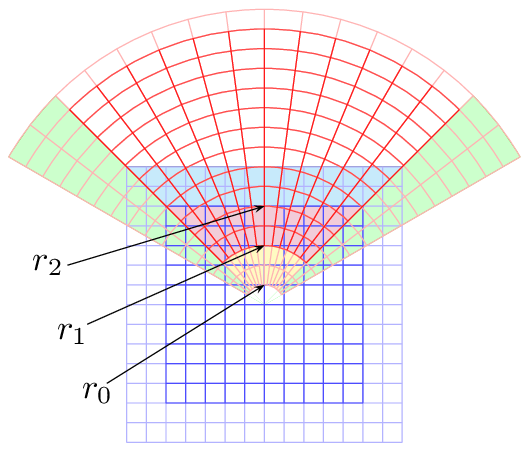}
  \caption{\label{fig:shells} 2D scheme of the spherical patches.
           The darker regions denote the physical grids, the translucent
           numerical ghostzones. See the text for details of the
           various radii and how the color scheme relates to the
           numerical method.}
\end{figure}

In this appendix the implementation of the spherical patches in the BAM
and AMSS-NCKU codes is described. We follow 
closely~\cite{Tho04,PolReiSch11}, to whom we refer for further details. 
We describe first the BAM implementation and then highlight the 
differences in the AMSS-NCKU code.

\paragraph*{The grid structure.} The BAM grid is made of a hierarchy 
of nested Cartesian grid boxes, each level is labeled by the 
integer~$l=0,1,...$. The level~$l=0$ (outermost box) is replaced 
with six patches with local coordinates $a_j=\{R,\phi,\theta\}$. 
The grid of each patch is uniformly spaced in local coordinates. 
The maps between the local coordinates and the Cartesian coordinates 
are
\begin{align}
  \pm x\, & \text{patches:} \quad \phi = \arctan(y/x), \ \theta =
  \arctan(z/x) \\
  \pm y\, & \text{patches:} \quad \phi = \arctan(x/y), \ \theta =
  \arctan(z/y) \\
  \pm z\, & \text{patches:} \quad \phi = \arctan(x/z), \ \theta =
  \arctan(y/z) \,
\end{align}
where~$(\phi,\theta)\in(-\pi/4:\pi/4)\times(-\pi/4:\pi/4)$.
The local radial coordinate~$R$ is a fisheye 
coordinate~\cite{AlcBruDie02,BakCamLou02,BakBruCam00}
\begin{align}
  R = f(r)-f(0) \ ,
\end{align}
as function of the radius $r=\sqrt{x^2+y^2+z^2}$.
The particular function function $f(r)$ implemented is
\begin{eqnarray}
  f(r) &=& B r + A \epsilon \log( \cosh[(r-R)/\epsilon)] \\
  \p_r f(r) &=& A \tanh[(r-R)/\epsilon] + B \ ,
\end{eqnarray}
where the parameters~$A$ and~$B$ determine the stretching of the
coordinate and $\epsilon$ the transition region. The condition
$f(r)=r$ at interface between spherical patches and the boxes is
important to avoid a step behavior and minimize interpolation
errors. In this work we did not employ the fisheye coordinate,
i.e.~we set $A=0$, $B=1$.

\paragraph*{Spatial derivatives.} Field derivatives at a grid point 
inside the patches are calculated by finite  differences on the 
uniformly spaced local grid. The derivatives in Cartesian coordinates 
are obtained using the chain rule and the Jacobian of the 
transformation~$x_i(a_j)$,
\begin{align}
\frac{\p}{\p x_i}&= \left(\frac{\p r_j}{\p x_i}\right)
\frac{\p}{\p r_j}\,, \\
\frac{\p^2}{\p x_i\p x_j}&= \left(\frac{\p r_k}{\p x_i} 
\frac{\p r_l}{\p x_j}\right)
\frac{\p^2}{\p r_k \p r_l}
+\left(\frac{\p^2 r_k}{\p x_i\p x_j}\right)\frac{\p}{\p r_k}\,,
\end{align}
note that the last term in these equations fixes a typo 
in equation~(5b) in~\cite{PolReiSch11}.

\paragraph*{Grid schematic.} The grid structure is sketched in 
Fig.~\ref{fig:shells}. The dark solid lines represent the physical 
grid, the lighter lines denote the ghost points which are needed 
for the finite differences. The green shaded regions denote the 
ghost zones populated by inter-patch interpolation, and overlap 
with the neighbor patches. The other colored areas overlap with 
the~$l=0$ level, and the resolution of the Cartesian box is the same 
as the one of the radial one in the patch.  The number of grid points 
in all the ghost zones (cyan, green, yellow) is equal. The  
distance between the points~$r_0$ and~$r_1$ is equal to that 
between~$r_1$ and~$r_2$.

\paragraph*{Data communication.} Spherical patches overlap on ghosts 
zones. Two neighboring patches share the radial coordinate and the 
angular coordinate perpendicular to the mutual boundary. Therefore 
only a 1D interpolation parallel to the boundary has to be performed 
in the green regions of Fig.~\ref{fig:shells}. A Lagrange interpolation
(sixth order in this work) which uses the most centered possible
stencil is employed. Interpolation between patches and the~$l=0$ 
level is performed with Lagrangian polynomial interpolation 
in 3D (colored regions of Fig.~\ref{fig:shells}). The various 
interpolations are done in the following order:
i)~interpolate from box to shell (yellow region),
ii)~interpolate between shells (green region),
iii)~interpolate from shell to box (cyan and red regions),
iv)~set symmetries in box.
For simplicity, grid symmetries are not applied in the shells during
evolution. Each patch is evolved entirely and only afterward values at
symmetry points are overwritten by copying them. We note that because the 
interpolation of the red region in Fig.~\ref{fig:shells} depends on points 
in the cyan region, the ordering of the different interpolation can give, 
in principle, different results. On the other hand, not interpolating the 
red region results in a ``double evolution'', which we find leads to high 
frequency oscillations.

\paragraph*{Dissipation.} Artificial dissipation is necessary for numerical 
stability during the evolution. In particular to maintain stable 
box-shell interface regions. Experimentally we found it important 
to apply different amounts of dissipation in box and spherical patches. 
In particular we use a {\it lower} dissipation on the spherical patches 
than in the box. For this work we tested only sixth order Kreiss-Oliger 
dissipation operators, using artificial dissipation coefficients~$\sigma=0.5$ 
in the bulk and~$\sigma=0.1$ in the shells (see~\cite{BruGonHan06} for 
our terminology). The number of angular points has been chosen according 
to the size of the~$l=0$ box,~$n_{\theta,\, \phi}\sim n/2$. Placing the 
box-shell interface close to the strong field region may also cause large 
growth of the error, even causing the code to crash, or affect the choice 
of the dissipation parameters required for stability.

\paragraph*{Parallelization.} In order to reduce global MPI communication 
during shells synchronization, we use an optimized parallelization. The 
computations on the patches are distributed only in the radial direction.
Every processor has six patch parts with the same radial extension,
hence the synchronization in the angular directions is performed locally by 
each MPI job. Note that this method implies a minimum number of radial 
points is necessary for a given number of processors. In some early tests 
we found that avoiding interpolation in this way can result in computations 
that are an order of magnitude faster in the shells.

\paragraph*{Variations of the numerical method in the AMSS-NCKU 
code.} There are several differences in the implementation of shells 
in AMSS-NCKU with respect to BAM. In AMSS-NCKU there is no option 
to use a radial fish-eye coordinate. Relating to the interpolation
between box and shell, the scheme is somewhat different to BAM. We 
set the length between the point~$r_0$ and~$r_1$ as the length of 
six points and we take these points as buffer points. We set the 
distance between~$r_1$ and~$r_2$ larger than the distance between~$r_0$
and~$r_1$. We take this part of region as the double cover region 
of both box and shell. We set six points for the box outside of~$r_2$ and
take them as buffer points. Similar to the treatment of the mesh
refinement interface, we fill the buffer points for box and shell
only at the end of a full RK4 step. As opposed to BAM we do not
interpolate the points between~$r_1$ and~$r_2$. Due to this double 
cover region, we can interpolate between box and shell in parallel 
fashion. But before that we have to synchronize the data between 
different shell patches. In the MPI communication, we divide the 
data in both radial direction and angular directions but try 
to make resulting blocks as cubic as possible for minimization of 
inter processor data change.

\begin{figure*}[t]
  \includegraphics[width=0.49\textwidth]{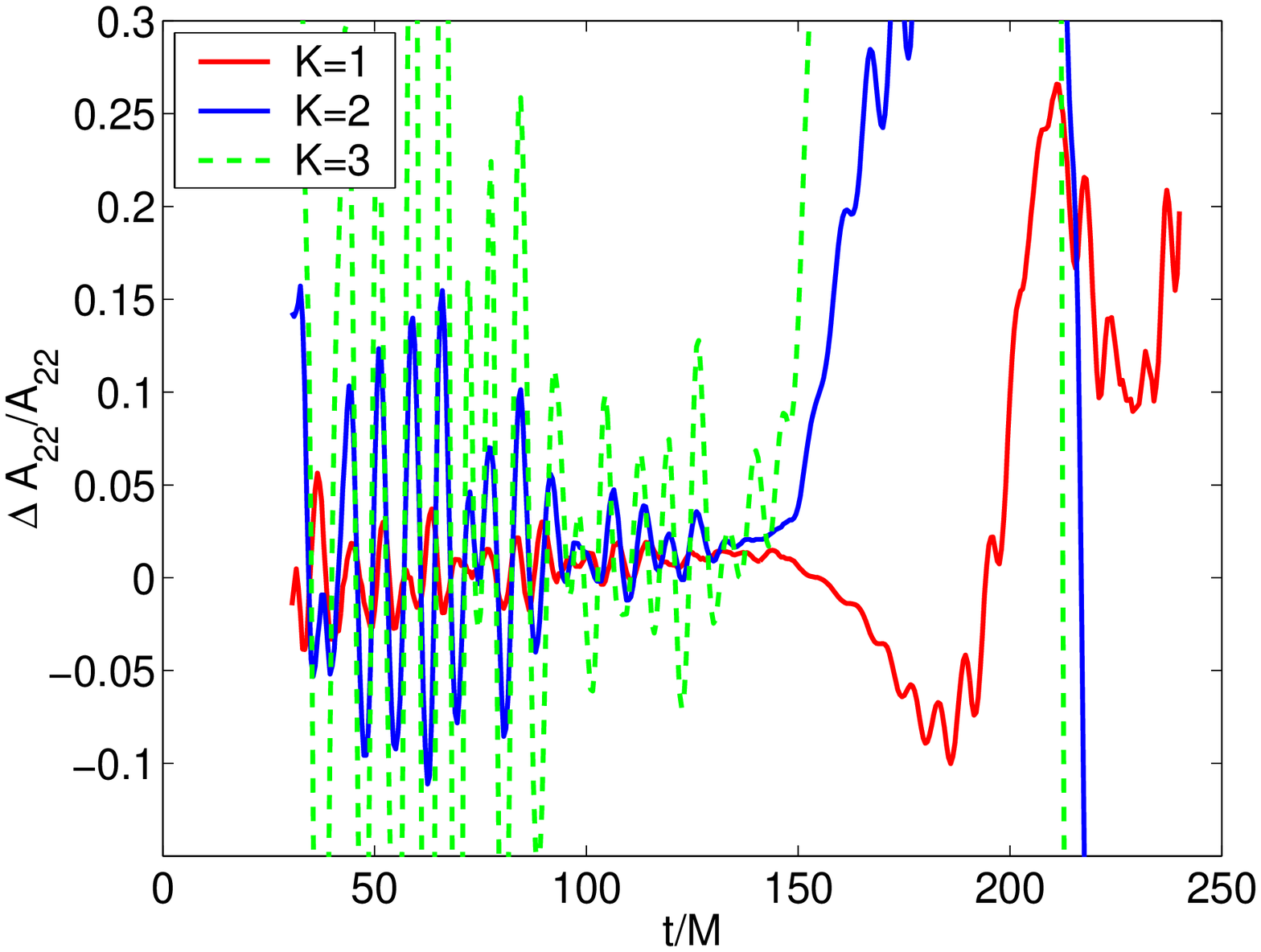}
  \includegraphics[width=0.49\textwidth]{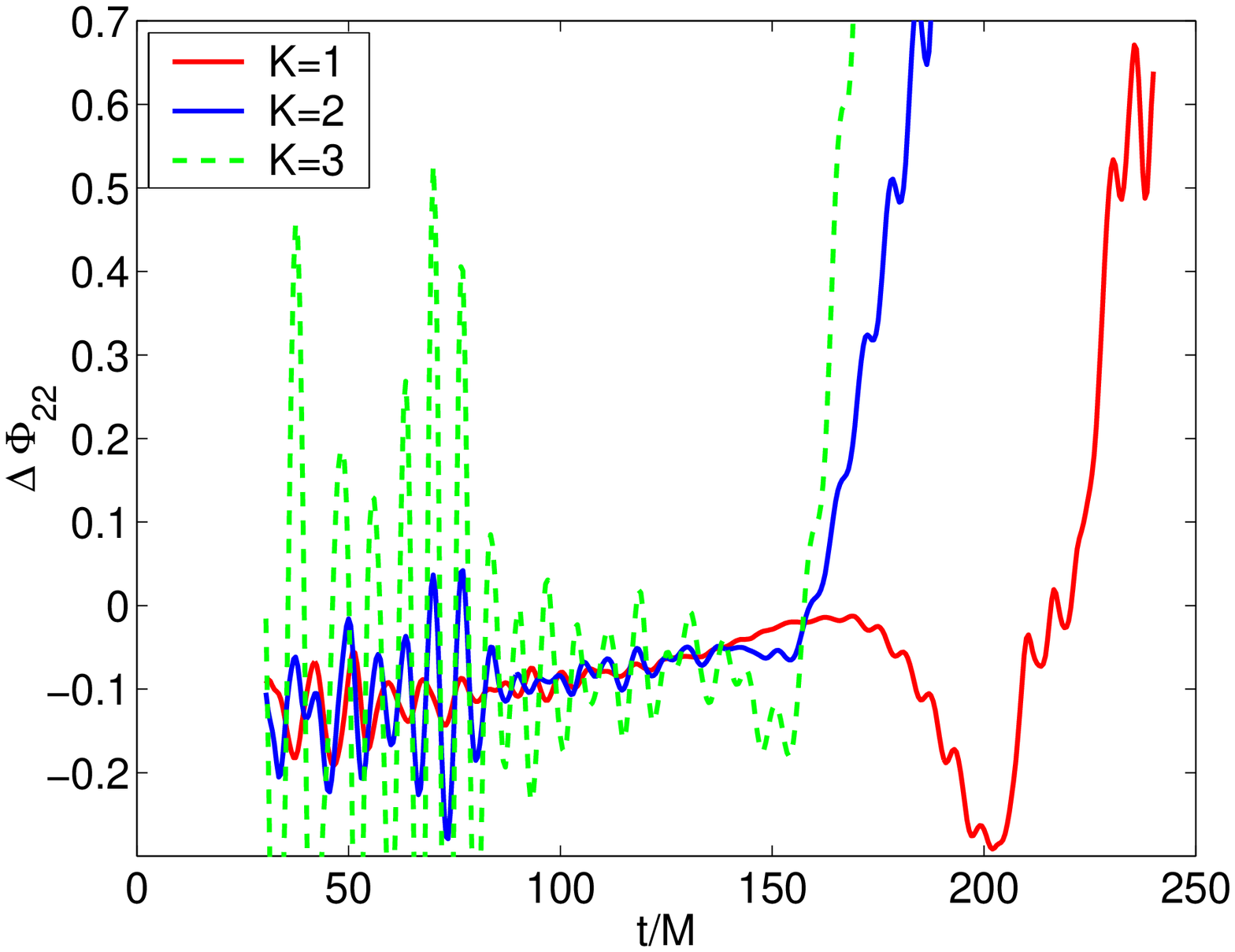}\\
  \includegraphics[width=0.49\textwidth]{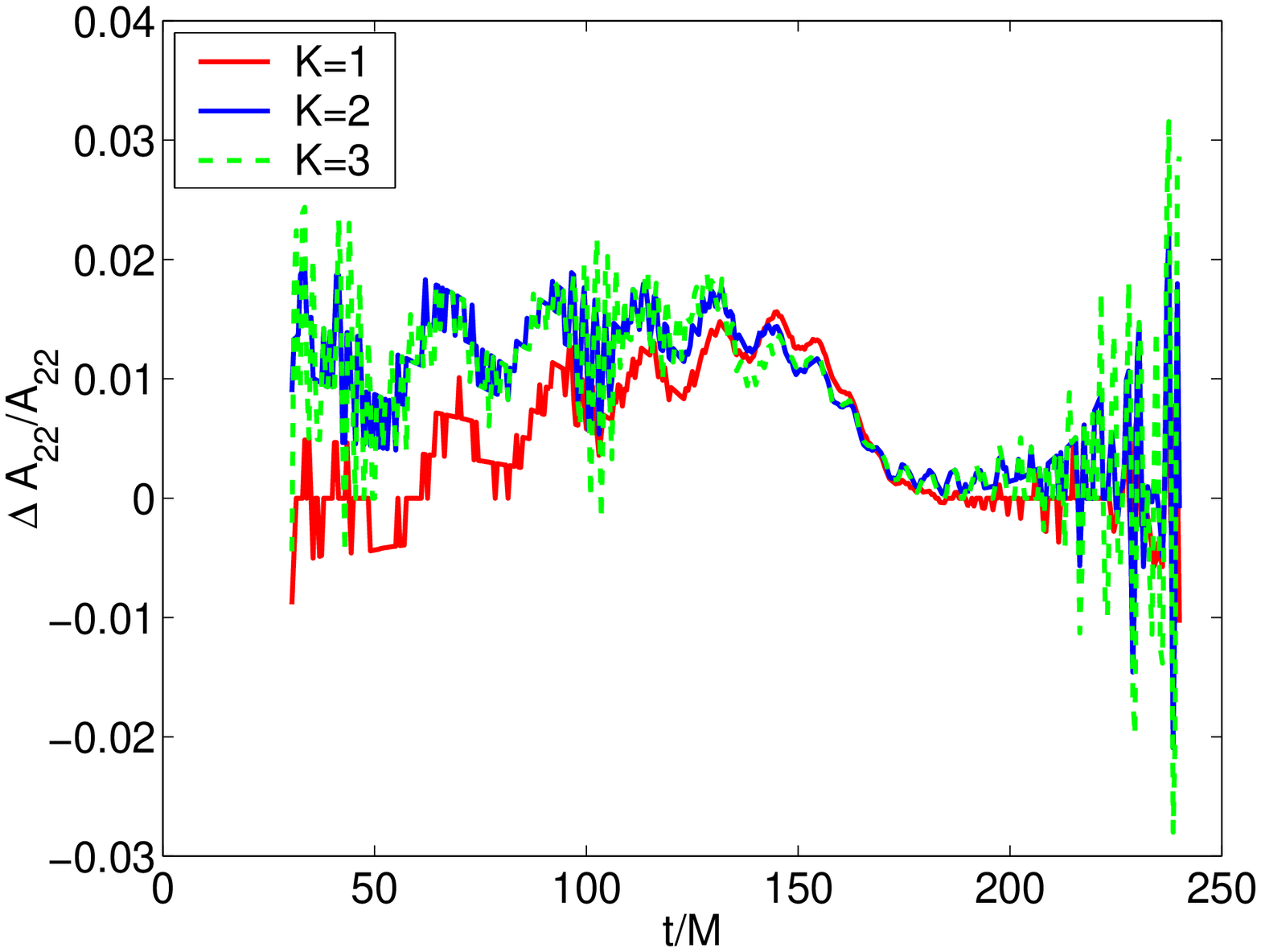}
  \includegraphics[width=0.49\textwidth]{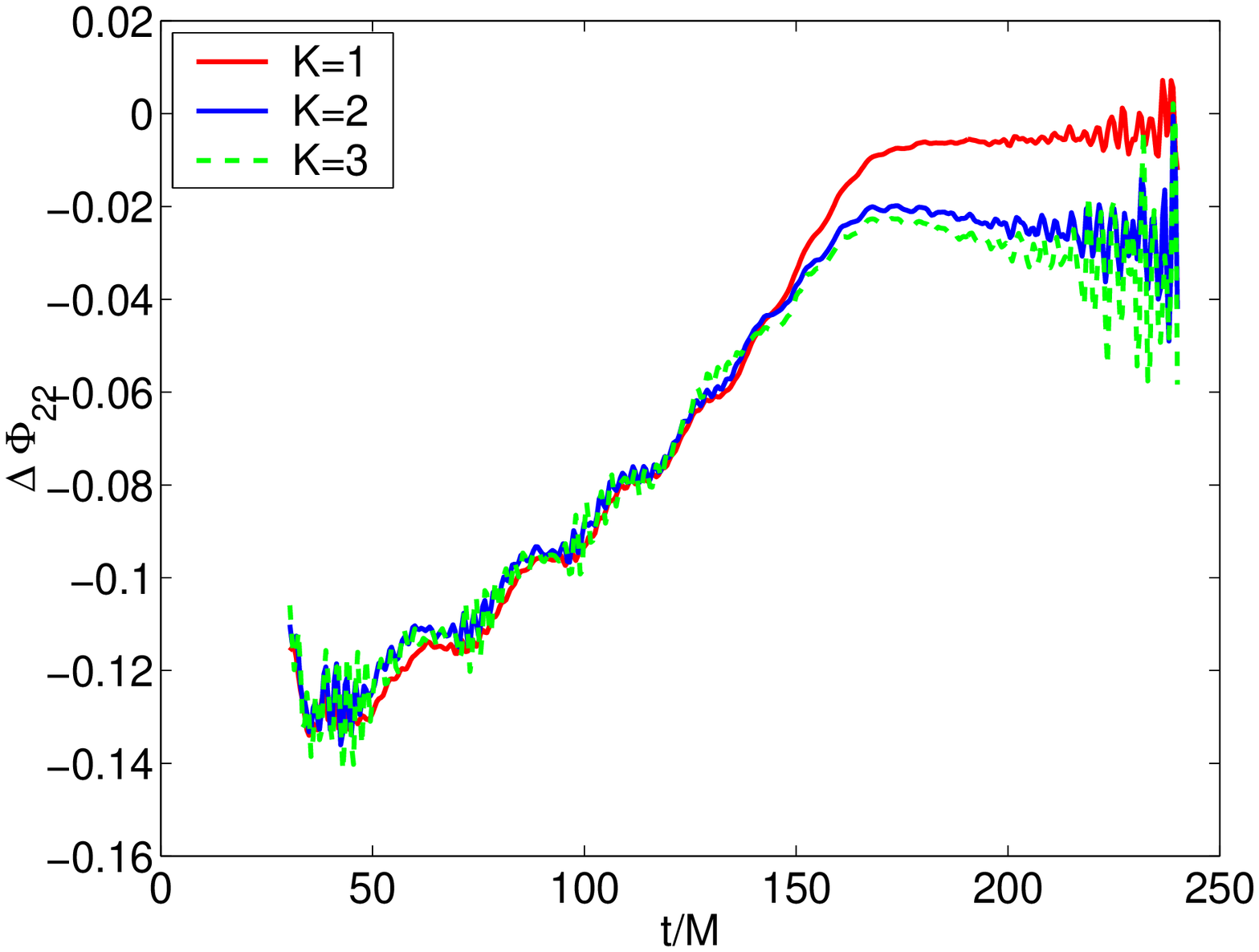}
  \caption{ \label{fig:extrapolation_boxshell} The BSSNOK box-shell 
    comparison. The top row shows the errors in extrapolated 
    waveforms when the wave zone is covered by Cartesian boxes as described 
    in the text. The lower row shows the shell results. The left panel shows 
    the amplitude and the right the phase of the~$(2,2)$ mode of $r\,\psi^4$.
    The amplitude is improved by at least an order of magnitude and the phase 
    by a factor around five.}
\end{figure*}

\paragraph*{Box-Shell comparison.} Let us briefly discuss code 
performance and waveform quality by comparing runs which employ 
spherical patches (``shell runs'' hereafter) in the wave zone to 
those that do not (``box runs'' hereafter). For shell runs, radial and 
angular resolution can be adjusted separately, at linear scaling in 
the number of grid points in radius and quadratic scaling for both 
angular directions. Nested boxes imply effectively constant angular 
resolution for increasing radius and decreasing radial resolution.  
Increasing radial resolution leads to a cubic scaling in the number of 
grid points, because angular resolution increases simultaneously.  
This can be partially compensated by choosing larger boxes in the 
wave zone, which are comparatively cheap in run-time because of 
Berger-Oliger time adaptivity, and typical runs are limited by 
run-time rather than memory.
Constant radial resolution is helpful for tracking waves traveling to
infinity, on the other hand decreasing radial resolution is a recipe
sometimes employed deliberately for filtering features going to and
coming from the outer boundary. An entirely different type of filter
is effectively in use when extracting only $(2,2)$ modes. Features
due to rectangular outer boundaries and rectangular refinement
boundaries (cmp.\ Fig.~\ref{fig:ss:rho_ham})
are not visible at $l=2$, but start to show at $l=4$. 
A spherical outer boundary has the advantage that a unique
normal vector to the boundary exists, in fact our early (and incomplete) 
experiments with the new Z4c boundary conditions failed for a cubical
outer boundary. However, a spherical boundary can also lead to a
focused inward reflection of outgoing waves, while a rectangular
boundary scatters spherical waves leading to a diffuse reflection.
For example, the back-reflection feature in Fig.~\ref{fig:sp:wav} is 
much smaller for cubical outer boundaries, as are 
oscillations of a neutron star induced by boundary reflections.
A clean treatment of a spherical boundary is nevertheless preferable
since the reflections can be minimized, and even though a constraint
violation from a cubical outer boundary may not be as visible, it is 
still present in the inner domain. 
The choice between box or shell runs actually depends on the waveform
accuracy goal, and requires a balance between accuracy and
computational cost. We discuss two examples employing the BSSNOK
formulation.

\paragraph*{BAM example.} As a first example, we find that BAM BBH 
box runs with~$L=9$, $L^{\rm mv}=4$,~$n^{\rm mv}=\{48,\,56,\,64\}$, 
$n=\{82,\,96,\,110\}$, and~$h_9=\{0.0365,\,0.0313,\,0.0293\}$,
give results comparable to shell runs with~$L=7$, $L^{\rm  mv}=2$, 
$n^{\rm mv}=\{48,\,56,\,64\}$,
$n=\{82,\,96,\,110\}$, $n_r=\{686,\,800,\,914\}$,
$n_{\theta,\phi}=\{14,\,16,\,18\}$, and
$h_6=\{0.0365,\,0.0313,\,0.0293\}$. Both series of simulations have
the same resolution in the Cartesian boxes and a similar computational
cost. The resolution in the wave zone differs a factor two in the wave
zone (lower in the box runs). The waveform errors and convergence
properties are the same in both series.

\paragraph*{AMSS-NCKU example.} As a second example we compare 
AMSS-NCKU BBH box runs with~$L=11$,~$L^{\rm   mv}=8$,~$n^{\rm mv}=
\{ 56 ,\, 64 \, 72 \}$, $n=\{ 112,\, 128 \, 144 \}$ 
and~$h_{11}=\{0.505/112,\, 0.505/128\, 0.505/144\}\,M$. With shell 
runs when the four coarsest levels are substituted by spherical 
patches that approximately extend to the same outer radius. The 
computational cost of the shell runs is approximately twice the 
box runs. Using this triplet, third order self-convergence is 
achieved in the shell runs at all the extraction radii, while in 
the box runs self-convergence is lower than three at small radii and
progressively degrades when the extraction is performed on coarser
levels. The waveform in box runs is more noisy. Phase and amplitude
of the waveform can be extrapolated according to
\begin{equation}
f(u,R)= \sum_{k=0}^{K} f(u) R^{-k}
\end{equation}
where~$f(u,R)$ is the quantity to extrapolate extracted at radius~$R$,
$u$ is a retarded time, and $R$ the Schwarzschild radius, see 
e.g.~\cite{BerThiBru11,BerNagZen11}. A measure of the extrapolation error is
the difference between the extrapolated function and the last radius
values. This error is reported in the upper row of 
Fig.~\ref{fig:extrapolation_boxshell} for box runs and the lower row for 
shell runs, using the radii $R = \{150,\, 140,\, 130,\, 120,\, 110,\, 100,\, 
90,\, 80,\,70,\, 60,\, 50\}$ and different values of~$K$, i.e.~different
polynomials. As evident from the figures the extrapolation for shell
runs is more accurate, and results for different $K$ are consistent
with each other (no oscillations and overshooting). The ``zig-zag''
behavior in the left panel of the lower row of 
Fig.~\ref{fig:extrapolation_boxshell} is due to the limited number of 
digits used in the output. For these data the choice~$K=1$ seems to be 
optimal.

\section{Evolution of Teukolsky waves}
\label{app:teukolsky}

\begin{figure*}[t]
  \begin{center}
    \includegraphics[width=0.49\textwidth]{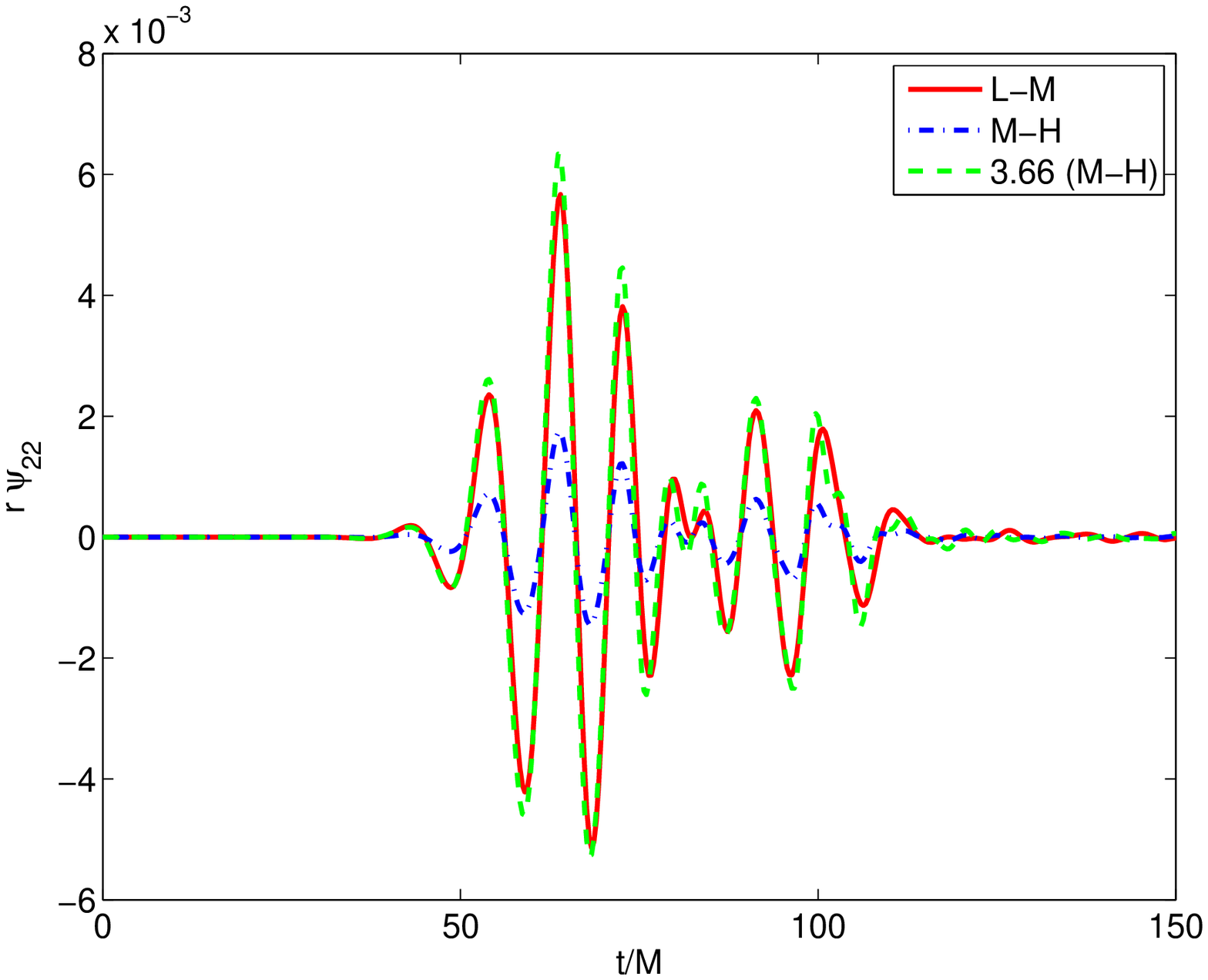}
    \includegraphics[width=0.49\textwidth]{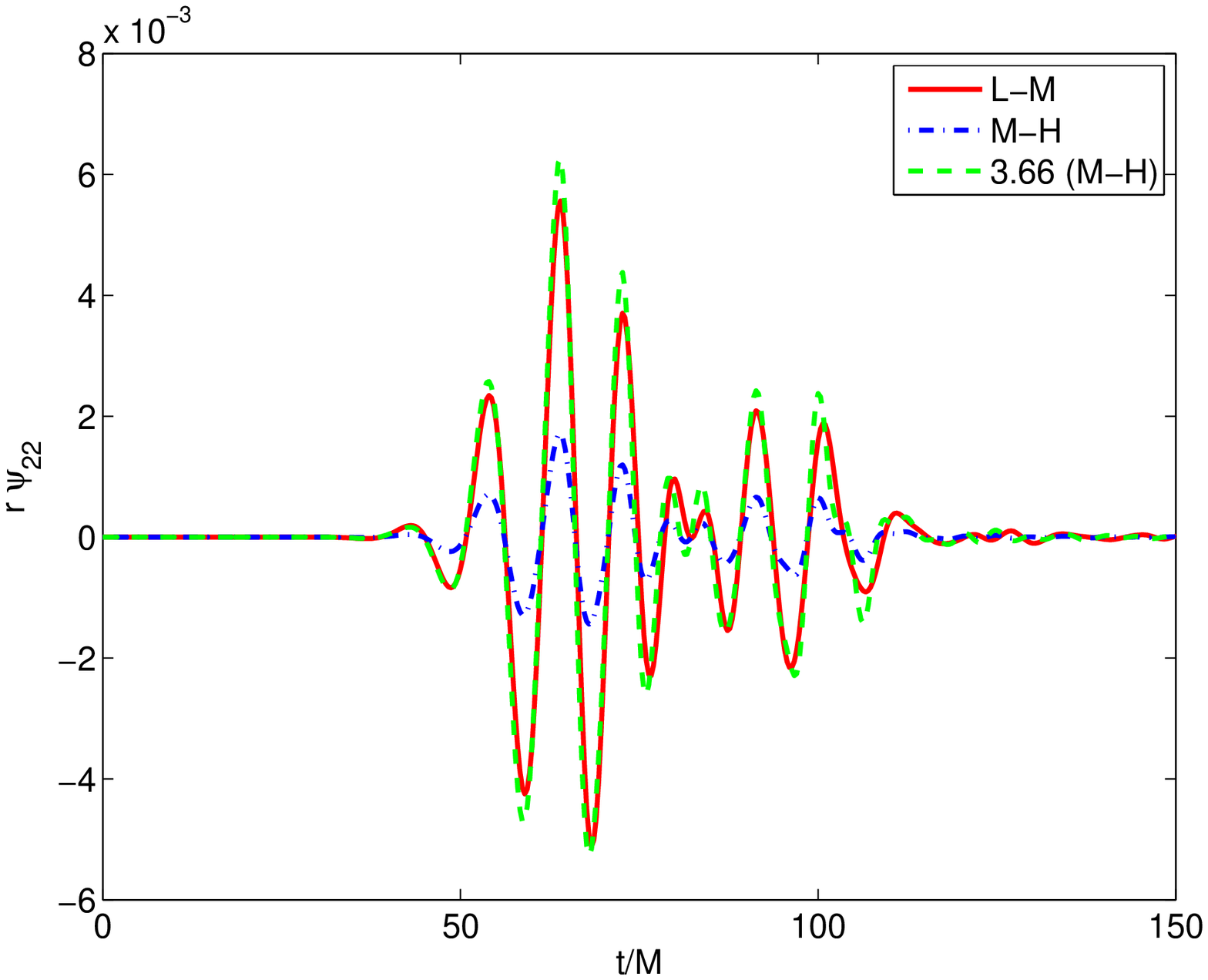}
    \caption{ \label{fig:teukwav1} Convergence plot for Teukolsky waves
      of the~$(2,2)$ mode of~$r\,\psi^4$ scaled for fourth order
      convergence. Convergence series with~$n=\{48,\,64,\,80\}$,
      and~$h_{4}=\{ 0.167, 0.125, 0.100 \}$. The left panel shows the
      BSSNOK data, the right the Z4c. Extraction radius is at~$r=80$.
    }
  \end{center}
\end{figure*}

In this appendix we evolve, for code validation, Teukolsky's wave
solution with the BAM code.

\paragraph*{Motivation.} The codes used in this work have sources
of error of different polynomial order in the grid spacing, but the
operations performed most often, finite differencing and time integration,
are performed at fourth order. It is perhaps not surprising that we 
never find clear fourth order convergence in our evolutions of compact 
binaries.

For the hydrodynamics simulations this behavior can be excused because the 
HRSC scheme we employ is only second order accurate. But for our vacuum 
simulations the situation is less clear. We therefore also studied the 
evolution of Teukolsky's solution~\cite{Teu82} to the linearized Einstein 
equations. This solution represents a weak gravitational wave propagating 
on a flat background. It satisfies the constraint equations of General 
Relativity at linear order. We use this solution as initial data and then 
evolve forward in time with the BAM code. Note that extensive convergence 
testing demonstrating fourth order convergence, in test cases, was shown 
for the AMSS-NCKU code in~\cite{CaoHil11}.

\paragraph*{Teukolsky wave initial data.} The metric has the form
\begin{align}
ds^2 &= -dt^2 + (1 + A f_{rr} )dr^2 + 2B f_{r\theta} r dr d\theta
\nonumber \\
&\quad  + 2B f_{r\phi} r \sin\theta dr d\phi
 +(1 + C f_{\theta\theta}^{(1)} + A f_{\theta\theta}^{(2)}) r^2 d\theta^2
\nonumber \\
&\quad + 2(A - 2C)f_{\theta\phi} r^2 \sin\theta d\theta d\phi
\nonumber \\
&\quad +(1 + C f_{\theta\theta}^{(1)} + A f_{\theta\theta}^{(2)})
   r^2 \sin^2\theta d\phi^2
\end{align}
where we have
\begin{align}
f_{rr}  &= \sin^2\theta  (\cos^2\phi - \sin^2\phi)\,, \\
f_{r\theta}  &= \sin\theta  \cos\theta  (\cos^2\phi - \sin^2\phi)\,, \\
f_{r\phi}  &= -2\sin\theta \sin\phi\cos\phi\,,
\end{align}
and also
\begin{align}
f_{\theta\theta}^{(1)} &= (1 + \cos^2\theta) (\cos^2\phi - \sin^2\phi)\,, \\
f_{\theta\theta}^{(2)} &= -(\cos^2\phi - \sin^2\phi)\,, \\
f_{\theta\phi}  &= 2\cos\theta  \sin\phi\cos\phi\,, \\
f_{\phi\phi}^{(1)} &= -f_{tt}^{(1)}\,, \\
f_{\phi\phi}^{(2)} &= \cos\theta \cos\theta (\cos^2\phi - \sin^2\phi)\,,
\end{align}
with
\begin{align}
A &= -3\left(F^{(2)}\frac{\lambda^5}{r^3} + 3 F^{(1)}\frac{\lambda^5}{r^4}
  + 3F\frac{\lambda^5}{r^5}\right)\,, \\
B &= \Big(F^{(3)}\frac{\lambda^5}{r^2} + 3F^{(2)}\frac{\lambda^5}{r^3}
  + 6F^{(1)}\frac{\lambda^5}{r^4} + 6F\frac{\lambda^5}{r^5}\Big)\,, \\
C &= -\frac{1}{4}\Big(F^{(4)}\frac{\lambda^5}{r}
  + 2F^{(3)}\frac{\lambda^5}{r^2} + 9F^{(2)}\frac{\lambda^5}{r^3} \nonumber \\
  & \quad + 21F^{(1)}\frac{\lambda^5}{r^4} + 21F\frac{\lambda^5}{r^5}\Big) .
\end{align}
This corresponds to an outgoing wave that is a pure~$m=2$ mode. The
generating function $F$ is given by
\begin{align}
F = a \frac{(t-r)^N}{\lambda^N}\exp(-\frac{(t-r)^2}{\lambda^2}) .
\end{align}
Derivatives of order~$n$ of~$F$ are denoted by~$F^{(n)}$. Apart from
choosing a different generating function our wave is very similar
to what was used in~\cite{FisBakMet05}. For our runs we have
chosen~$a=10^{-4}$, $N=10$ and $\lambda=10$.

\paragraph*{Setup.} We have evolved such a wave with an initial lapse
of one and a shift of zero. For the evolution, as in the rest of the 
work, we have chosen the standard 1+log and Gamma driver evolution 
equation for lapse and shift~(\ref{eqn:BMlapse}-\ref{eqn:Gamdriver}) 
with $\mu_L=2/\alpha\,,\mu_S=1/\alpha^2$. We use four levels of box mesh 
refinement, and attach the spherical grids at~$r\sim 30$. In the lowest 
resolution runs each box has~$n=48$ points per direction, and the 
resolution of level~$l=4$ is~$h_4=0.167$. We choose~$n_{\theta,\,\phi}=n/2$ 
angular and~$n_r=n$ radial points in each spherical patch so that the 
outer boundary is located at~$r=100$. Runs at resolutions~$n=48,\,64,\,80$ 
(with the grid spacing scaled in order to maintain the same grid setup) 
are performed.

\paragraph*{Results.} We find that both the BSSNOK and the
Z4c system can successfully be used to evolve these waves.
For Z4c we have set $\kappa_1=0.02$ and $\kappa_2=0$.
Since there are no strong gravitational fields
we can extract the waves at any radius.
Figure~\ref{fig:teukwav1} shows a convergence plot for waves extracted at
a radius of $r=80$. The solid lines shows the difference between
$\psi^4$ at low and medium resolution ($L-M$),
while the dash dotted lines show the difference between medium
and high resolution ($M-H$). When scaled with the proper factor of 3.66 for
fourth order convergence, we see that $M-H$ coincides approximately
with $L-M$. For this data set the observed order of convergence is 
around~$3.5$. The left panel shows the results for BSSNOK, the right 
panel shows those for Z4c.

\bigskip

\bibliographystyle{apsrev}

\end{document}